\documentclass[aps,prd,nofootinbib,showpacs,superscriptaddress,twocolumn,floatfix]{revtex4-2}

\usepackage[utf8]{inputenc}
\usepackage{mathrsfs}
\usepackage{bm}
\usepackage{bbm}
\usepackage{indentfirst}
\usepackage{epstopdf}
\usepackage{color}
\usepackage{xcolor}
\usepackage{amssymb, amsmath,color, hyperref, graphicx}
\usepackage{braket}
\usepackage{placeins}
\usepackage{multirow}
\usepackage{slashed}
\usepackage{physics}
\usepackage{booktabs}

\begin{document}

\title{Chiral properties of $(2\!+\!1)$-flavor QCD in magnetic fields at zero temperature}

\author{Heng-Tong Ding}
\affiliation{Key Laboratory of Quark and Lepton Physics (MOE) and Institute of
    Particle Physics, \\
    Central China Normal University, Wuhan 430079, China}
\author{Dan Zhang}
\affiliation{Key Laboratory of Quark and Lepton Physics (MOE) and Institute of
    Particle Physics, \\
    Central China Normal University, Wuhan 430079, China}

% \date{\today}

\begin{abstract}
We present a lattice QCD study of the chiral properties of $(2\!+\!1)$-flavor QCD in background magnetic fields at zero temperature with physical pion masses. Simulations are performed using the highly improved staggered quark action across four different lattice spacings to enable a controlled continuum extrapolation. We compute the renormalized chiral condensates together with pseudoscalar meson masses and decay constants for pions, kaons, and the fictitious $\eta^0_{s\bar{s}}$ pseudoscalar as functions of the magnetic-field strength $eB$ up to $eB\simeq1.2$~$\mathrm{GeV}^2$. 
The chiral condensates exhibit clear magnetic catalysis, increasing monotonically with the field strength. In the meson sector, neutral pseudoscalar masses decrease steadily with $eB$, whereas charged pseudoscalar masses display a nonmonotonic response: They rise at small fields, consistent with the lowest-Landau-level expectation, but then saturate and slightly decrease at larger fields, signaling sizable internal-structure effects. At the same time, neutral pseudoscalar decay constants are strongly enhanced by the magnetic field. To quantify deviations from chiral symmetry relations, 
we isolate the magnetic-field-induced shift in the Gell-Mann--Oakes--Renner corrections and find it to remain small for the neutral pion but to become sizable for the neutral kaon.  To elucidate the origin of the magnetic response, we separately analyze the sea- and valence-quark contributions to both neutral and charged meson masses, finding that valence effects dominate at zero temperature. These results provide new insights into the interplay between QCD chiral symmetry breaking and strong magnetic fields.
\end{abstract}

\maketitle

\section{Introduction}
\label{sec:intro}
Quantum chromodynamics (QCD) in background magnetic fields has attracted sustained interest over the past decade.
Strong fields can be created transiently in off-central heavy-ion collisions~\cite{Kharzeev:2007jp,Skokov:2009qp,Deng:2012pc}, and they may also play a role in the early Universe~\cite{Vachaspati:1991nm} and in compact stars such as magnetars~\cite{Duncan:1992hi}.
These settings motivate first-principles studies of how a magnetic field modifies QCD dynamics across energy scales, from the chiral condensate in the vacuum to hadron properties and, at finite temperature, the nature of the QCD transition; see also the recent reviews~\cite{Hattori:2023egw,Endrodi:2024cqn,Yamamoto:2021oys}.

A central theme is the magnetic-field dependence of chiral symmetry breaking.
At zero temperature, lattice QCD calculations established that the chiral condensate is enhanced by the magnetic field, a phenomenon known as magnetic catalysis~\cite{DElia:2010abb,Shovkovy:2012zn}.
It is then tempting to expect that stronger chiral symmetry breaking in the vacuum should be accompanied by an increase of the pseudocritical temperature ($T_{pc}$) with the magnetic-field strength ($eB$), and such a trend was indeed reported in early finite-cutoff studies using standard staggered fermions~\cite{DElia:2010abb,Ding:2020inp}.
However, continuum-extrapolated simulations with improved staggered fermions later revealed the opposite behavior: $T_{pc}$ decreases with $eB$~\cite{Bali:2011qj}, and the condensate is suppressed in the vicinity of the transition region, giving rise to the so-called inverse magnetic catalysis~\cite{Bali:2011qj,Bali:2012zg}.
This behavior is now widely understood to be driven predominantly by sea-quark effects, has been corroborated in subsequent studies employing improved discretizations~\cite{Ilgenfritz:2013ara,Bornyakov:2013eya,Bali:2014kia,Tomiya:2019nym} and has stimulated extensive model and theoretical work aimed at understanding the underlying mechanism and its relation to deconfinement and the thermal transition~\cite{DElia:2011koc,Shovkovy:2012zn,Andersen:2014xxa,Kojo:2012js,Bruckmann:2013oba,Fukushima:2012kc,Ferreira:2014kpa,Yu:2014sla,Feng:2015qpi,Li:2019nzj,Mao:2016lsr,Gursoy:2016ofp,Xu:2020yag,Abreu:2022cgm,Li:2025wqb,Mao:2024rxh}.
Effective descriptions also continue to refine the magnetic-field dependence of the condensate beyond leading order, e.g., within chiral perturbation theory~\cite{Hofmann:2020ism}.
It has also been suggested that inverse magnetic catalysis is not necessarily tied one to one to the reduction of $T_{pc}$ and may be more directly connected to deconfinement-related effects~\cite{DElia:2018xwo,Endrodi:2019zrl,Bonati:2016kxj}.
A detailed understanding of the magnetic response, therefore, requires simultaneously tracking the chiral order parameter and the hadronic channels that encode the low-energy realization of chiral symmetry.

In the vacuum, the interplay between chiral symmetry breaking and pseudoscalar mesons is encapsulated by the Gell-Mann--Oakes--Renner (GMOR) relation~\cite{Gell-Mann:1968hlm}.
At vanishing magnetic field, the leading-order relation links the light-quark condensate to the pion mass and decay constant, with well-studied higher-order corrections~\cite{Jamin:2002ev,Bordes:2010wy,Bordes:2012ud}, and analogous relations exist in the three-flavor theory~\cite{Gasser:1984gg}.
Lattice QCD has confirmed the GMOR relation in the vacuum at $eB = 0$~\cite{Boucaud:2007uk,Engel:2014cka}.
In a background magnetic field, the GMOR relation becomes an especially incisive diagnostic: It ties the magnetic catalysis of the condensate to the magnetic response of pseudoscalar masses and decay constants, and it tests whether magnetic-field effects can be organized within (extended) chiral dynamics.
Chiral perturbation theory supports the robustness of the neutral-pion GMOR relation in the weak-field regime, both at low temperature and at $T=0$~\cite{Gasser:1986vb,Shushpanov:1997sf,Agasian:2001ym}, and recent effective theory analyses further constrain pion matrix elements and decay observables in a magnetic background~\cite{Adhikari:2024vhs}.
It remains nontrivial to assess the quantitative validity of GMOR-type relations at stronger fields and at physical quark masses, where additional dynamical scales enter and higher-order effects may become important. 
Given that lighter pions generally correlate with a lower transition temperature at $eB=0$~\cite{Ding:2019prx,Ding:2020rtq,Bazavov:2017xul,Kotov:2021rah,Aarts:2020vyb,Bhattacharya:2014ara,Umeda:2016qdo,Li:2020wvy}, it is natural to ask to what extent the magnetic-field dependence of (neutral) Goldstone modes may help elucidate the reduction of $T_{pc}$ at nonzero $eB$.

Hadron spectroscopy in magnetic fields is also of intrinsic interest~\cite{Hattori:2023egw,Endrodi:2024cqn,Yamamoto:2021oys,Andersen:2014xxa,Hattori:2015aki,Avancini:2016fgq,Wang:2017vtn,Mao:2018dqe,Avancini:2018svs,Coppola:2019uyr,Cao:2019res,Xu:2020sui,Xu:2020yag,Kojo:2021gvm,Xing:2021kbw,Mei:2022dkd,Coppola:2023mmq,Wen:2023qcz,Dominguez:2018njv,Ayala:2018zat,Colucci:2013zoa}.
For pointlike charged particles, Landau-level quantization implies a monotonically increasing energy with $eB$, while pointlike neutral particles would be insensitive to the field at the level of their rest mass.
QCD bound states, however, are neither pointlike nor insensitive: The magnetic field couples directly to quarks, breaks isospin explicitly, and distorts the internal structure of mesons.
As a result, both neutral and charged pseudoscalar channels provide controlled settings to study how compositeness competes with the point-particle Landau-level expectation.
In particular, the behavior of charged pseudoscalar mesons at intermediate and strong fields can reveal when the point-particle picture breaks down and internal quark dynamics becomes essential.
Moreover, via QCD inequalities, the masses extracted from the flavor-resolved connected contributions to the neutral-pion correlator constrain other channels; for instance, the sum $M_{\pi^0_u}+M_{\pi^0_d}$ provides a lower bound on the charged $\rho$-meson mass~\cite{Hidaka:2012mz}, which is relevant in discussions of possible $\rho$-condensation scenarios at strong fields~\cite{Chernodub:2010qx,Chernodub:2011mc}.

Lattice studies of mesons in background magnetic fields have been actively pursued over the past decade, but several limitations have persisted.
Many investigations of the meson spectrum were carried out in the quenched approximation and/or with valence-only magnetic fields, using Wilson or overlap fermions~\cite{Luschevskaya:2012xd,Hidaka:2012mz,Luschevskaya:2014lga,Luschevskaya:2015cko,Bali:2017ian}.
Earlier quenched results exhibited discrepancies for the neutral-pion mass, which were later understood to be tied to discretization-specific issues (notably the magnetic-field dependence of the Wilson hopping parameter)~\cite{Hidaka:2012mz,Bali:2017ian}.
In full $N_f=2+1$ simulations, existing work has largely focused on the pion sector and on selected observables~\cite{Bali:2011qj,Bali:2017ian}.
Studies of decay constants in magnetic fields have also revealed additional structure, such as the appearance of an extra charged-pion decay constant related to the vector current in the presence of $B$~\cite{Bali:2018sey}, and the general decomposition of pion-to-vacuum matrix elements into vector and axial-vector components has been discussed and modeled~\cite{Coppola:2018ygv,Coppola:2019uyr,Coppola:2019idh,Adhikari:2024vhs}.
Nevertheless, continuum-extrapolated spectroscopy and decay-constant results at the physical point have remained scarce; see also Refs.~\cite{Endrodi:2024cqn,Yamamoto:2021oys} for broader summaries of recent lattice progress in external fields.
In our earlier work at a single lattice spacing and heavier-than-physical pion mass~\cite{Ding:2020hxw}, we explored chiral properties and pseudoscalar spectroscopy in strong magnetic fields; the natural next step is to remove these systematics and establish continuum-extrapolated benchmarks at the physical point.

In this paper, we present a lattice QCD study of the chiral properties of $(2\!+\!1)$-flavor QCD in background magnetic fields at $T\simeq 0$, performed with the highly improved staggered quark (HISQ) action at physical quark masses and four lattice spacings to enable controlled continuum extrapolations.
We compute renormalized chiral condensates and determine pseudoscalar meson masses and decay constants for pions, kaons, and the hidden-strangeness $\eta^0_{s\bar s}$ channel as functions of $eB$.
We extract ground-state masses with both the small-sample corrected Akaike information criterion (AICc)-selected multistate fits~\cite{cavanaugh1997unifying,Akaike:1974vps} and the oblique Lanczos method~\cite{Wagman:2024rid,Ostmeyer:2024qgu}, and we include the spread between these determinations as a systematic uncertainty.
Furthermore, we assess magnetic-field-induced corrections to the GMOR relation and disentangle valence- and sea-quark contributions to the magnetic response of both neutral and charged observables.

The remainder of this paper is organized as follows.
In Sec.~\ref{sec:basics} we introduce the basic observables and definitions used throughout this work.
Section~\ref{sec:setup} describes the lattice setup, the implementation of the background magnetic field, and our analysis strategy for extracting ground-state quantities.
In Sec.~\ref{sec:results} we present our main results: Sec.~\ref{subsec:mass} discusses the masses of pseudoscalar mesons, including the role of valence- and sea-quark effects; Sec.~\ref{subsec:neutral-dc} presents the decay constants of neutral pseudoscalar mesons; and Sec.~\ref{subsec:pbp-GMOR} reports results for the chiral condensates and analyzes the GMOR relation and its magnetic-field dependence.
We finally conclude in Sec.~\ref{sec:conclusion}.
Additional details on the mass extraction using two methods and on the continuum extrapolations are collected in Appendix~\ref{app:aicc_lanczos_comparison} and \ref{app:cont_extr}, respectively, and tables of our numerical results are compiled in Appendix~\ref{app:summary_data_tables}.

\section{Temporal correlators, masses of pseudoscalar mesons, decay constants and chiral condensates}
\label{sec:basics}
In lattice QCD, hadronic observables are obtained from two-point correlation functions evaluated in Euclidean spacetime. For a given hadronic channel $H$, the meson correlator $G_H(B;\tau)$ is defined as the expectation value of a hadronic operator $\mathcal{M}_H$ evaluated at Euclidean time $\tau$ and its Hermitian conjugate at time 0:
\begin{equation}
  G_H(B;\tau)=\sum_{\vec{x}}
  \left\langle \mathcal{M}_H(\vec{x},\tau)\,
  \mathcal{M}_H^\dagger(\vec{0},0)\right\rangle_{B},
  \label{eq:GH_def}
\end{equation}
where $\langle\cdots\rangle_{B}$ denotes the expectation value with respect to
the $B$-dependent path-integral measure,
\begin{equation}
\begin{aligned}
    \langle \mathcal{O}\rangle_{B} &\equiv
        \frac{1}{Z(B)}\int\!\mathcal{D}U\, e^{-S_g[U]}\,
        \det D(B)\,\mathcal{O}[U], \\
    \det D(B) &\equiv \prod_{f=u,d,s}\det D_f(U;q_fB,m_f).
\end{aligned}
\label{eq:expB_def}
\end{equation}
Here $Z(B)$ is the corresponding partition function, and $D_f$ denotes the lattice Dirac (fermion) matrix for quark flavor $f$. The interpolating operator $\mathcal{M}_H(\vec{x},\tau) = \bar{\psi}(\vec{x},\tau) \Gamma_H \psi(\vec{x},\tau)$ creates a meson state with quantum numbers specified by $\Gamma_H$. For example, $\Gamma_H=\gamma_5$ corresponds to the pseudoscalar channel, while $\Gamma_H=\gamma_\mu$ corresponds to the vector channel. 

Inserting a complete set of energy eigenstates yields the following spectral representation:
\begin{equation}
    G_H(B;\tau) = \sum_{n} |\langle \Omega |\mathcal{M}_H(0)|n\rangle|^2 e^{-E_n(B)\tau},
      \label{eq:GH_spectral}
\end{equation}
where the sum runs over all intermediate states $|n\rangle$ with the same quantum numbers as $\mathcal{M}_H$, and $E_n(B)$ denotes the corresponding energy.
Assuming the vacuum energy is normalized to zero, at large Euclidean times, excited-state contributions are exponentially suppressed, and the correlator is dominated by the ground state,
\begin{equation}
    \lim_{\tau \to \infty} G_H(B;\tau) \sim e^{-M_H(B) \tau},
    \label{eq:GH_asympt}
\end{equation}
where $M_H(B)$ is the mass of the lightest state in the channel specified by $\Gamma_H$.

For staggered fermions, the meson operator takes the form $\bar{\psi}(x)(\Gamma_D\otimes\Gamma_T^*)\psi(x)$, where $\Gamma_D$ and $\Gamma_T$ are Dirac and taste matrices, respectively. Because of the mixing of spin and taste degrees of freedom on the hypercube, the fermion field $\psi$
becomes a 16-component object. In this work, we focus on local meson operators, setting $\Gamma_D=\Gamma_T\equiv\Gamma_H$. The corresponding operator can then be written in terms of the staggered spinor field $\chi$, as $\mathcal{M}_H=\zeta_H(\vec{x})\bar{\chi}(\vec{x})\chi(\vec{x})$, where $\zeta_H(\vec{x})$ is a position-dependent phase factor determined by the choice of $\Gamma_H$. 
To evaluate \autoref{eq:GH_def} numerically, we first perform the Grassmann integration over the quark fields on a fixed gauge configuration $U$. This yields the configuration-specific correlator, denoted as $\mathcal{G}_H(B;\tau; U)$, which involves the trace of quark propagators. For the connected part of the correlator, this is given by
\begin{equation}
\mathcal{G}_H(B;\tau; U) = \sum_{\vec{x}} \zeta_H(\vec{x}) \left\| S(\vec{x},\tau; \vec{0},0; U,B) \right\|^2,
\label{eq:Gconf_def}
\end{equation}
where $S(\vec{x},\tau;\vec{0},0;U,B)$ is the staggered fermion propagator from the source $(\vec{0},0)$ to the sink $(\vec{x},\tau)$ on the background field $U$ in the presence of magnetic field $B$, and the notation $\left\| \cdot \right\|^2$ denotes the squared Frobenius norm, $\left\| A \right\|^2 \equiv \operatorname{Tr}(A^\dagger A)$, with trace taken over color space. 

The physical correlator $G_H(B;\tau)$ is then recovered by averaging $\mathcal{G}_H(B;\tau; U)$ over the gauge ensemble:
\begin{equation}
  G_H(B;\tau)=\big\langle \mathcal{G}_H(B;\tau;U)\big\rangle_{B}.
  \label{eq:Gavg_def}
\end{equation}
Here $\langle\cdots\rangle_B$ denotes the gauge-field average over $U$ with the
$B$-dependent path-integral measure defined in \autoref{eq:expB_def}.

We consider mesons in the pseudoscalar channel, constructed from flavor combinations $\bar{q}_iq^j$ with $i,j\in\{u,d,s\}$. For these states, the phase factor simplifies to $\zeta_H(\vec{x}) = 1$~\cite{Kilcup:1986dg,Bazavov:2019www}.
The presence of background magnetic fields explicitly breaks isospin symmetry due to the differing electric charges of the $u$ and $d$ quarks, which allows mixing between the $u\bar{u}$ and $d\bar{d}$
components of the neutral pion. As a result, the $\pi^0$ correlation function can, in general, receive both connected and disconnected contributions. 
In this work, we take the up- and down-quark masses to be degenerate at zero magnetic field and neglect the disconnected diagrams. Earlier exploratory lattice studies in background magnetic fields found the disconnected part of the neutral-pion correlator to be consistent with zero within their uncertainties in a quenched setup~\cite{Luschevskaya:2015cko}. The quantitative impact of disconnected contributions in the present setup is not addressed here, and the neutral-pion results presented below, therefore, refer to the connected correlators.

A distinctive feature of staggered fermions is an alternating contribution in Euclidean time arising from opposite-parity states. The correlator is, thus, parameterized as~\cite{Bazavov:2019www,Ding:2020hxw}
\begin{equation}
    G_H(n_\tau) = \sum_{i=1}^{N_{\rm nosc}} A_{\rm nosc,i} e^{-M_{\rm nosc,i} n_\tau} - (-1)^{n_\tau} \sum_{i=0}^{N_{\rm osc}} A_{\rm osc,i} e^{-M_{\rm osc,i} n_\tau},
\label{eq:oscillatory}
\end{equation}
where $n_\tau=\tau/a$  is the temporal separation in lattice units, $N_{\rm nosc}$ and $N_{\rm osc}$ are the numbers of nonoscillating and oscillating states, respectively, $A_{\rm nosc,i}$ and $A_{\rm osc,i}$ are the corresponding positive amplitudes, and $M_{\rm nosc,i}$ and $M_{\rm osc,i}$ denote their masses. Note that $N_{\rm nosc} \ge 1$ is required to describe the ground state, while $N_{\rm osc} \ge 0$. In our analysis, the mass $M_{\rm nosc,1}$ is identified as the ground-state pseudoscalar meson mass.

We note that, in a background magnetic field, the reduced symmetry of the system can, in principle, allow mixing between channels that are orthogonal at $B=0$. In particular, pseudoscalar states may mix with vector states with spin projection $s_z=0$, as discussed, for example, in Ref.~\cite{Bali:2017ian}. Such mixing primarily affects the extraction of the heavier mixed partner, while the lightest pseudoscalar state still governs the leading large-Euclidean-time falloff of the pseudoscalar correlator. Since in the present work we determine only the lowest pseudoscalar ground state, we do not observe a visible complication from this effect within our current uncertainties.

To extract decay constants from the fit \textit{Ansatz} in \autoref{eq:oscillatory}, we first recall the relevant definitions. In the presence of a background magnetic field, Lorentz and isospin symmetries are explicitly broken. Consequently, the weak decay of pions receives contributions from both axial-vector and vector currents, leading to the introduction of additional decay constants~\cite{Fayazbakhsh:2013cha,Bali:2018sey,Coppola:2018ygv,Adhikari:2024vhs}. However, in this work, we focus solely on the conventional decay constants for the neutral pion and kaon, associated with the axial-vector current parallel to the magnetic field at zero spatial momentum. Their definitions remain identical to those at zero magnetic field~\cite{Kilcup:1986dg,Ding:2020hxw}:
\begin{equation}
\begin{aligned}
\sqrt{2} M_{\pi^0}^2 f_{\pi^0} = 
&(m_u + m_d) \times\\
&\left\langle 0 \left| \frac{1}{\sqrt{2}}(\bar u \gamma_5 u - \bar d \gamma_5 d) \right| \pi^0(\vec{p}=0) \right\rangle,
\end{aligned}
\label{eq:fpi_def_modified}
\end{equation}
\begin{equation}
\sqrt{2} M_{K^0}^2 f_{K^0} = (m_d + m_s) \left\langle 0 \left| \bar{d} \gamma_5 s \right| K^0(\vec{p}=0) \right\rangle.
\label{eq:fk_def_modified}
\end{equation}

Since disconnected contributions are omitted in this work, the connected neutral-pion correlator is evaluated as the average of the $\bar{u}\gamma_5 u$ and $\bar{d}\gamma_5 d$ channels. It is, thus, convenient to introduce the flavor components $\pi^0_u$ and $\pi^0_d$, whose decay constants are defined by\begin{equation}
\sqrt{2} M_{\pi_{u}^{0}}^2 f_{\pi_{u}^{0}} = 2m_u \left\langle 0 \left| \bar{u} \gamma_5 u \right| \pi_{u}^0(\vec{p}=0) \right\rangle,
\label{eq:fu_define}
\end{equation}
\begin{equation}
\sqrt{2} M_{\pi_{d}^{0}}^2 f_{\pi_{d}^{0}} = 2m_d \left\langle 0 \left| \bar{d} \gamma_5 d \right| \pi_{d}^0(\vec{p}=0) \right\rangle.
\label{eq:fd_define}
\end{equation}
These states, $|\pi^0_u\rangle$ and $|\pi^0_d\rangle$, are not eigenstates of QCD but are defined implicitly via the long-distance behavior of the corresponding connected correlators.

The matrix elements in Eqs.~\eqref{eq:fpi_def_modified}--\eqref{eq:fd_define} are extracted from the amplitudes of the two-point correlation functions. For a pseudoscalar operator $\mathcal{O}_P$ coupling to a state $|P(\vec{p}=0)\rangle$, the spectral decomposition implies that the amplitude $A_P$ obtained from fitting $G_P(\tau) \sim A_P e^{-M_P \tau}$ is~\cite{Gupta:1997nd}
\begin{equation}
    A_P = \frac{\big|\big\langle \Omega \big| \mathcal{O}_P \big| P(\vec{p}=\vec{0}) \big\rangle\big|^2}{2M_P},
    \label{eq:AP_def}
\end{equation}
where $M_P$ is the mass of the state. Combining this with the definitions above yields the expressions used to compute the decay constants on the lattice:
\begin{equation}
    f_{P^\pi} = 2m_l\sqrt{\frac{1}{4}}\sqrt{\frac{A_{P^\pi}}{M_{P^\pi}^3}},
    \label{eq:fPpi_lat}
\end{equation}
\begin{equation}
    f_{K^0} = \left(m_d+m_s\right)\sqrt{\frac{1}{4}}\sqrt{\frac{A_{K^0}}{M_{K^0}^3}},
    \label{eq:fK0_lat}
\end{equation}
where $P^\pi$ denotes $\pi^0$, $\pi^0_u$, or $\pi^0_d$, and $m_l \equiv m_u = m_d$. Both amplitudes $A_P$ and masses $M_P$ are determined from the fits to \autoref{eq:oscillatory}. The factor $\sqrt{1/4}$ accounts for the normalization of local staggered fermion operators, specifically the factor of $1/4$ arising from the fourth-root trick and taste multiplicity.
We emphasize that $\pi_u^0$ and $\pi_d^0$ are not physical particles but rather constructs corresponding to the flavor components of the connected neutral-pion correlator. Nevertheless, their masses $M_{\pi_u^0}$ and $M_{\pi_d^0}$ are well-defined quantities that provide a lower bound on the mass of the charged $\rho$ meson through QCD inequalities~\cite{Hidaka:2012mz}. Their associated decay constants $f_{\pi_u^0}$ and $f_{\pi_d^0}$, defined via \autoref{eq:fu_define} and \autoref{eq:fd_define}, offer insight into the magnetic catalysis effect on individual quark flavors.

The energy spectrum of a pointlike charged particle in a uniform magnetic field at zero temperature is quantized into Landau levels, given by~\cite{Johnson:1949wmq,Canuto:1968apg}
\begin{equation}
    E_{n}^{2} = M^{2} + (2n + 1)|eB| - g s_{z} qB + p_{z}^{2}, \quad n \in \mathbb{Z}_{0}^{+},
\end{equation}
where $M$ is the particle mass in the absence of a magnetic field, $q$ is the electric charge, and $s_{z}$ is its spin polarization along the direction of the magnetic field (assumed to be the $z$ axis). The gyromagnetic ratio $g$ is 0 for pseudoscalar mesons and 2 for vector mesons. For a charged pointlike pseudoscalar meson ($g = 0$) in the lowest Landau level (LLL, $n = 0$) and with zero momentum along the $z$ direction ($p_z = 0$), the mass evolves with the magnetic field as
\begin{equation}
    M_{\mathrm{ps}}^{\pm}(B) = \sqrt{\left(M_{\mathrm{ps}}^{\pm}(B = 0)\right)^{2} + |eB|}\,.
    \label{eq:LLLapproximation}
\end{equation}
This expression implies that charged pointlike particles in the pseudoscalar channel exhibit a monotonic increase in mass with increasing magnetic field, whereas neutral pointlike particles are expected to remain unaffected.

The chiral condensate is the standard order parameter for spontaneous chiral
symmetry breaking in the chiral limit and remains a sensitive diagnostic at
finite quark masses. The (bare) chiral condensate for flavor $f$ is obtained from the derivative of the logarithm of the partition function with respect to the quark mass:
\begin{equation}
\langle\bar{\psi}\psi\rangle_f \equiv \frac{T}{V} \frac{\partial \ln {Z}(B, T)}{\partial m_f},
\end{equation}
where $V$ is the spatial volume, $T$ is the temperature, and $m_f$ denotes the mass of a specific quark flavor---up, down, or strange.
While the chiral condensate suffers from both additive and multiplicative ultraviolet (UV) divergences, these divergences are independent of the background magnetic field $B$ at zero temperature~\cite{Bali:2011qj}. To isolate the magnetic-field response, we consider subtracted and normalized condensates that are free of these UV artifacts.

For the light sector we define~\cite{Bali:2012zg}
\begin{equation}
    \begin{aligned}
        \Delta \Sigma_{ud}(B) =\frac{m_l}{M_{\pi^0}^2 F_{\pi^0}^2}  \left[ \left(\langle\bar{\psi}\psi\rangle_u(B) - \langle\bar{\psi}\psi\rangle_u(B=0)\right) 
        \right.\\
        + \left. \left(\langle\bar{\psi}\psi\rangle_d(B) - \langle\bar{\psi}\psi\rangle_d(B=0)\right) \right],        
    \end{aligned}
    \label{eq:DeltaSigma_l}
\end{equation}
where $m_l\equiv m_u=m_d$ and $M_{\pi^0}$ and $F_{\pi^0}$ denote the pion mass and decay constant at $B = 0$, respectively.

Analogously, for the kaon sector we define
\begin{equation}
    \begin{aligned}
        \Delta \Sigma_{sd}(B) = \frac{m_s+m_d}{2M_{K^0}^2 F_{K^0}^2} \left[ \left(\langle\bar{\psi}\psi\rangle_s(B) - \langle\bar{\psi}\psi\rangle_s(B=0)\right) 
        \right.\\
        + \left. \left(\langle\bar{\psi}\psi\rangle_d(B) - \langle\bar{\psi}\psi\rangle_d(B=0)\right) \right],        
    \end{aligned}
    \label{eq:DeltaSigma_sd}
\end{equation}
with $M_{K^0}$ and $F_{K^0}$ again taken at $B=0$.

\section{Numerical setup}
\label{sec:setup}
In this study, we perform lattice QCD simulations with $N_f=2+1$ flavors using HISQ~\cite{Follana:2006rc} fermions and a tree-level improved Symanzik gauge action~\cite{Luscher:1984xn,Weisz:1982zw} in the presence of external magnetic fields. 
The magnetic field couples only to the quarks and is represented by a phase factor $u_\mu(n)$ of the $ \rm U(1)$ field  introduced along the $z$ direction, i.e., $\vec{B}=\left(0, 0, B\right)$. The phase factor $u_\mu(n)$, expressed in the Landau gauge, is given by~\cite{Smit:1986fn,Al-Hashimi:2008quu}
\begin{equation}
    \begin{aligned}
        & u_x(n_x, n_y, n_z, n_\tau) =\\ 
        & \qquad\qquad\begin{cases} 
            \exp[-i q a^2 B N_x n_y] & (n_x = N_x - 1) \\ 
            1 & (\text{otherwise})
        \end{cases}, \\
        & u_y(n_x,n_y,n_z,n_\tau)
       =\exp[iqa^2Bn_x], \\
        & u_z(n_x,n_y,n_z,n_\tau)=u_t(n_x,n_y,n_z,n_\tau)=1.
    \end{aligned}
    \label{eq:u_mu_factor}
\end{equation}
Here, $q$ is the electric charge of the quark, and the lattice size is denoted as $\left(N_x, N_y, N_z, N_\tau \right)$, where the coordinates are labeled $n_\mu=0,\dots,N_\mu-1\left(\mu=x, y, z, \tau\right)$. The external magnetic field applied along the $z$ direction is quantized as follows~\cite{DElia:2010abb,Bali:2012zg,Ding:2020hxw}:
\begin{equation}
    qB = \frac{2\pi N_b}{N_x N_y}a^{-2},
    \label{eq:B-field}
\end{equation}
where $N_x\ (N_y)$  is the number of lattice sites in the $x\ (y)$ direction. In our setup, the lattices are isotropic in space, i.e., $N_x = N_y = N_z \equiv N_\sigma$.
The electric charges of the quarks are $q_d=q_s=-q_u/2=-e/3$, with $e$ representing the elementary electric charge. To satisfy the quantization condition for all quark flavors simultaneously, the magnetic flux quantum is set by the greatest common divisor of their charges, leading to $\left\lvert q_d\right\rvert = \left\lvert q_s \right\rvert = e/3$. Consequently, the magnetic-field strength is given by $eB = 6\pi N_b/(N_\sigma^2a^2)$, where $N_b$ is an integer denoting the number of magnetic flux quanta piercing the entire $x$-$y$ plane and $a$ is the lattice spacing. 

The strange-quark mass $m_s$ is tuned to its physical value by adjusting the mass of the fictitious $\eta^0_{s\bar{s}}$ pseudoscalar meson~\cite{Bazavov:2014cta}. This tuning uses the leading-order chiral perturbation theory relation $M_{\eta^0_{s\bar{s}}} = \sqrt{2M_K^2-M_\pi^2}$. The light-quark mass is set to $m_l = m_s/27$, yielding a near-physical pion mass in the vacuum. We adopt the scale setting procedure common to the HotQCD Collaboration~\cite{Bazavov:2011nk,HotQCD:2014kol}; further specifics regarding the HISQ discretization in magnetic fields are available in Ref.~\cite{Ding:2020hxw}.

In our simulations, all gauge configurations were generated using a modified version of the SIMULATeQCD software suite~\cite{Mazur:2021zgi,Mazur_2024}, with configurations saved every ten trajectories. We employed four different lattice spacings: $a=$0.112, 0.084, 0.067, and 0.056~fm, corresponding to lattice sizes $24^3\times 48$, $32^3\times 64$, $40^3\times 80$, and $48^3\times 96$, respectively. For all ensembles, we employ a zero-temperature finite-$B$ setup consistent with Ref.~\cite{DElia:2021tfb}:
The spatial extent is kept fixed at $L=aN_\sigma \simeq 2.7~\mathrm{fm}$ and the aspect ratio is chosen as $N_\tau=2N_\sigma$.
Fixing $L$ also makes the mapping $eB = 6\pi N_b/L^2$ for a given flux quantum $N_b$ nearly identical across lattice spacings. We selected seven different values of $N_b \in \left\{0,2,\dots ,12\right\}$ corresponding to $eB$ values ranging from 0 to 1.22~$\mathrm{GeV}^2$. 

To ensure the reliability of the continuum extrapolation, we also comment on the potential discretization errors introduced by the magnetic field itself. A key requirement to mitigate such effects is $a^2 |qB| \ll 1$, which is equivalent to $N_b / N_\sigma^2 \ll 1$. Adopting the benchmark from Refs.~\cite{DElia:2021tfb, Endrodi:2019zrl}, we ensure that  $N_b / N_\sigma^2 < 5\%$. In our setup, the most stringent case occurs for the coarsest lattice ($a = 0.112$~fm, $N_\sigma = 24$) at the largest magnetic field ($N_b = 12$), yielding $N_b / N_\sigma^2 \approx 2.1\%$. This value is well within the accepted tolerance, confirming that lattice discretization errors associated with the magnetic field remain under control throughout our parameter space. The simulation parameters are summarized in~\autoref{tab:statistics}.
\begin{table}[!htpb]
   \centering  	
    \begin{tabular}{*{10}{c}}
        \toprule  
        \toprule  
        \multirow{2.3}*{$N_\sigma^3\times N_\tau$} 
        & \multirow{2.3}*{$\beta$}  
        & \multirow{2.3}*{$am_s$} 
        & \multicolumn{7}{c}{\# of conf at $eB$ [GeV$^2$]} \\
        \cmidrule(lr){4-10}
        & 
        & 
        & 0.0 & 0.2 & 0.41 & 0.61 & 0.81 & 1.02 & 1.22 \\
        \midrule 
        $24^3\times 48$  
        & 6.722 & 0.0484
        & 600 & 701 & 1401 & 1350 & 1350 & 1350 & 1350 \\
        $32^3\times 64$ 
        & 7.016 & 0.0353
        & 600 & 800 & 800 & 800 & 800 & 800 & 800 \\
        $40^3\times 80$ 
        & 7.254 & 0.0274
        & 475 & 555 & 555 & 555 & 555 & 321 & 316 \\
        $48^3\times 96$ 
        & 7.455 & 0.0222
        & 350 & 266 & 249 & 248 & 222 & 306 & 216 \\
        \bottomrule  		
    \end{tabular}
    \caption{Summary of the zero-temperature lattice parameters: gauge coupling $\beta$, strange-quark mass $am_s$ in lattice units, and the number of gauge configurations used for each magnetic field $eB$ (corresponding to $N_b = 0, 2, 4, 6, 8, 10, 12$ from left to right). The strange-to-light-quark mass ratio is fixed at $m_s/m_l=27$.}
    \label{tab:statistics}
\end{table}

The two-point correlation functions were computed using smeared multicorner wall sources~\cite{Bazavov:2019www,Ding:2020hxw} to suppress excited-state contamination and enhance ground-state overlap. These sources are placed on the corners of all $2^3$ spatial cubes on a given time slice. To balance statistical precision against computational cost, we varied the temporal sampling density across different ensembles. For $N_b=0$, we employed a spacing of 8 (4) slices for $N_\tau = 48,64\ (80,96)$. In contrast, for $N_b>0$ where statistical noise is more prominent, the sampling frequency was increased to every time slice, except for the $48^3\times 96$ lattices where a spacing of 2 was adopted. 

\begin{figure}[!htp]
  \centering
  \includegraphics[width=0.45\textwidth]{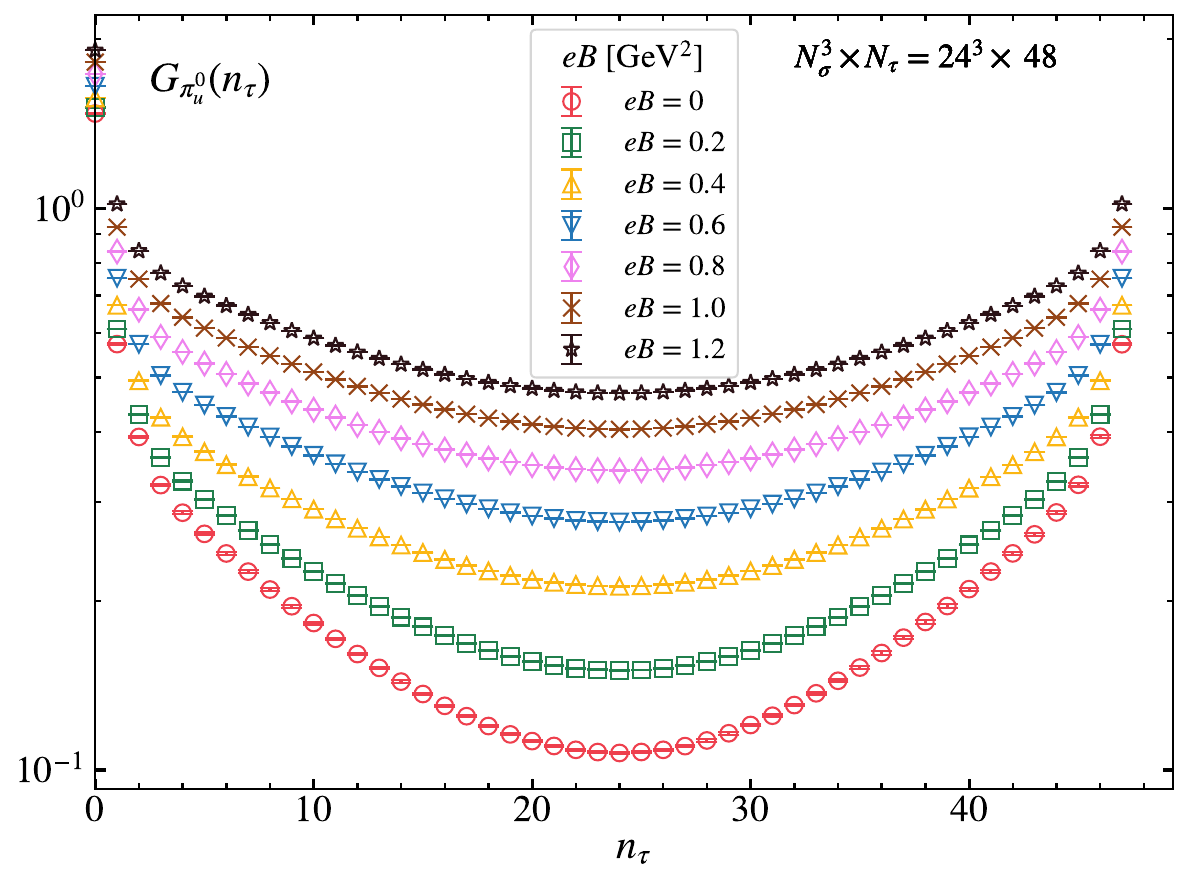}
  \includegraphics[width=0.45\textwidth]{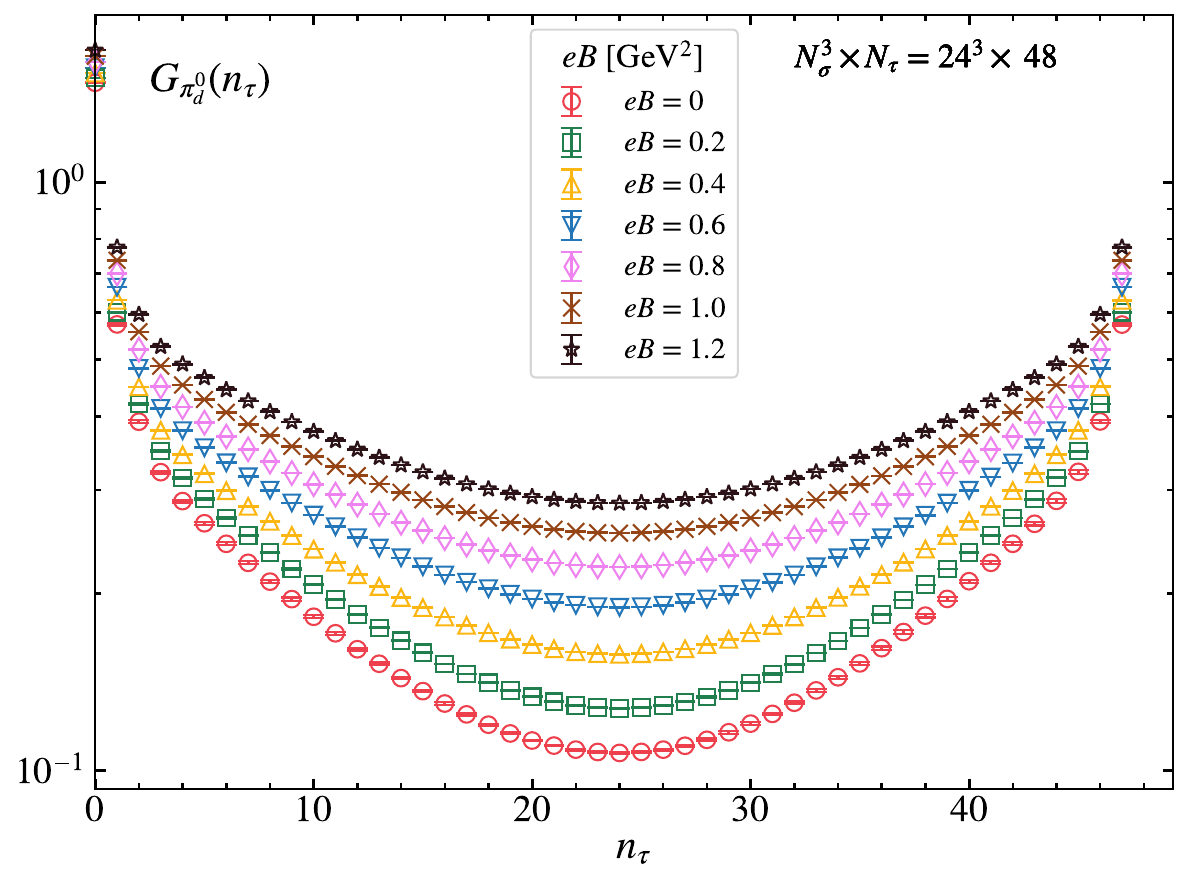}
  \caption{Temporal correlators $G(n_\tau)$ for the neutral pion's $\bar u u$ (top) and $\bar d d$ (bottom) components at various values of $eB$ measured on the $24^3\times 48$ ensemble.}
  \label{fig:corr_ud}
\end{figure}

\begin{figure}[!htpb]
  \centering
  \includegraphics[width=0.45\textwidth]{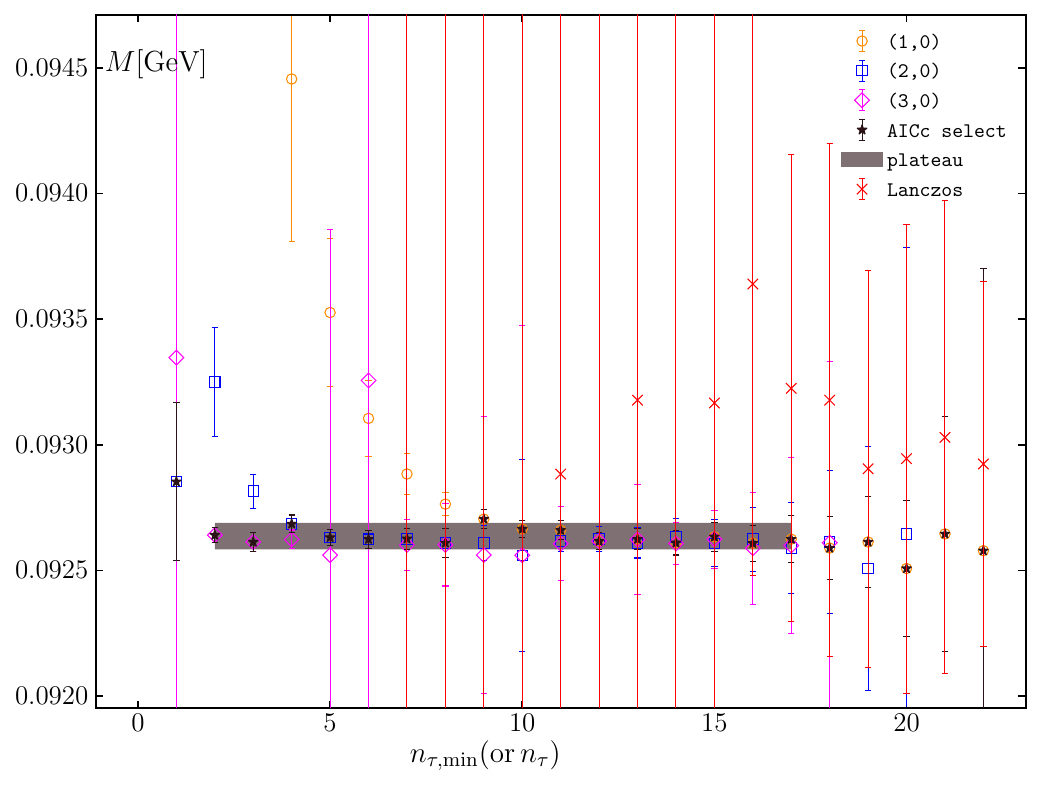}
  \includegraphics[width=0.45\textwidth]{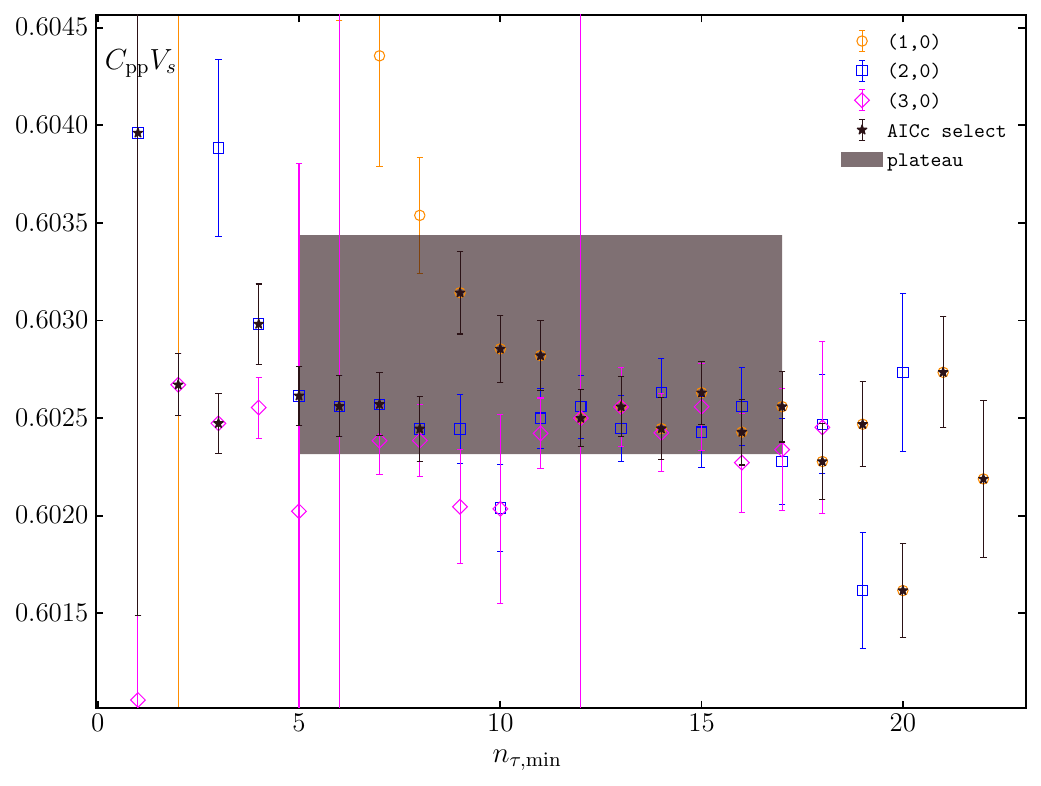}
  \caption{Top: ground-state mass extraction for $\pi^0_u$ on the $24^3\times 48$ ensemble at $eB=0.81~\mathrm{GeV}^2$. 
Results from correlated fits with $(N_{\rm nosc},N_{\rm osc})=(1,0),(2,0),(3,0)$ are shown as open symbols versus the fit start time $n_{\tau,\min}$. 
Black stars denote the AICc-selected fits, and the gray band indicates the final plateau value. 
Masses extracted using the oblique Lanczos method (red crosses) are also plotted against Euclidean time $n_\tau$. Bottom: the corresponding matrix element $C_{PP} V_s$ obtained from the same set of fits; this quantity is not accessible in the Lanczos method.
}
  \label{fig:plateau_example}
\end{figure}

As representative examples, we present the temporal correlators for the flavor-resolved neutral-pion components
$\pi^0_u(\bar u u)$ and $\pi^0_d(\bar d d)$ in \autoref{fig:corr_ud}, for magnetic-field strengths $eB \in [0,1.22]~\mathrm{GeV}^2$, on $24^3\times 48$ ensembles. Both channels exhibit an increasing correlator magnitude with rising magnetic field, with the $\bar u u$ component displaying a stronger enhancement due to the larger electric charge of the $u$ quark.
Meson masses are extracted by fitting the correlators to the form in \autoref{eq:oscillatory}. We systematically consider fits with combinations $(N_{\rm nosc},N_{\rm osc})$=(1,0), (1,1), (2,0), (2,1), (3,0), while varying the fit window $n_\tau\in[n_{\tau,\min},\,N_\tau/2]$ with $n_{\tau,\min}$ ranging from 1 to $N_\tau/2-d$ (where $d$ ensures sufficient degrees of freedom).
The preferred model and window are selected by minimizing the AICc~\cite{cavanaugh1997unifying,Akaike:1974vps}:
\begin{equation}
  \mathrm{AICc}=2k - 2\ln\hat L + \frac{2k(k+1)}{n-k-1}\,,
\end{equation}
where $k$ is the number of fit parameters, $\hat L$ is the maximum likelihood, and $n$ is the number of data points. 
Once the AICc-optimal fits yield a stable mass plateau as a function of $n_{\tau,\min}$, the ground-state mass and its associated uncertainty are determined via bootstrap resampling over that plateau region~\cite{Ding:2020hxw,Bazavov:2019www}. As demonstrated in \autoref{fig:plateau_example}, this approach yields clear plateaus for both mass and amplitude extractions.

Furthermore, as part of our systematic studies, we include masses from the oblique Lanczos method~\cite{Ostmeyer:2024qgu} at various $n_\tau$ values. The method and the detailed comparisons with the multistate fits are presented in Appendix \ref{app:aicc_lanczos_comparison}.

\section{Results}
\label{sec:results}

\subsection{Masses of pseudoscalar mesons}
\label{subsec:mass}
In this subsection, we present our results for the masses of pseudoscalar mesons at seven different values of the magnetic-field strength, from $eB=0$ to $1.22~\mathrm{GeV}^2$, using four lattice spacings and performing a controlled continuum extrapolation. 
To improve the robustness of the mass determination, we use two complementary ground-state extraction methods (AICc-selected multistate fits and the oblique Lanczos method) together with two continuum-extrapolation \textit{Ans\"atze} (linear and quadratic). 

For each observable, we combine the continuum estimates from the different analysis choices within the bootstrap procedure to define the final continuum estimator. The quoted central value is taken as the median of the resulting bootstrap distribution, and the uncertainty is determined from its central 16th--84th percentile interval. This uncertainty, therefore, represents the total uncertainty of the final combined estimator, obtained after propagating both the bootstrap fluctuations and the variation among the accepted analysis choices. A more detailed description of this procedure is given in Appendix~\ref{app:cont_extr}. For transparency, the continuum central values and bootstrap uncertainties obtained from the individual analysis choices are listed in Appendix~\ref{app:summary_data_tables} together with the final quoted results.

\begin{figure}[!htpb]
    \includegraphics[width=0.45\textwidth]{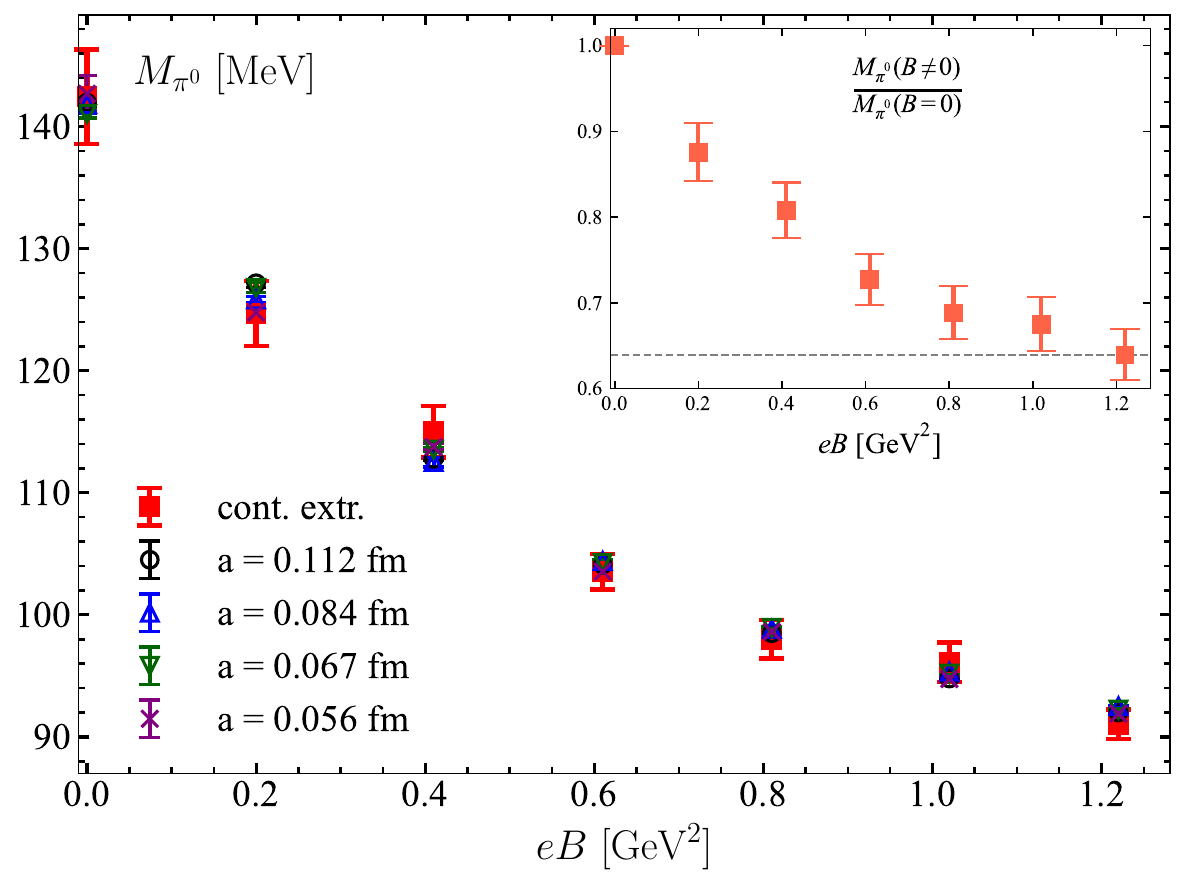} \\
    \includegraphics[width=0.45\textwidth]{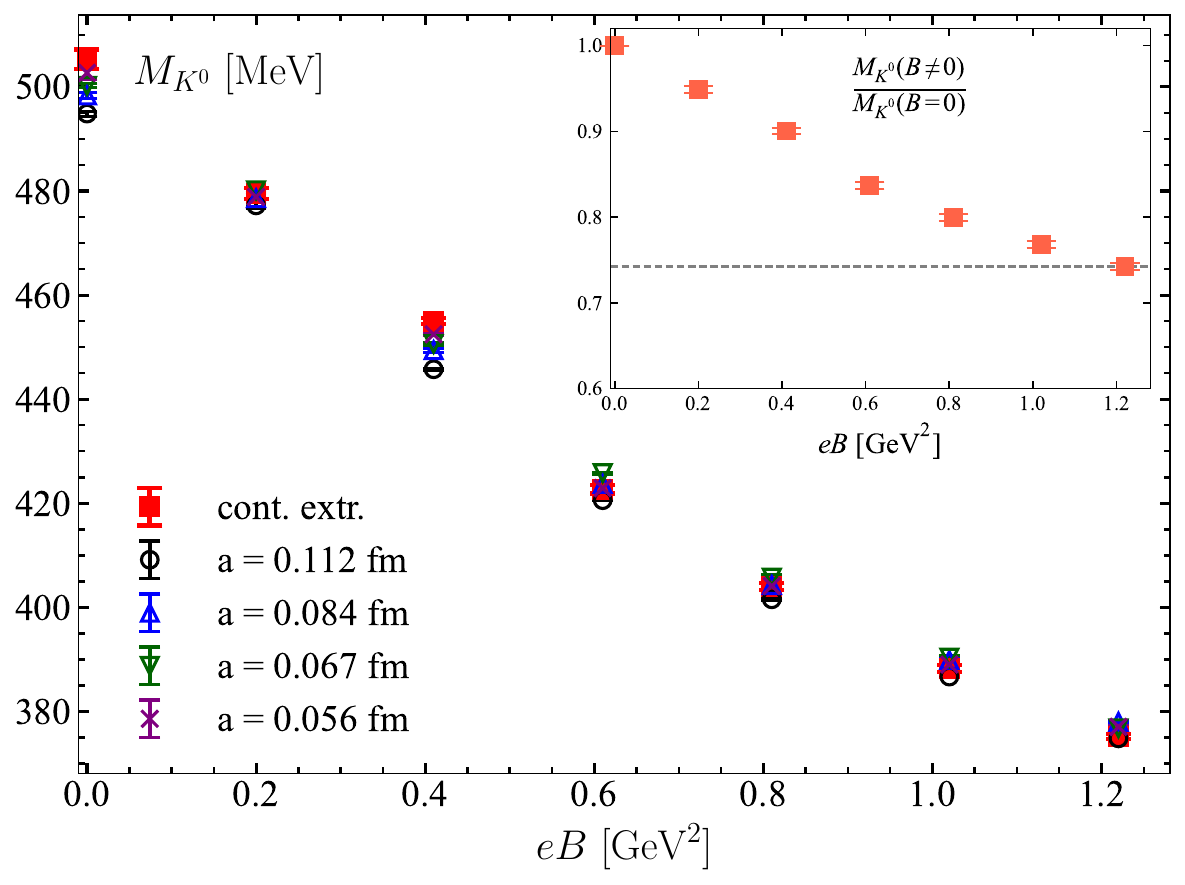} \\
    \includegraphics[width=0.45\textwidth]{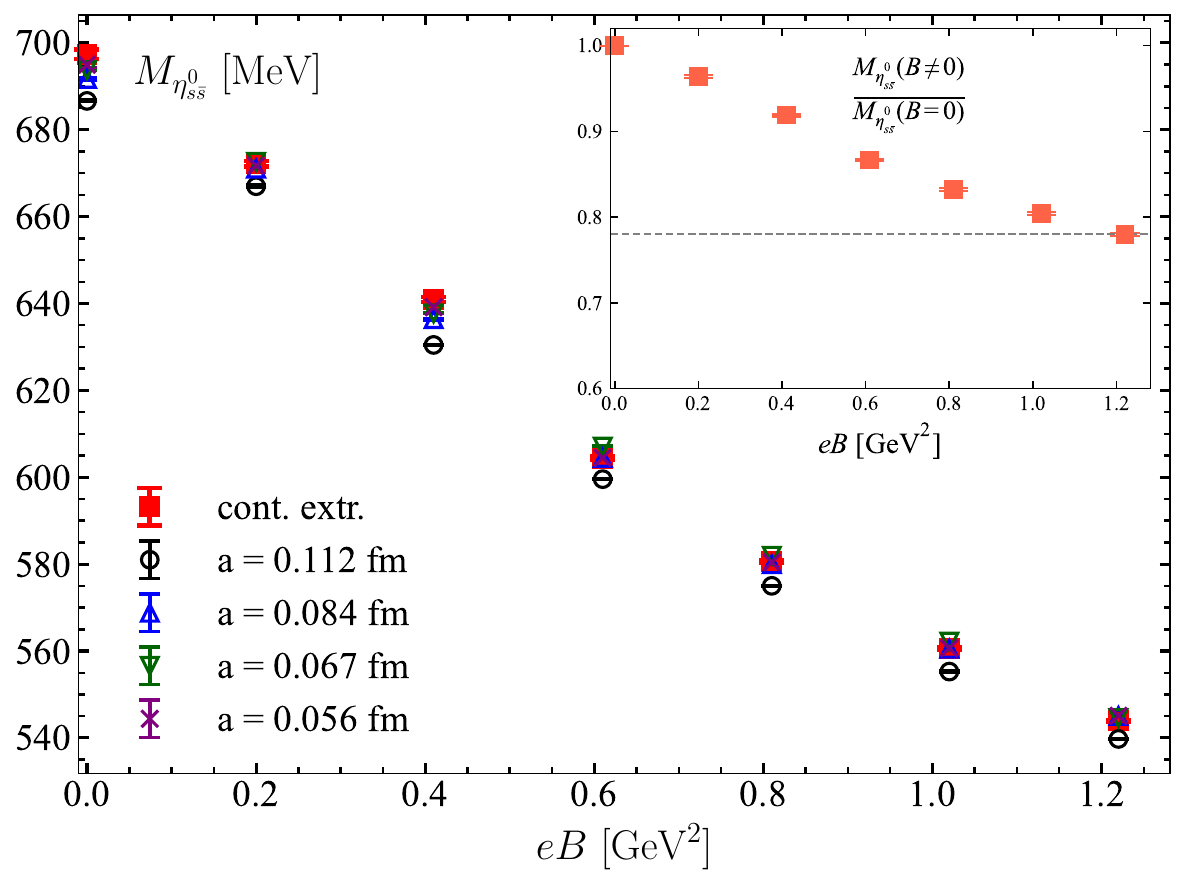}
    \caption{Masses of neutral pseudoscalar mesons as functions of the background magnetic field $eB$. The top, middle, and bottom panels show results for $\pi^0$, $K^0$, and $\eta^0_{s\bar{s}}$, respectively. Open symbols denote lattice measurements on the individual ensembles, while the red filled symbols show the corresponding continuum-extrapolated results. The insets display the meson masses normalized to their $eB=0$ values to highlight the relative changes.}
    \label{fig:contband_SIM24_3particle}
\end{figure}

\begin{figure}[!htbp]
\centering
\includegraphics[width=0.45\textwidth]{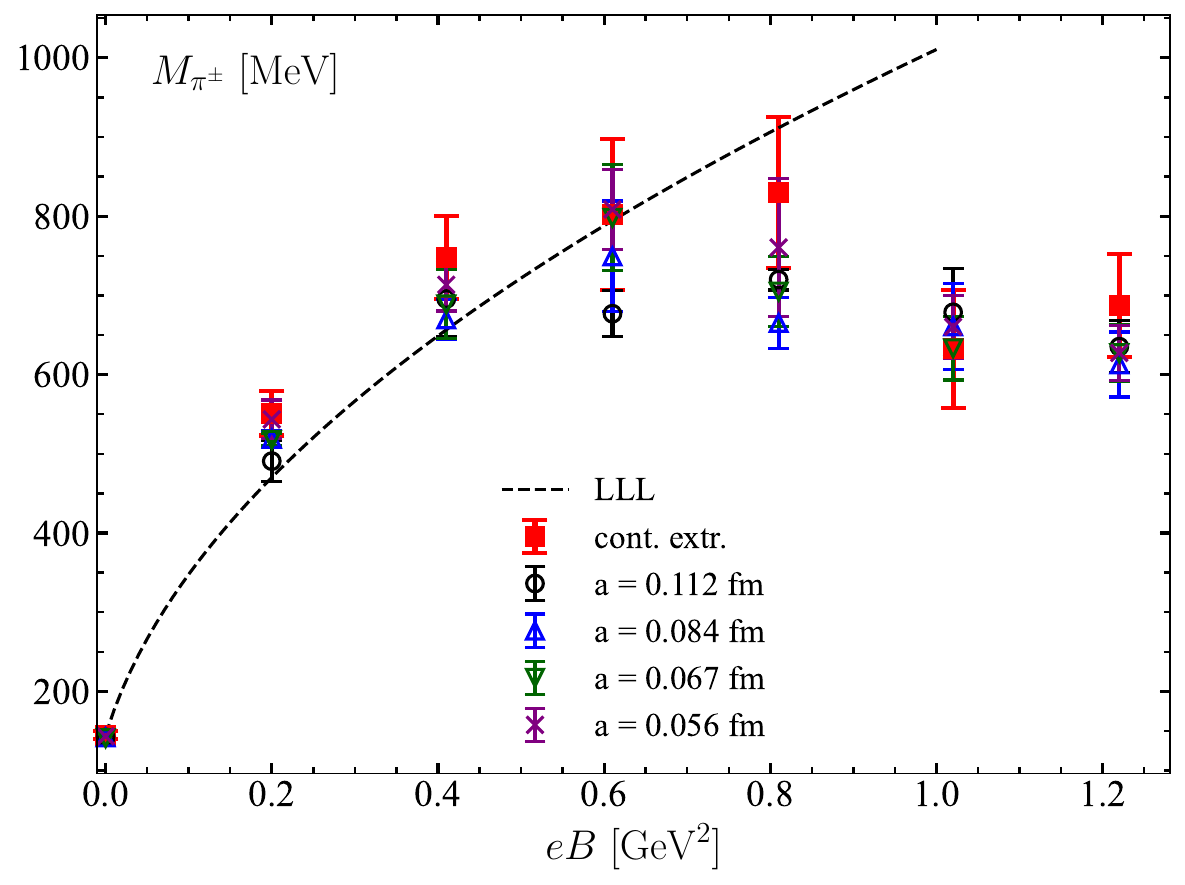}\\
\includegraphics[width=0.45\textwidth]{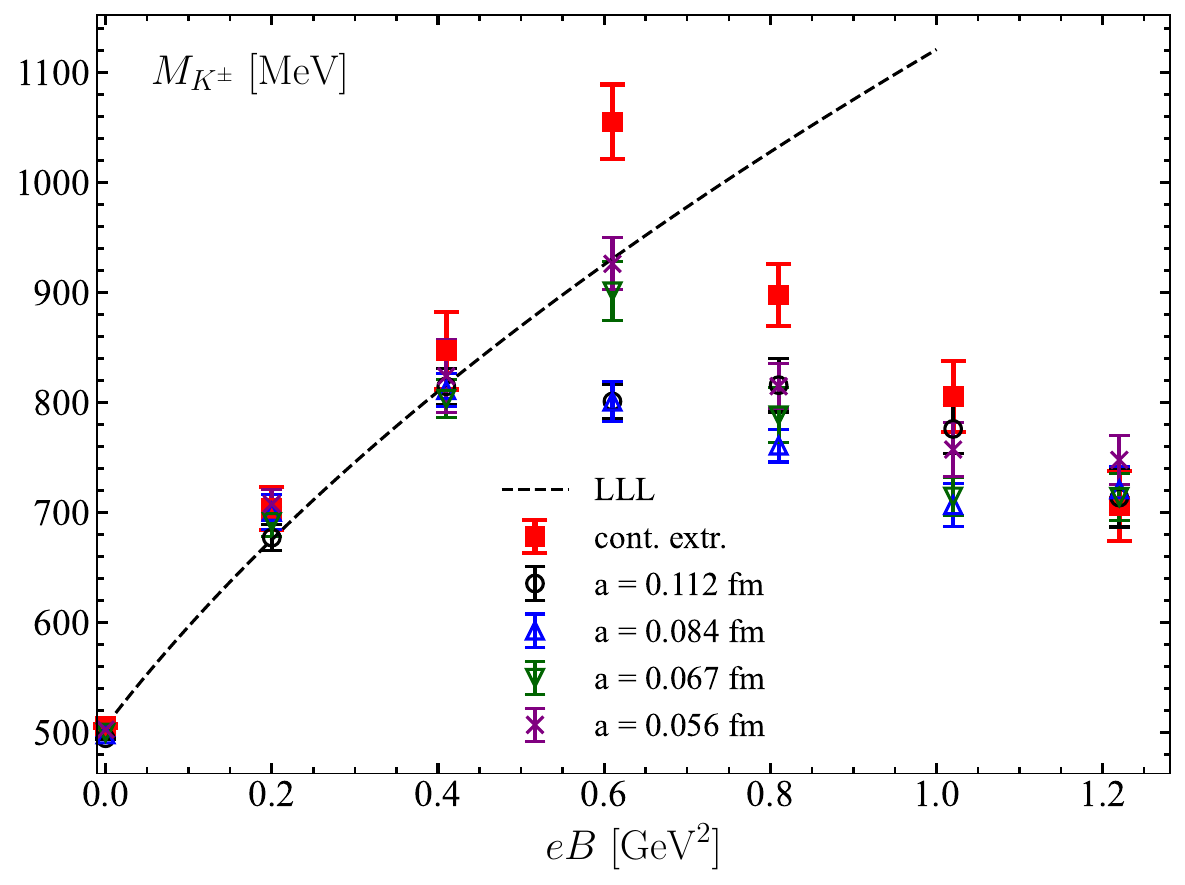}
\caption{Masses of charged pseudoscalar mesons as functions of the background magnetic field $eB$. The top and bottom panels show results for $\pi^\pm$ and $K^\pm$, respectively. Open symbols denote lattice QCD results on the individual ensembles, while the red filled symbols show the continuum-extrapolated results. In each panel, the dashed line represents the LLL prediction for a pointlike charged particle with the corresponding zero-field mass.}
\label{fig:charged_mass}
\end{figure}

\begin{figure}[!htpb]
        \includegraphics[width=0.45\textwidth]{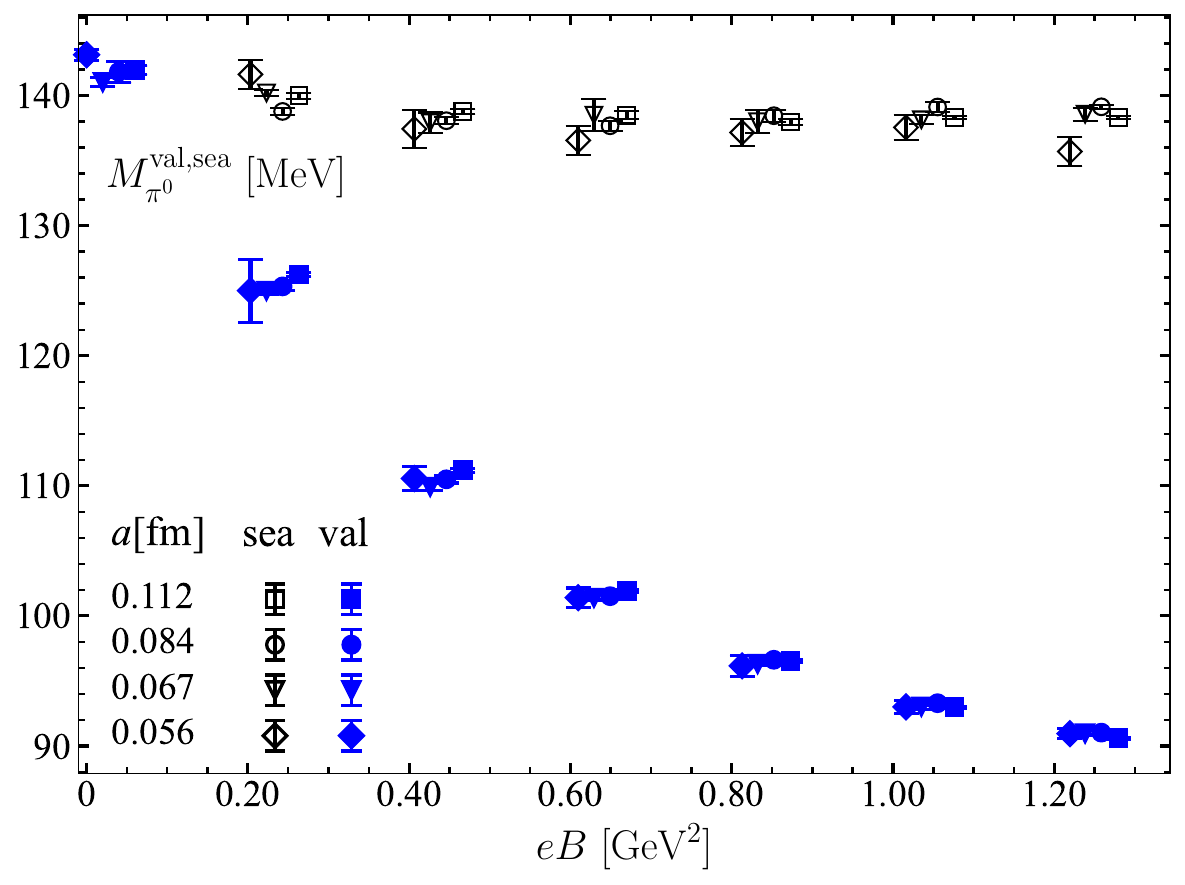} \\
        \includegraphics[width=0.45\textwidth]{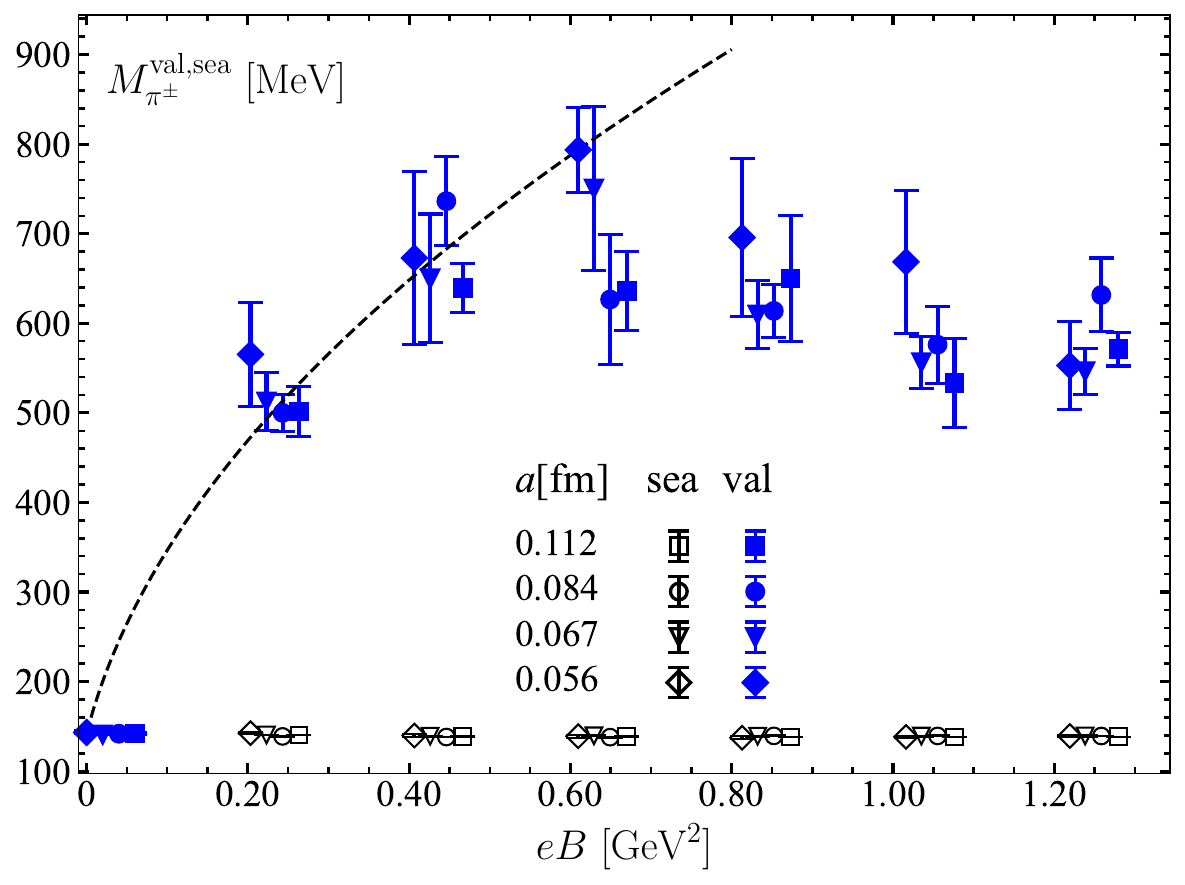} \\
        \caption{Valence- and sea-quark contributions to pseudoscalar meson masses as functions of the magnetic field $eB$ at four lattice spacings. The top (bottom) panel shows the neutral (charge) pion mass. Contributions from valence and sea quarks are denoted by filled and open symbols, respectively.  Different symbol shapes denote lattice spacings. Data are horizontally offset for clarity. The dashed line in the bottom panel indicates the LLL prediction for a pointlike particle.
        }
        \label{fig:sea_valence_pi}
\end{figure}

\begin{figure}[!htpb]
        \includegraphics[width=0.45\textwidth]{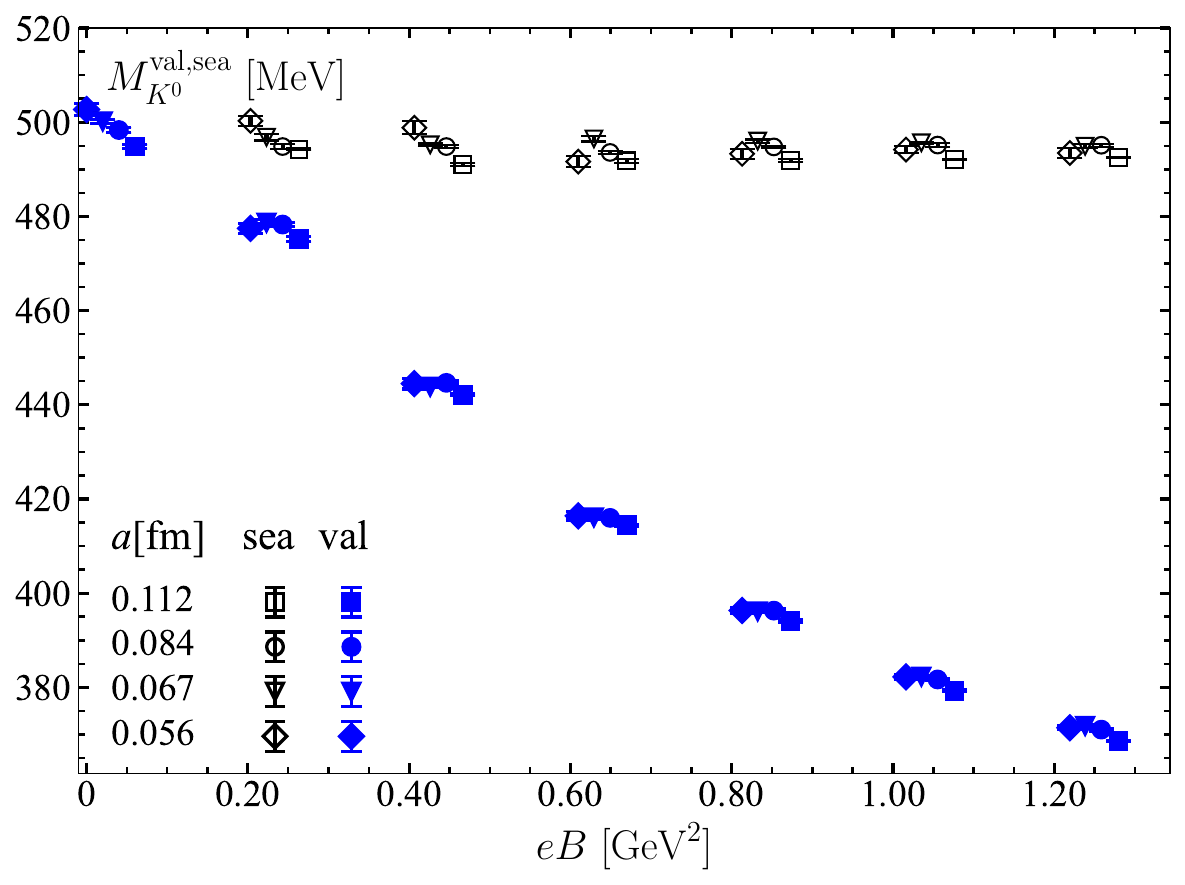} \\
        \includegraphics[width=0.45\textwidth]{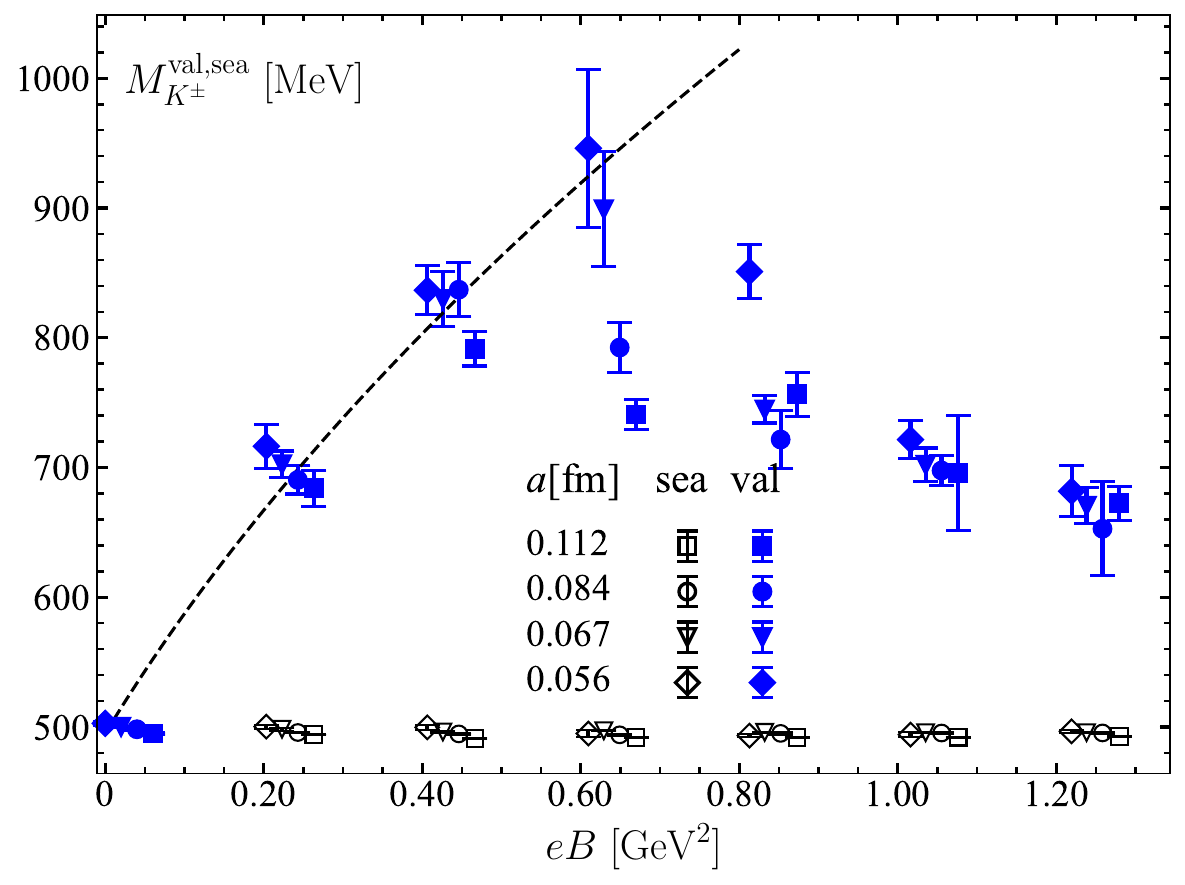} \\
        \caption{The same as~\autoref{fig:sea_valence_pi} but for kaons.}
        \label{fig:sea_valence_K}
\end{figure}

In~\autoref{fig:contband_SIM24_3particle}, we show the results for the neutral-pion, kaon, and $\eta^0_{s\bar{s}}$ masses in the top, middle, and bottom panels, respectively, with open symbols indicating finite-$a$ data and filled squares denoting the continuum-extrapolated values.  While the results in the continuum limit are obtained using both the AICc-selected fit analysis and the oblique Lanczos method, for visibility we show at nonzero lattice spacing only the AICc determinations; the same convention is used in~\autoref{fig:charged_mass},~\autoref{fig:sea_valence_pi} and~\autoref{fig:sea_valence_K}.

From \autoref{fig:contband_SIM24_3particle} we observe that in all three channels the neutral meson masses decrease monotonically with increasing $eB$. The effect is largest for the neutral pion, milder for the neutral kaon, and weakest for the $\eta^0_{s\bar{s}}$. 
This hierarchy is made more transparent by the insets in the upper-right corner of each panel, which directly show the ratios of the meson masses at nonzero $eB$ to their values at $eB=0$, $M(B)/M(B=0)$. At our strongest magnetic field ($eB=1.22~\mathrm{GeV}^2$), we find that $M_{\pi^0}$, $M_{K^0}$, and $M_{\eta^0_{s\bar{s}}}$ are reduced to about 64\%, 74\%, and 78\% of their $eB=0$ values, respectively. This pattern is consistent with our earlier
fixed-lattice-spacing study in Ref.~\cite{Ding:2020hxw}, performed on
$32^3\times96$ lattices with a heavier-than-physical pion mass at $eB=0$
($\sim 220~\mathrm{MeV}$), and it extends those results to physical quark masses
and the continuum limit. Moreover, the observed ordering---heavier mesons being
less affected by the magnetic field---is in line with finite-temperature
screening-mass studies, where heavier states exhibit a smaller $B$-induced
shift~\cite{Ding:2025pbu,Ding:2022tqn}.

As discussed in Sec.~\ref{sec:basics}, neutral pointlike particles would have  masses that are independent of the magnetic field. 
The pronounced decrease of the neutral pseudoscalar meson masses with $eB$, therefore, reflects their internal quark-antiquark structure and the distortion of the bound state by the external field.
This neutral-sector behavior provides a useful baseline for interpreting the charged pseudoscalars, where a pointlike picture
instead predicts an increasing energy with $eB$ through Landau-level quantization.

We now turn to the masses of the charged pseudoscalar mesons, $M_{\pi^\pm}$ and $M_{K^\pm}$. 
Our lattice results in \autoref{fig:charged_mass} show that the pointlike expectation captures only the weak-field regime: For small $eB$, both masses rise in qualitative agreement with LLL estimate (cf. \autoref{eq:LLLapproximation}), shown by the dashed lines. For stronger fields, however, the trend changes; starting around $eB \approx 0.6~\mathrm{GeV}^2$, the meson masses deviate from the trend of the LLL prediction and instead approach a saturation regime, with a slight downward trend at the largest fields. 
This deviation signals that charged pseudoscalar mesons cannot be treated as pointlike particles in strong magnetic fields, and their internal structure and strong dynamics become essential.
Furthermore, we observe that at large magnetic fields the mass of the kaon is less affected than that of the pion. This relative insensitivity of $M_{K^\pm}$ compared to $M_{\pi^\pm}$ may be attributed to the larger vacuum mass of the kaon, consistent with the pattern also seen in the neutral sector.

\begin{figure}[!htpb]
    \centering
    \includegraphics[width=0.45\textwidth]{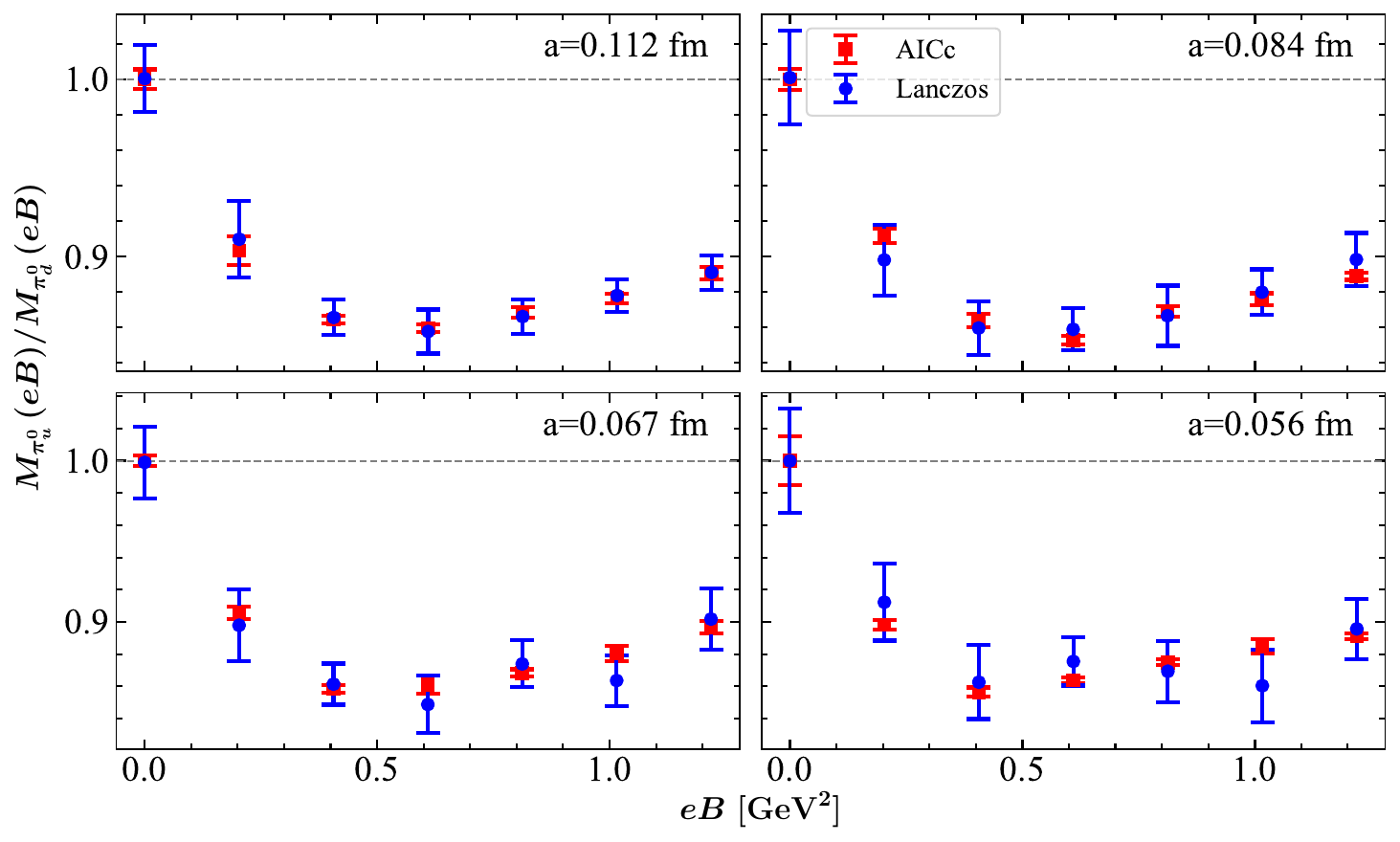}
    \caption{Ratio of the neutral-pion mass components, $M_{\pi_u^0}/M_{\pi_d^0}$, as a function of the magnetic-field strength $eB$ for four different lattice spacings. Results obtained from the AICc-selected multistate fits and the oblique Lanczos method are denoted by red squares and blue circles, respectively.}
    \label{fig:mass_ratio_ud}
\end{figure}

The nonmonotonic behavior of the charged pseudoscalar masses in \autoref{fig:charged_mass} already highlights that, beyond the pointlike Landau-level expectation, the magnetic field probes the internal quark structure of the mesons. A complementary and more microscopic handle is provided by the neutral-pion sector. 
In a magnetic field, isospin is explicitly broken and the connected $\pi^0$ correlator naturally separates into $u\bar u$ and $d\bar d$ components. 
Since these components carry different electric charges, $q_u=+2/3\,e$ and $q_d=-1/3\,e$, they experience different effective field strengths and, thus, develop different mass shifts, motivating a direct comparison of $M_{\pi_u^0}$ and $M_{\pi_d^0}$.

The ratio $M_{\pi_u^0}/M_{\pi_d^0}$ as a function of $eB$ is shown in \autoref{fig:mass_ratio_ud}. Here, the comparison is made at the same external field strength but for quarks of different electric charges. We display results from four lattice spacings separately. To ensure the robustness of our mass extractions, we employ two independent techniques: AICc-selected multistate fits and the oblique Lanczos method. As shown in the figure, the results obtained from these two complementary methods are in excellent agreement across all lattice spacings and magnetic-field strengths. At vanishing magnetic field, the ratio is consistent with unity within statistical errors, although the uncertainties are relatively large due to limited statistics. As the magnetic field increases, the ratio systematically deviates from unity: It decreases up to $eB\approx 0.6~\mathrm{GeV}^2$ and then rises slightly at larger $eB$. This implies that $M_{\pi_u^0}$ decreases more rapidly than $M_{\pi_d^0}$ at moderate fields, consistent with the stronger coupling of the $u$ quark to the magnetic field, while at higher fields the behavior becomes more intricate, reflecting a nontrivial interplay of quark charges and magnetic effects. 

\begin{figure}[!htbp]
    \centering
    \includegraphics[width=0.45\textwidth]{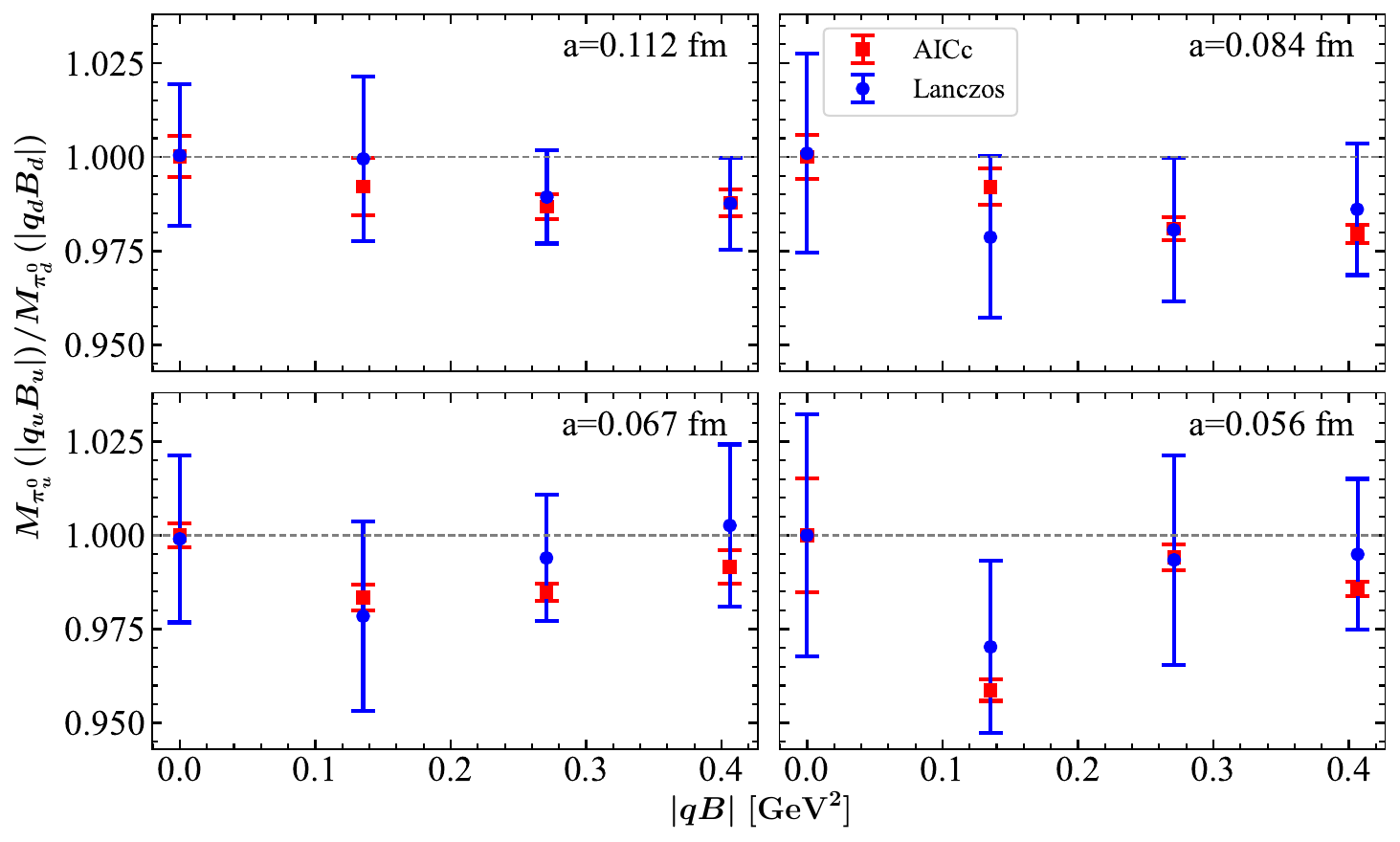}
    \caption{Similar as~\autoref{fig:mass_ratio_ud} for $M_{\pi^0_u}/M_{\pi^0_d}$ but plotted as a function of the rescaled field strength $|qB|$. }
    \label{fig:mass_ratio_ud_qBscale}
\end{figure}

\begin{figure}[!htbp]
    \centering
    \includegraphics[width=0.45\textwidth]{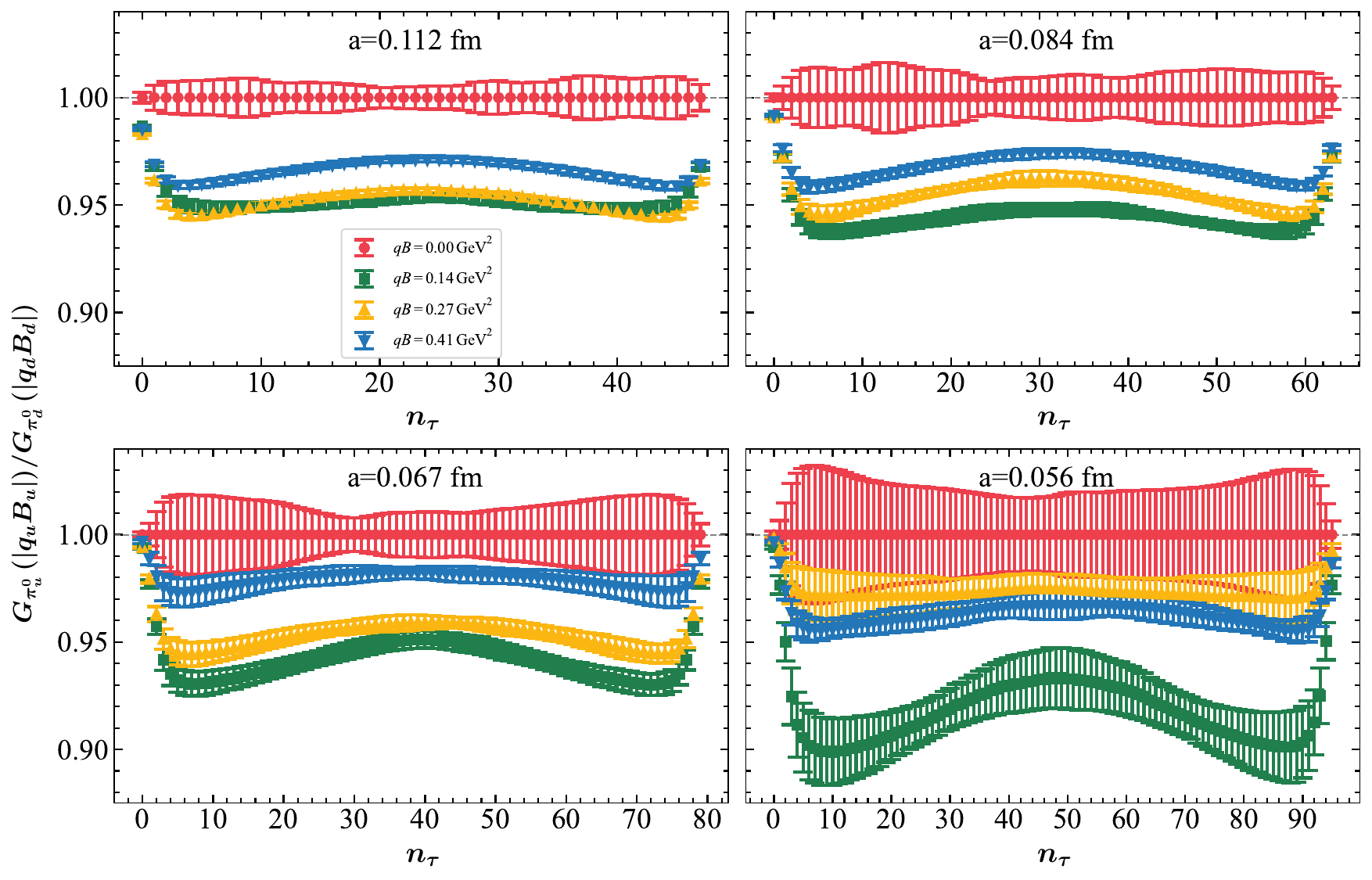}
    \caption{Ratio of the temporal correlation functions,
$G_{\pi^0_u}(n_\tau)/G_{\pi^0_d}(n_\tau)$, as a function of Euclidean time $n_\tau$ at fixed values of $|qB|$, for the lattice spacings indicated in the panels. The different colors correspond to the $|qB|$ values listed in the legend.}
    \label{fig:corr_ratio_ud_qBscale}
\end{figure}

In Ref.~\cite{Ding:2020hxw}, the terminology $qB$ scaling was introduced to refer to the observation that the $u$- and $d$-quark connected components
of the neutral-pion mass depend only on the product of the quark electric charge and the magnetic-field strength,
i.e., $qB$. Such a scaling would be expected to hold exactly in the quenched limit, because in that case $eB$
enters solely through the Dirac operator, without contributions from sea quarks. Consequently, $M_{\pi_u^0}$ and
$M_{\pi_d^0}$ should coincide when plotted against $qB$. In full QCD, however, dynamical sea quarks carry different
charges and enter the quark determinant, potentially breaking $qB$ scaling.
To test whether $qB$ scaling holds in our simulations with dynamical quarks, we show the ratio $M_{\pi_u^0}/M_{\pi_d^0}$ as a function of $qB$ in~\autoref{fig:mass_ratio_ud_qBscale}. After rescaling the magnetic field by the quark charges, $|qB|=|q_uB_u|=|q_dB_d|$, the results from different lattice spacings deviate from unity by at most $\sim 5\%$, using two different mass extraction methods. This indicates that $qB$ scaling is only mildly broken in our full QCD simulations. A more granular view of the scaling behavior is provided in~\autoref{fig:corr_ratio_ud_qBscale}, which displays the ratio of correlators constructed from the $u$- and $d$-quark sectors as a function of $n_\tau$. The correlator ratios display a similar pattern: At $qB=0$ they are consistent with unity within statistical errors, while at nonzero fields we observe deviations up to $\sim 8\%$ at intermediate temporal separations. Nevertheless, at large Euclidean times ($n_\tau\sim N_\tau/2$), where the ground-state contribution dominates, the ratios tend to return toward unity.
Thus, the correlator ratio provides the most direct evidence for the origin of the deviations of $M_{\pi_u^0}$ and $M_{\pi_d^0}$ from exact $qB$ scaling. Our results suggest that although dynamical quarks introduce noticeable corrections, the breaking of $qB$ scaling remains moderate in the parameter region explored here.

To pinpoint the dynamical origin of the magnetic-field dependence of pseudoscalar meson masses, we separate the response into ``valence'' and ``sea'' contributions, following the strategy introduced in Refs.~\cite{DElia:2011koc,Bruckmann:2013oba,Ding:2022tqn}. 
The key point is that the external field enters QCD in two distinct ways: It modifies the valence-quark propagators that appear in the hadronic contractions, and it also modifies the sampling weight of gauge configurations through the fermion determinant. 
Using the $B$-dependent
ensemble average $\langle\cdots\rangle_B$ defined in Sec.~\ref{sec:basics}, we introduce two
auxiliary correlators in which only one of these effects is switched on~\cite{DElia:2011koc,Bruckmann:2013oba,Ding:2022tqn}.

The \emph{valence} correlator switches on the magnetic field only in the valence propagators, while keeping the sea-quark determinant (and, hence, the gauge-field sampling) at $B=0$,
\begin{align}
    G_H^{\rm val}(B;t) &= \big\langle \mathcal{G}_H(B;t;U)\big\rangle_{0}.
    \label{eq:CH_val_compact}
\end{align}
Conversely, the \emph{sea} correlator switches on the magnetic field only in the determinant, while evaluating the hadronic contraction with $B=0$ propagators:
\begin{align}
    G_H^{\rm sea}(B;t) &= \big\langle \mathcal{G}_H(0;t;U)\big\rangle_{B}.
    \label{eq:GH_sea_compact}
\end{align}
By construction, the $B$ dependence of $G_H^{\rm val}$ originates solely from the valence sector, whereas that of $G_H^{\rm sea}$ originates solely from the sea sector.
From $G_H(B;t)$, $G_H^{\rm val}(B;t)$, and $G_H^{\rm sea}(B;t)$ we extract the corresponding masses using identical fit procedures and fit windows, yielding $M_H(B)$, $M_H^{\rm val}(B)$, and $M_H^{\rm sea}(B)$.

The results are summarized in~\autoref{fig:sea_valence_pi} and~\autoref{fig:sea_valence_K} for pions and kaons, respectively, where we compare the valence-only and sea-only counterparts for both charged and neutral pseudoscalar channels. 
A clear pattern emerges: Within uncertainties, the full $eB$ dependence is almost entirely reproduced by the valence-only determination, $M_H(B)\simeq M_H^{\rm val}(B)$, whereas the sea-only masses remain close to their $B=0$ values, $M_H^{\rm sea}(B)\approx M_H(0)$, over the entire field range explored. 
This demonstrates that, at the physical point and for the values of $eB$ considered here, magnetic-field effects on pseudoscalar meson masses are dominated by valence dynamics, while sea-quark effects are quantitatively subleading for both charged and neutral states. 
In particular, this valence dominance provides a dynamical explanation for the near-validity of $qB$ scaling observed above: Once the leading valence response is organized in terms of $|q_f B|$, only a small residual breaking---consistent with the minor sea-quark contribution---remains.

We emphasize that the above conclusion---namely that the magnetic-field dependence of pseudoscalar meson masses is overwhelmingly carried by the valence sector, with only a minor sea-quark contribution---pertains to the vacuum (zero-temperature) ensembles studied here. 
At finite temperature, the situation is known to change qualitatively. 
In particular, in the vicinity of the pseudocritical temperature, sea-quark effects become dominant and provide the primary driving mechanism behind the so-called inverse magnetic catalysis. 
This sea-quark-driven behavior has been observed both in the nonmonotonic temperature and field dependence of the chiral condensate and in the corresponding behavior of pseudoscalar screening masses (or screening correlators) in the vicinity of the transition region~\cite{DElia:2011koc,Bruckmann:2013oba,Ding:2022tqn}.
\begin{figure}[!htpb]
        \includegraphics[width=0.45\textwidth]{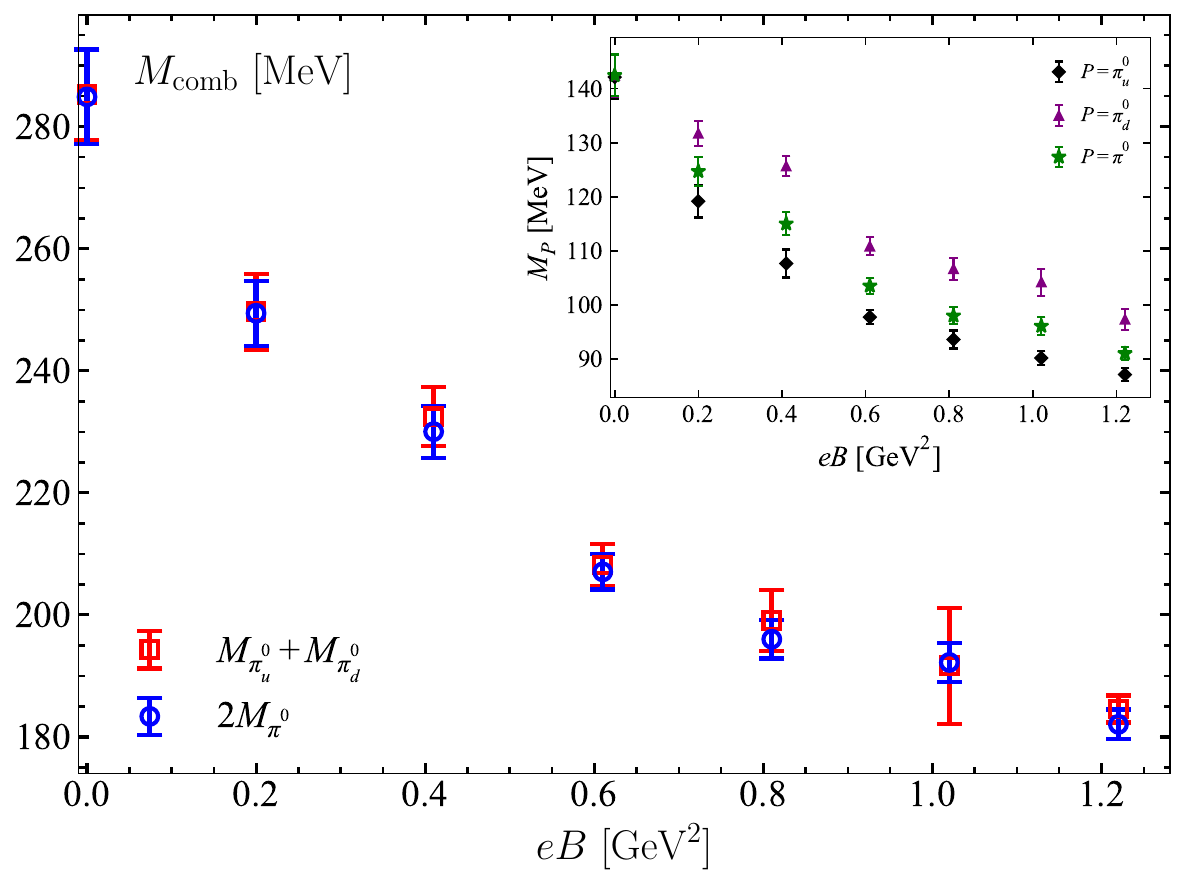} \\
        \caption{continuum-extrapolated results for combination of masses, $M_{\rm comb}$, i.e., the sum of quark components $M_{\pi^0_u}+M_{\pi^0_d}$ (red squares) and twice the neutral-pion mass $2M_{\pi^0}$ (blue circles) as a function of the magnetic field $eB$. The inset displays the individual flavor components $M_{\pi^0_u}$ and $M_{\pi^0_d}$ alongside the neutral-pion mass $M_{\pi^0}$ using filled symbols as indicated in the legend.}
        \label{fig:sum_mass_u_d}
\end{figure}

At the end of this subsection, we show the magnetic-field dependence of the combination
$M_{\pi^0_u}+M_{\pi^0_d}$, which, through QCD inequalities, provides a lower bound on the charged $\rho$-meson mass.
As discussed in~\autoref{sec:basics}, the flavor-resolved connected contributions to the neutral-pion correlator define the masses
$M_{\pi^0_u}$ and $M_{\pi^0_d}$.
While these connected-component masses are not associated with physical asymptotic states, they are well-defined quantities extracted from gauge-invariant correlators and can be used in rigorous inequality relations. In particular, $M_{\rho^\pm}(eB)\ \ge\ M_{\pi^0_u}(eB) + M_{\pi^0_d}(eB)\,$.
In \autoref{fig:sum_mass_u_d}, we present our continuum-extrapolated results for this bound.
We observe a monotonic decrease of $M_{\pi^0_u}+M_{\pi^0_d}$ with increasing $eB$, reaching $\sim 180~\mathrm{MeV}$ at our
largest field strength, $eB = 1.22~\mathrm{GeV}^2$.
Remarkably, within uncertainties this sum remains consistent with $2M_{\pi^0}$ over the entire field range.
The inset further illustrates the expected charge hierarchy of magnetic effects: The $u\bar u$ component shows the
largest shift, followed by the neutral pion, while the $d\bar d$ component is least affected.

\subsection{Decay constants of neutral pseudoscalar mesons}
\label{subsec:neutral-dc}

\begin{figure}[!htpb]
    \centering
    \includegraphics[width=0.45\textwidth]{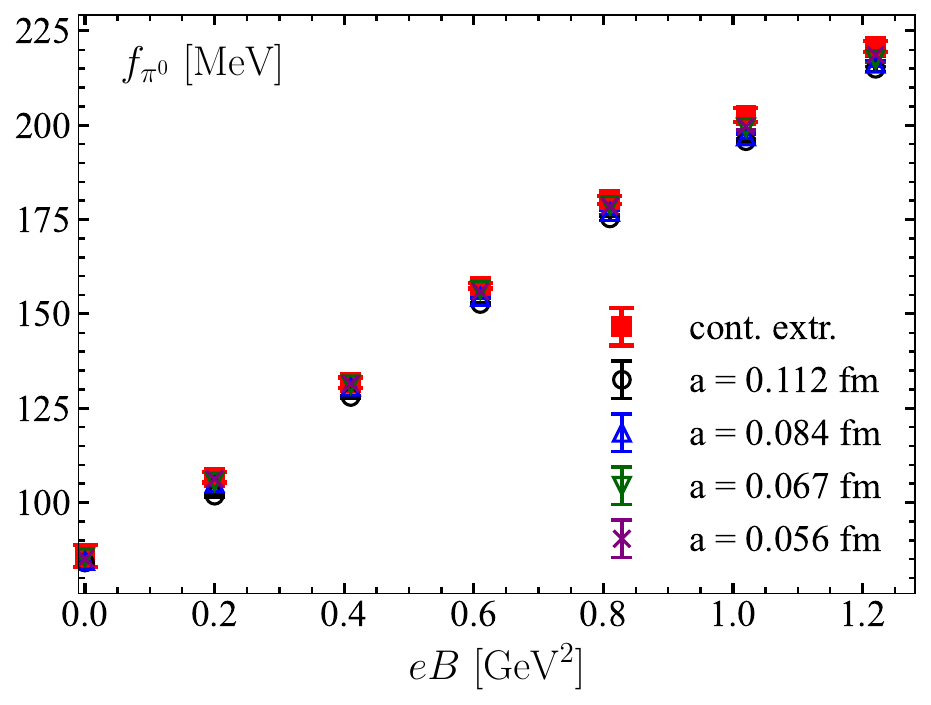} \\
    \includegraphics[width=0.45\textwidth]{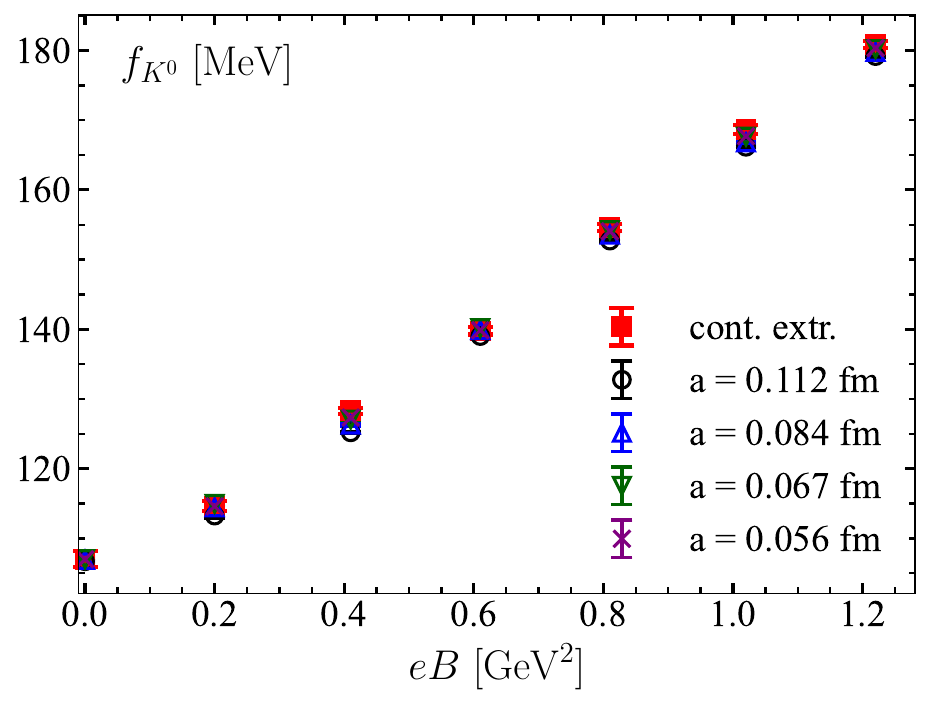} \\
    \includegraphics[width=0.45\textwidth]{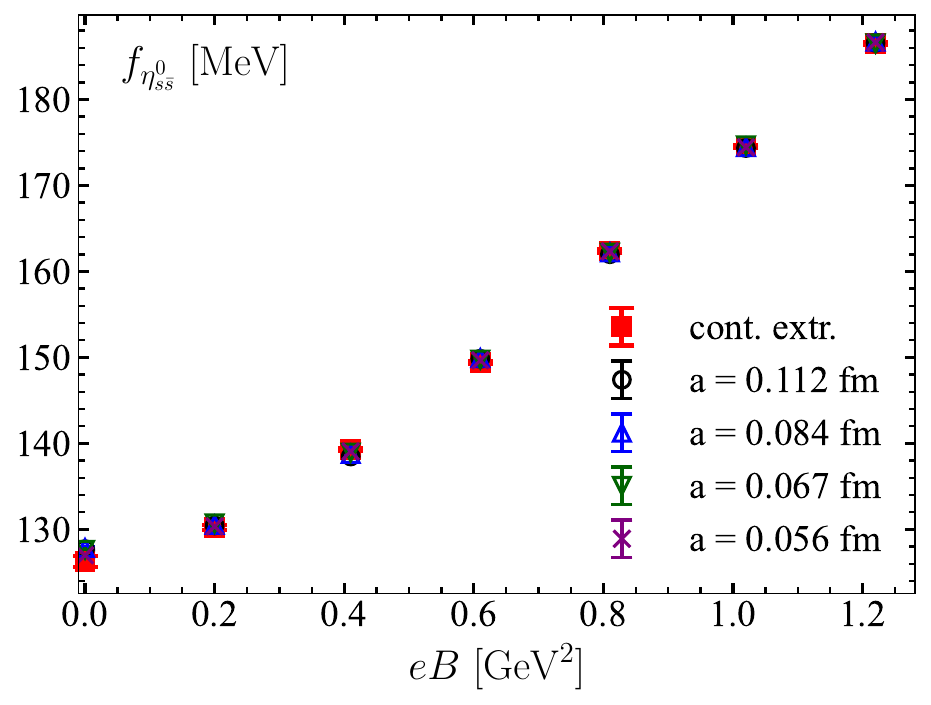} \\
    \caption{Decay constants of neutral pseudoscalar mesons as functions of the magnetic field $eB$. From top to bottom: $\pi^0$, $K^0$, and $\eta^0_{s\bar{s}}$. Open symbols show results at finite lattice spacing, while filled symbols denote the corresponding continuum extrapolations.}
    \label{fig:dc_mesons}
\end{figure}

We present the decay constants of neutral pseudoscalar mesons as functions of the magnetic-field strength $eB$ in~\autoref{fig:dc_mesons}. The open symbols denote lattice results at finite lattice spacing, while the filled symbols correspond to continuum extrapolations. At vanishing magnetic field, we obtain $f_{\pi^0}=85.8(2.9)$~MeV and $f_{K^0}=107.0(1.2)$~MeV, yielding $f_{K^0}/f_{\pi^0}=1.247(44)$. The ratio is reasonably close, within uncertainties, to the continuum $N_f=2{+}1$ benchmark
$f_K/f_\pi=1.1916(34)$ quoted by FLAG 2024~\cite{FlavourLatticeAveragingGroupFLAG:2024oxs}. 
The small
separation between finite-$a$ data and continuum-extrapolated points indicates
that discretization effects are mild for $f_{\pi^0}$, $f_{K^0}$, and
$f_{\eta^0_{s\bar s}}$ over the entire $eB$ range.

When the magnetic field is switched on, the decay constants of all three neutral pseudoscalar mesons ($\pi^0$, $K^0$, and $\eta^0_{s\bar{s}}$) increase monotonically with increasing $eB$. $f_{\pi^0}$ grows most rapidly and reaches 220.8(1.4) MeV at $eB = 1.22$~$\mathrm{GeV}^2$, corresponding to an increase of about 157\% relative to $eB=0$. The kaon decay constant $f_{K^0}$ increases more moderately to 180.9(5) MeV, i.e., by about 69\%. The hidden-strangeness channel $f_{\eta^0_{s\bar{s}}}$ exhibits the mildest response, increasing by roughly 48\% over the same field range. The ordering
$\pi^0 > K^0 > \eta^0_{s\bar s}$ is consistent with the general expectation that
observables dominated by lighter valence flavors are more susceptible to
magnetic-field effects.

\begin{figure}[!htbp]
    \centering
    \includegraphics[width=0.45\textwidth]{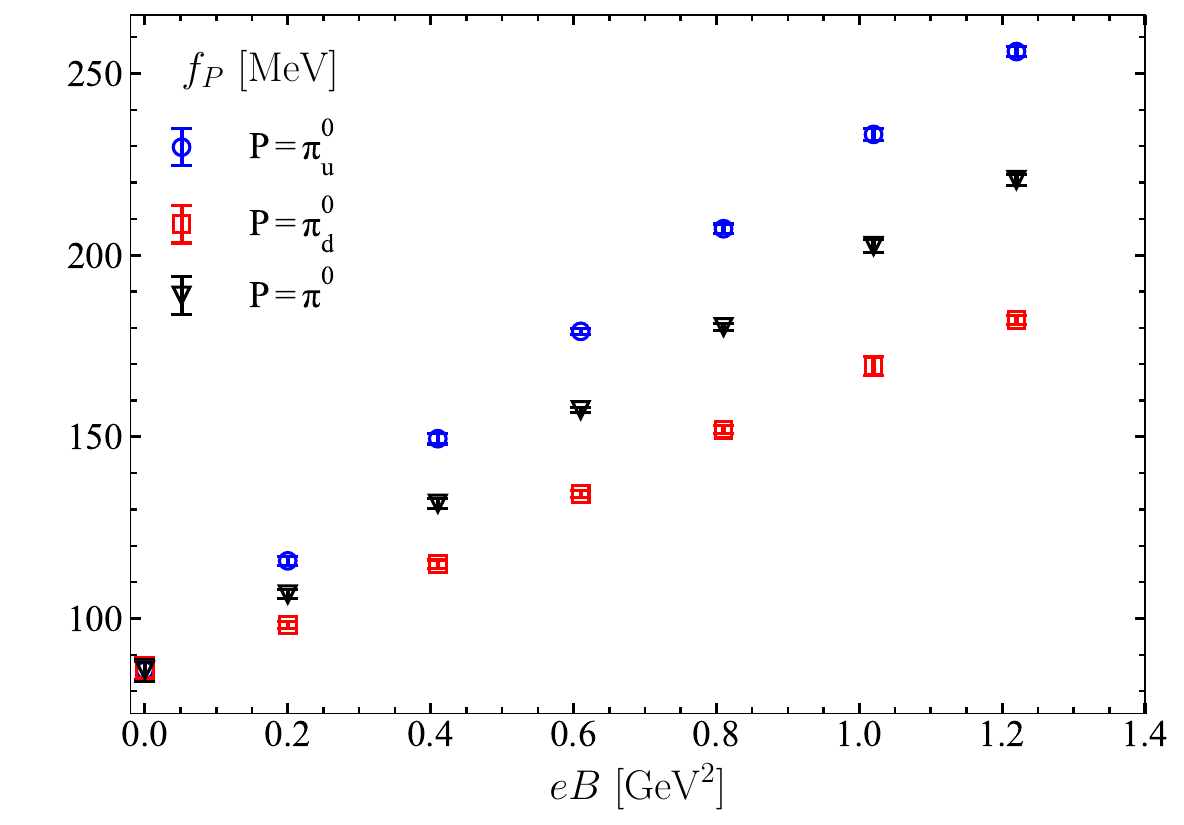}\\
    \includegraphics[width=0.45\textwidth]{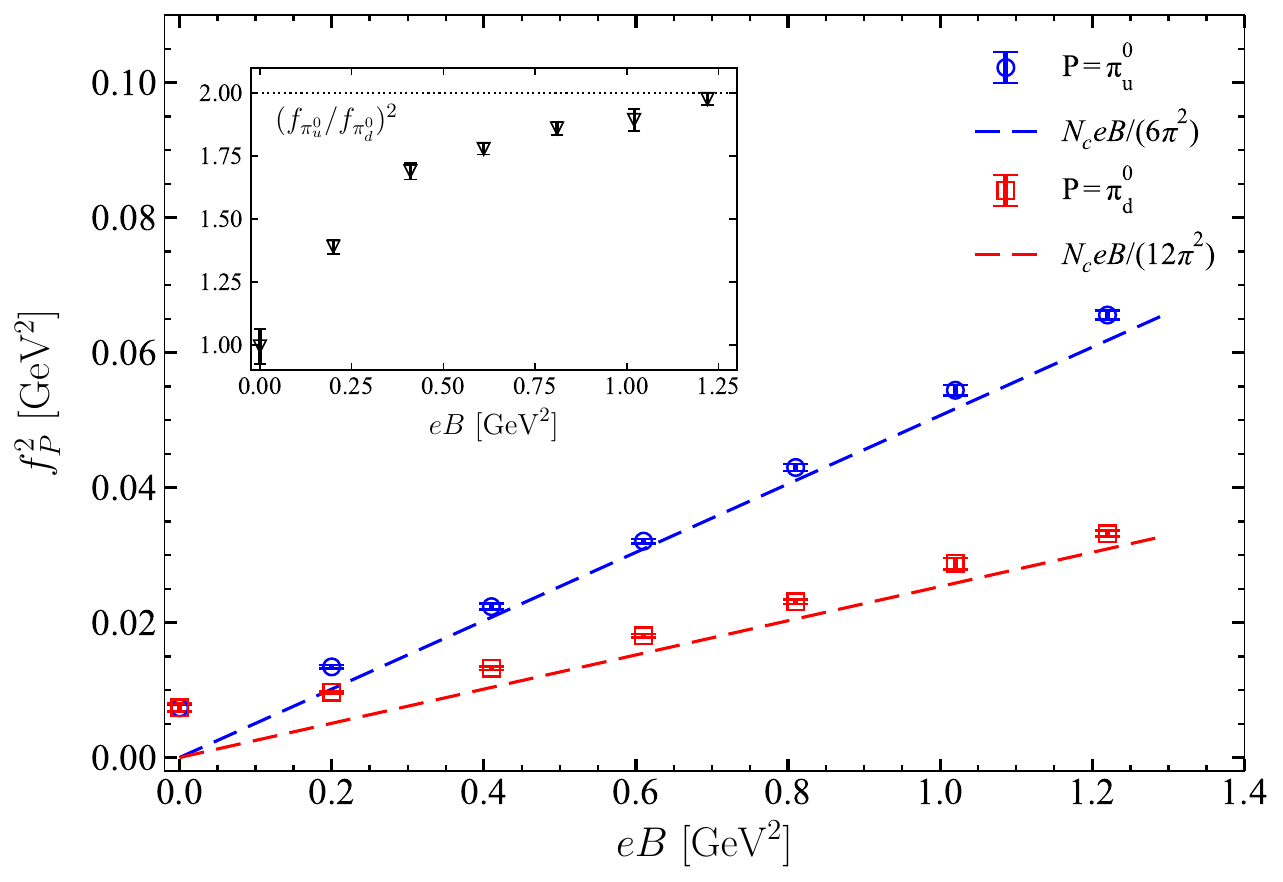}
    \caption{Flavor decomposition of the connected neutral-pion decay constant as a function of $eB$. Top: comparison of the individual flavor components $f_{\pi_u^0}$ and $f_{\pi_d^0}$ with the physical neutral pion $f_{\pi^0}$. Bottom: squared decay constants $f_{\pi_u^0}^2$ and $f_{\pi_d^0}^2$ versus $eB$. The dashed lines indicate the
    LLL expectation for the asymptotic slopes, $N_c eB/(6\pi^2)$ and $N_c eB/(12\pi^2)$ for the $u\bar u$ and $d\bar d$ channels,
    respectively~\cite{Miransky:2002rp}. The inset shows the ratio $(f_{\pi_u^0}/f_{\pi_d^0})^2$.}
    \label{fig:dc_pi0}
\end{figure}

To further investigate the charge dependence of the valence response, \autoref{fig:dc_pi0} shows a flavor decomposition of the connected neutral-pion channel. In this analysis we work throughout in the purely connected sector: The
$\pi^0$ decay constant $f_{\pi^0}$ and the flavor-resolved quantities
$f_{\pi_u^0}$ and $f_{\pi_d^0}$ are extracted in the same way from the corresponding connected correlators, following~\autoref{eq:fPpi_lat}, with
$f_{\pi_u^0}$ obtained from the connected $u\bar u$ correlator and $f_{\pi_d^0}$
from the connected $d\bar d$ correlator. 
The top panel in~\autoref{fig:dc_pi0} shows that $f_{\pi_u^0}$
increases substantially faster with $eB$ than $f_{\pi_d^0}$, consistent with the
larger magnitude of the $u$-quark electric charge, $|q_u|=2|q_d|$; the connected
$\pi^0$ result lies between these two components.

In the bottom panel in~\autoref{fig:dc_pi0}, the squared decay constants $f_{\pi_u^0}^2$ and $f_{\pi_d^0}^2$ show an approximately linear dependence on $eB$ for field strengths exceeding approximately 0.4~$\mathrm{GeV}^2$. We compare this trend to the strong-field
expectation discussed in Ref.~\cite{Miransky:2002rp}, where the relevant dynamics
is governed by Landau quantization at the quark level: In the regime dominated
by the LLL, the LLL degeneracy is proportional to $|q_f B|$,
leading to a linear-in-$|q_f B|$ growth of the decay constants of the associated
Nambu-Goldstone modes and a characteristic charge dependence of the slope. 
In our normalization this expectation is shown by the dashed lines with slopes
$N_c eB/(6\pi^2)$ and $N_c eB/(12\pi^2)$ for the $u\bar u$ and $d\bar d$
components, respectively; here $N_c=3$ is the number of colors. Our lattice data
follow these perturbative slopes closely at larger fields. The most direct signature of
this charge-controlled scaling is the ratio $(f_{\pi_u^0}/f_{\pi_d^0})^2$ shown in
the inset: Starting from unity at $eB=0$, it increases monotonically and
approaches $\simeq 2$ at our largest field $eB=1.22~\mathrm{GeV}^2$, consistent
with the expected factor $|q_u/q_d|=2$.

Finally, we emphasize the conceptual distinction from charged pseudoscalar
mesons: For charged states, the magnetic field Landau quantizes the
center-of-mass motion and the LLL can often be interpreted (within limitations)
in a point-particle picture, whereas for the neutral $\pi^0$ the center of mass
does not Landau quantize. The evidence for an ``LLL regime'' in
\autoref{fig:dc_pi0} should, therefore, be understood as an onset of quark-level
LLL control of the connected correlator amplitudes---and, hence, of the extracted
decay constants---rather than a point-particle Landau-level interpretation of the
neutral meson itself.

\subsection{Chiral condensates and the GMOR relation}
\label{subsec:pbp-GMOR}
\begin{figure}[!htbp]
    \centering
    \includegraphics[width=0.45\textwidth]{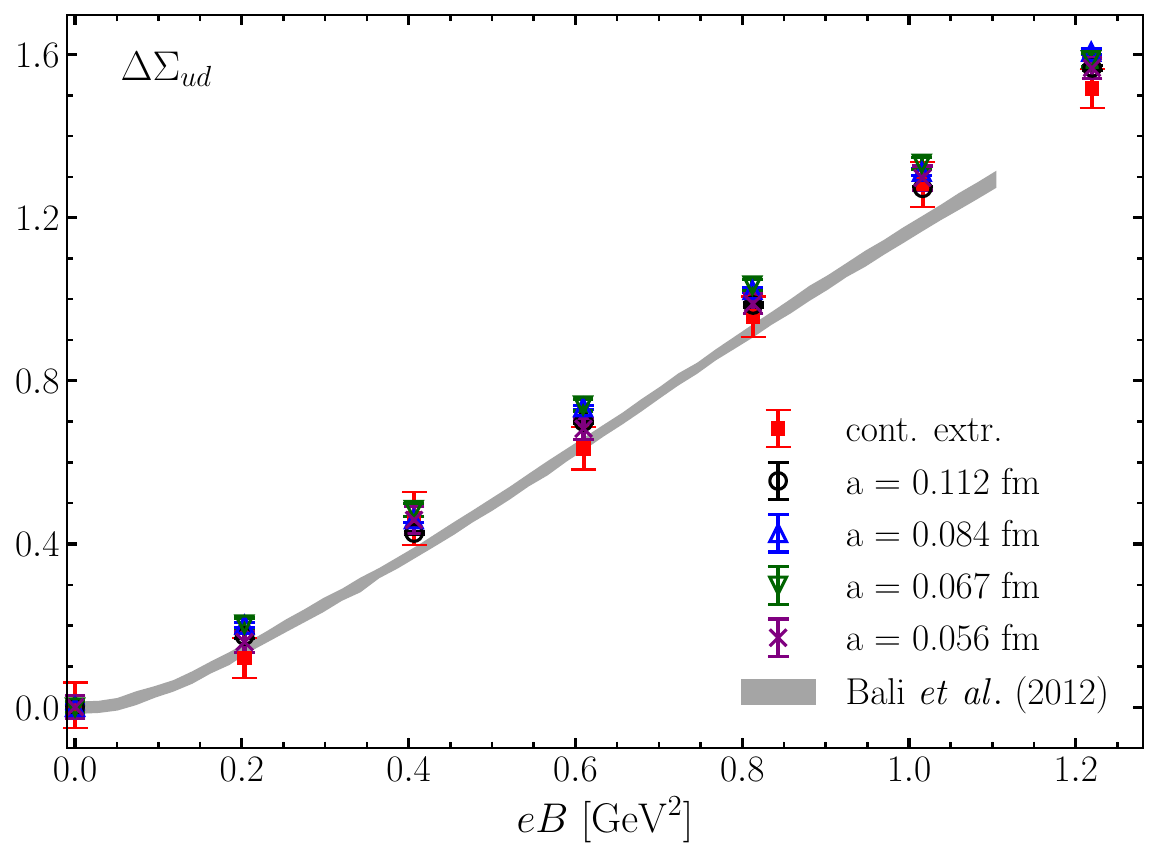}
    \includegraphics[width=0.45\textwidth]{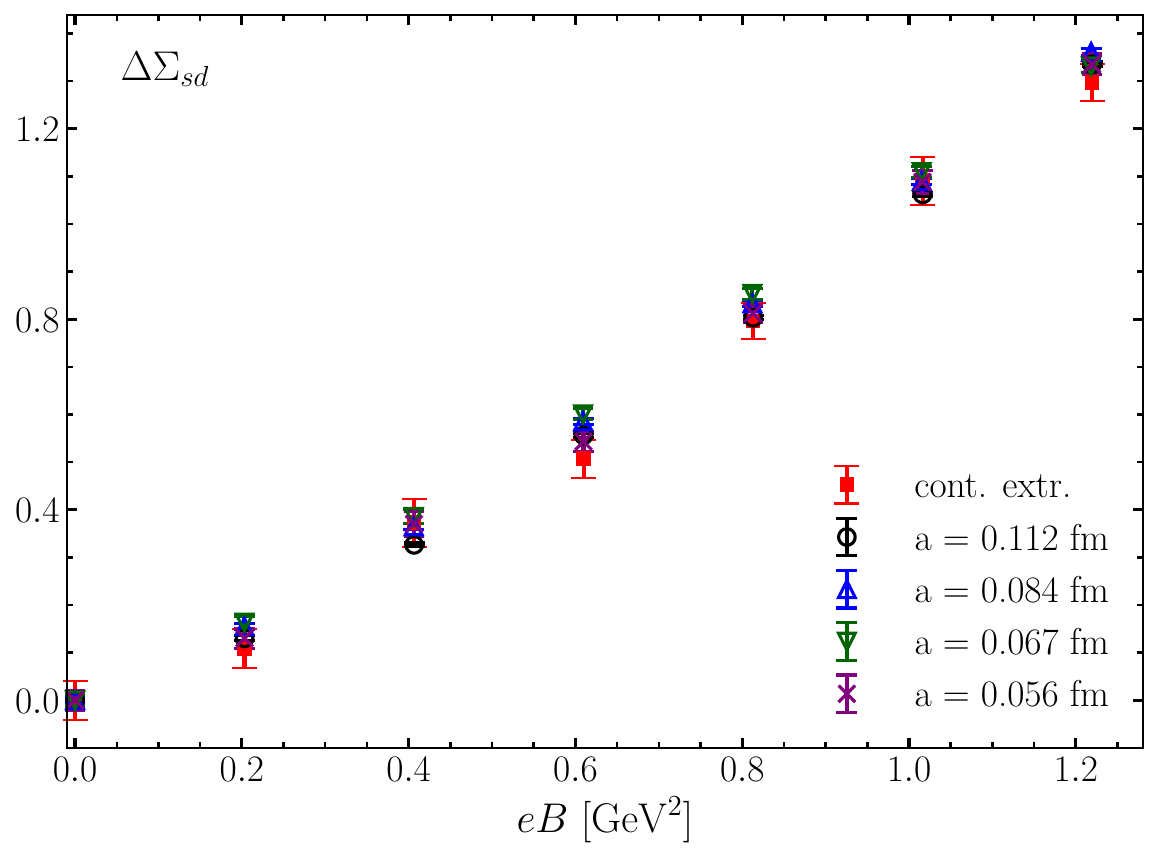}
    \caption{Renormalized chiral condensates as functions of $eB$. Top: $\Delta \Sigma_{ud}$. Bottom: $\Delta \Sigma_{sd}$. In the top panel, the gray band indicates the results of Bali \textit{et al.}~\cite{Bali:2012zg}.}
    \label{fig:PBPuds}
\end{figure}

The enhancement of the light-quark chiral condensate by a magnetic field, known as magnetic catalysis, is a fundamental phenomenon in QCD. In the following we use the renormalized, subtracted condensate
$\Delta \Sigma_{ud}(B)$ defined in Sec.~\ref{sec:basics}; it provides a dimensionless measure of
the magnetic-field response and is free of additive and multiplicative UV
divergences at $T\simeq 0$~\cite{Bali:2012zg}. For a direct comparison with Ref.~\cite{Bali:2012zg}, we adopt in \autoref{eq:DeltaSigma_l}
normalization with $M_{\pi^0}=m_{\pi^0}=135$~MeV
and $F_{\pi^0}=F_\pi=86$~MeV (SU(2) chiral limit), with $m_l\equiv m_u=m_d$.

Our continuum-extrapolated results for the renormalized light-quark condensate,
$\Delta \Sigma_{ud}$, are shown in the top panel in \autoref{fig:PBPuds}.  With this
convention, $\Delta \Sigma_{ud}$ increases monotonically with $eB$.
Quantitatively, our continuum-extrapolated results are consistent with the earlier lattice
determination of Bali \textit{et al.} within uncertainties while providing a
refined continuum control based on our set of ensembles.
This monotonic enhancement is the expected manifestation of magnetic catalysis at
$T\simeq 0$, i.e.,\ an increase of the chiral order parameter with $eB$.

Following the same strategy, we also analyze the strange-light combination
$\Delta \Sigma_{sd}(B)$ defined as \autoref{eq:DeltaSigma_sd} in Sec.~II.
For the normalization, we take the neutral-kaon mass from the PDG~\cite{ParticleDataGroup:2024cfk}, $M_{K^0}=m_{K^0}=498$~{MeV}, and use the decay constant in the SU(3) chiral limit, $F_{K^0}=F_{K}=71$~{MeV}, following Ref.~\cite{Bali:2022qja}. The corresponding continuum-extrapolated results are
shown in the bottom panel in \autoref{fig:PBPuds}. Compared to the light-quark
case, $\Delta \Sigma_{sd}$ exhibits a noticeably milder increase with $eB$.
This light-versus-strange hierarchy is qualitatively consistent with the pattern
seen in spectroscopy of pseudoscalar mesons in strong magnetic fields, where pion observables
typically show a stronger $eB$ dependence than kaon observables; see, e.g.,
Ref.~\cite{Ding:2020hxw}.

With the chiral condensate, pseudoscalar masses, and decay constants determined
at nonzero magnetic field, we can quantify the magnetic-field-induced violation
of the Gell-Mann--Oakes--Renner (GMOR) relation, beyond the explicit quark-mass
effects away from the chiral limit. The GMOR relation links the chiral order
parameter to pseudoscalar properties and, thus, provides a compact way to track
how magnetic catalysis and the field dependence of hadron observables combine
into deviations from leading-order chiral dynamics.
We write the GMOR relations
for the neutral pion and kaon in a form that applies at arbitrary $B$,
\begin{align}
    (m_u+m_d)\Big(\langle\bar{\psi}\psi\rangle_u(B)
    +\langle\bar{\psi}\psi\rangle_d(B)\Big)
    &= 2\,f_{\pi^0}^2(B)\,M_{\pi^0}^2(B)\,\nonumber\\
    &\qquad\times\Big[1-\delta_{\pi^0}(B)\Big],
    \label{eq:GMOR_pi_B}\\
    (m_s+m_d)\Big(\langle\bar{\psi}\psi\rangle_s(B)
    +\langle\bar{\psi}\psi\rangle_d(B)\Big)
    &= 2\,f_{K^0}^2(B)\,M_{K^0}^2(B)\,\nonumber\\
    &\qquad\times\Big[1-\delta_{K^0}(B)\Big],
    \label{eq:GMOR_K_B}
\end{align}
where $f_{P}(B)$ and $M_{P}(B)$ are the field-dependent decay constants and
ground-state masses determined on the lattice, respectively, and $\delta_P(B)$ parametrizes
the deviation from the leading-order GMOR relation. At $B=0$,
$\delta_{\pi^0}(0)\equiv\delta_{\pi}$ and $\delta_{K^0}(0)\equiv\delta_{K}$ reduce
to the usual vacuum next-to-leading-order (NLO) chiral corrections associated with low-energy constants.
At the physical point, estimates from chiral perturbation theory (ChPT) and QCD sum rules give
$\delta_{\pi}=(6.2\pm1.6)\%$ and $\delta_{K}=(55\pm5)\%$~\cite{Bordes:2010wy,Bordes:2012ud}.

To isolate the effect induced by the magnetic field $eB$ and to minimize sensitivity
to the absolute normalization of the condensate, we focus on the shift
\begin{equation}
    \Delta\delta_P(eB)\equiv \delta_P(eB)-\delta_P(0).
\end{equation}
Using the UV-free renormalized condensates defined in
Eqs.~\eqref{eq:DeltaSigma_l} and \eqref{eq:DeltaSigma_sd}, we can express
$\Delta\delta_P(eB)$ directly in terms of 
$f_P^2(eB)M_P^2(eB)$ and the renormalized condensate. For the pion
sector this yields
\begin{align}
    \Delta \delta_{\pi^0}(eB)
    =\;& 1-\delta_{\pi}
    -\frac{F_{\pi}^{2}m_{\pi^0}^{2}}
          {f_{\pi^0}^{2}(eB)\,M_{\pi^0}^{2}(eB)}
    \nonumber\\
    &\times
    \left( 1 + \Delta \Sigma_{ud} - \delta_{\pi} \right),
    \label{eq:delta_pi_final}
\end{align}
and, analogously, for the kaon sector,
\begin{align}
    \Delta \delta_{K^0}(eB)
    =\;& 1 - \delta_{K}
    -\frac{F_{K}^{2}m_{K^0}^{2}}
          {f_{K^0}^{2}(eB)\,M_{K^0}^{2}(eB)}
    \nonumber\\
    &\times
    \left( 1 + \Delta \Sigma_{sd} - \delta_{K} \right),
    \label{eq:delta_K_final}
\end{align}
where $(m_{\pi^0},F_{\pi})$ and $(m_{K^0},F_{K})$ are the fixed $eB=0$ reference
inputs used in the definitions of $\Delta \Sigma_{ud}$ and $\Delta \Sigma_{sd}$
, [see the discussion below Eqs.~\eqref{eq:DeltaSigma_l} and
\eqref{eq:DeltaSigma_sd}].

\begin{figure}[!htbp]
    \centering
    \includegraphics[width=0.4\textwidth]{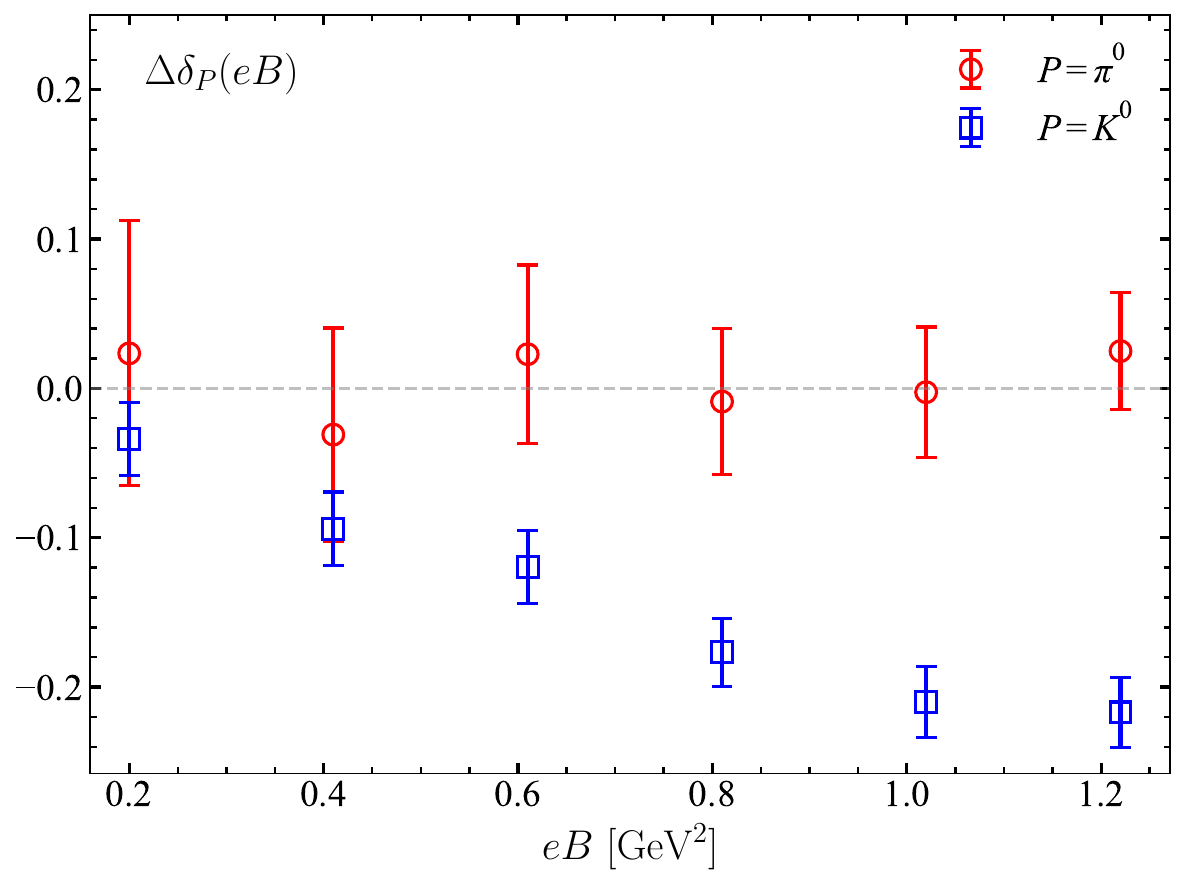}
    \caption{Magnetic-field-induced shift of the GMOR correction $\Delta\delta_P(eB)$ for the neutral pion (red circles) and neutral kaon (blue squares). }
    \label{fig:delta_difference}
\end{figure}

Our numerical results for $\Delta \delta_{\pi^0}(eB)$ and $\Delta \delta_{K^0}(eB)$
are shown in~\autoref{fig:delta_difference}. Over the full range of $eB$ studied,
$\Delta \delta_{\pi^0}(eB)$ remains small and is consistent with zero within
uncertainties, indicating that the magnetic-field dependence of
$f_{\pi^0}^2(eB)M_{\pi^0}^2(eB)$ closely tracks that of the renormalized light-quark condensate.
This trend is naturally interpreted in light of leading-order chiral perturbation
theory at $T=0$: In the weak-field expansion, the magnetic shifts of light-quark chiral condensate,
$f_{\pi^0}$, and $M_{\pi^0}$ conspire such that the GMOR relation for the neutral
pion is preserved; see, e.g., Ref.~\cite{Shushpanov:1997sf}. Remarkably, our
continuum-extrapolated results at the physical point show that this cancellation
persists up to $eB\simeq 1.2~\mathrm{GeV}^2$, i.e., well beyond the nominal regime
of applicability of the weak-field expansion. At such fields, standard ChPT is
expected to break down; complementary effective approaches that incorporate
explicit $q\bar q$ dynamics nevertheless suggest that the GMOR relation remains
robust for neutral Nambu-Goldstone mesons (in contrast to charged channels); see,
e.g., Ref.~\cite{Orlovsky:2013gha}. For completeness, we note that the same LO-ChPT
preservation mechanism has also been discussed at finite temperature in background
magnetic fields~\cite{Agasian:2001ym}, although our focus here is on $T\simeq 0$.

In contrast, the kaon sector exhibits a sizable violation, with
$\Delta \delta_{K^0}(eB)\simeq -0.22$ at $eB=1.22~\mathrm{GeV}^2$.
This qualitative difference between $\pi^0$ and $K^0$ is not unexpected:
Even at $eB=0$ the kaon receives large NLO corrections in SU(3) chiral dynamics,
as reflected by the sizable vacuum value $\delta_K$ quoted above, and the
magnetic field further accentuates the mismatch between the strange-sector
condensate combination $(\langle\bar{\psi}\psi\rangle_s+\langle\bar{\psi}\psi\rangle_d)$
and the pseudoscalar combination $f_{K^0}^2(eB)M_{K^0}^2(eB)$ entering
Eq.~\eqref{eq:delta_K_final}. Equivalently, unlike the light sector where
$f_{\pi^0}^2(eB)M_{\pi^0}^2(eB)$ closely tracks the corresponding condensate factor in
Eq.~\eqref{eq:delta_pi_final}, the strange sector responds more mildly.
In particular, the field dependence encoded in $\Delta \Sigma_{sd}$ does not keep
pace with the variation of $f_{K^0}^2(eB)M_{K^0}^2(eB)$, leading to the observed
negative shift in $\Delta\delta_{K^0}(eB)$. Our findings are consistent with earlier
lattice observations of a small (or absent) GMOR shift in the neutral-pion channel
and a more pronounced deviation in the neutral-kaon channel in background magnetic
fields~\cite{Ding:2020hxw}, and here we confirm this pattern at the physical point
with controlled continuum extrapolation.

We finally comment on how the observed reduction of the Goldstone mode may relate to the decrease of the
pseudocritical temperature in magnetic fields. The GMOR systematics underscore the special role of the Goldstone
sector in chiral dynamics; however, the decrease of $T_{pc}(eB)$ need not be tied one to one to inverse magnetic
catalysis. Indeed, lattice studies at heavier-than-physical pion masses show that the magnetic-field-induced
suppression of the chiral condensate around $T_{pc}$ can cease, while $T_{pc}$ nonetheless continues to decrease
with increasing $eB$~\cite{DElia:2018xwo,Endrodi:2019zrl}.

This apparent tension can be understood as reflecting different physical regimes. In the light-quark regime, where
chiral symmetry and its (approximate) restoration control the crossover, it is natural to correlate the reduction of
the chiral condensate with a downward shift of $T_{pc}$. By contrast, as the quark masses are increased toward the
heavy-quark limit (and ultimately toward the pure-gauge theory), the chiral condensate ceases to be the relevant
order parameter for the transition, and the connection between condensate suppression and the transition temperature
is no longer expected to hold.

This motivates a complementary interpretation in terms of hadronic thermodynamics and the magnetically modified
spectrum. At $eB=0$, it is well known that lighter pions correlate with a lower crossover temperature~\cite{Ding:2019prx,Ding:2020rtq,Bazavov:2017xul,Kotov:2021rah,Aarts:2020vyb,Bhattacharya:2014ara,Umeda:2016qdo,Li:2020wvy}.
In the same spirit, a lighter neutral pion at $eB\neq 0$ carries a larger thermodynamic weight at a fixed temperature, thereby enhancing its contribution to bulk observables and making
thermal excitation of the medium more efficient; this naturally shifts the crossover to lower temperatures.
This expectation is borne out in hadronic-model analyses in magnetic fields, where the $\pi^0$ contribution is found
to dominate the energy density over a broad range of parameters~\cite{Ding:2022uwj}.

\section{Conclusion}
\label{sec:conclusion}
In this work we investigated the chiral properties of $(2\!+\!1)$-flavor QCD in background magnetic fields at $T\simeq 0$ using the HISQ action at the physical point.
Using ensembles at four lattice spacings, we controlled discretization effects and obtained continuum-extrapolated results for renormalized chiral condensates and for pseudoscalar observables in both the neutral and charged sectors.
For the pseudoscalar masses we employed two complementary extraction strategies—an AICc-guided fit analysis and an oblique Lanczos spectral approach—which provide a valuable cross-check and a handle on systematic uncertainties.
In addition, we analyzed magnetic-field-induced corrections to the GMOR relation and disentangled valence- and sea-quark contributions to the magnetic response.

For the chiral sector, we find clear magnetic catalysis in the vacuum: The renormalized light-quark chiral condensate increases monotonically with $eB$.
For the strange sector, we consider a strange-light condensate combination, which also exhibits an enhancement with $eB$ but with a noticeably milder response, reflecting the reduced sensitivity of strange degrees of freedom to the magnetic field at fixed $T\simeq 0$.
These results sharpen the quantitative picture of chiral symmetry breaking and its flavor dependence in magnetized QCD in the vacuum.

Turning to the pseudoscalar spectrum, the neutral channel shows a pronounced decrease of $M_{\pi^0}$ with increasing $eB$, with analogous (but weaker) trends for $K^0$ and $\eta^0_{s\bar s}$.
In the charged channel, we observe a characteristic nonmonotonic behavior: $M_{\pi^\pm}$ and $M_{K^\pm}$ rise in the small-field region consistent with the LLL expectation, but deviate at intermediate fields and approach a saturation regime, with a slight downward trend at the largest fields studied.
This departure from the point-particle Landau-level picture highlights the onset of genuinely non-pointlike dynamics in strong magnetic fields.
By separating valence- and sea-quark contributions, we further showed that the magnetic-field dependence of pseudoscalar masses on the vacuum ensembles is overwhelmingly carried by valence-quark effects, with only a minor sea-quark contribution. This is also reflected by the approximate $qB$ scaling of the neutral-pion connected components: After rescaling the field by the quark charges, the $u\bar u$ and $d\bar d$ contributions nearly collapse onto a common curve, indicating that $qB$ scaling is only mildly broken in full QCD and, thus, that sea-quark effects remain small. Moreover, the sum $M_{\pi_u^0}(eB)+M_{\pi_d^0}(eB)$ remains finite at our largest field, implying a nonzero lower bound on $M_{\rho^\pm}(eB)$ via the QCD inequality and thereby disfavouring a $\rho$-condensation scenario up to the fields explored here.

For the decay constants, we presented continuum-limit results for the neutral pseudoscalars, $f_{\pi^0}$, $f_{K^0}$, and $f_{\eta^0_{s\bar s}}$.
All three increase monotonically with $eB$, with the strongest enhancement in the pion channel and a progressively milder response in the strange-containing channels, mirroring the hierarchy seen in the condensate sector.

Combining condensates, masses, and decay constants, we quantified the magnetic-field-induced shifts of the GMOR corrections.
In the neutral-pion channel, the GMOR correction remains small over the full field range studied, consistent with a substantial cancellation between the $eB$ dependence of $M_{\pi^0}$ and $f_{\pi^0}$ that persists well beyond the nominal weak-field regime.
In contrast, the neutral-kaon channel exhibits a sizable violation, reflecting the stronger role of SU(3) corrections and the comparatively mild strange-sector response of the condensate combination entering the kaonic GMOR relation.

The results presented here provide continuum-limit benchmarks for effective-theory descriptions and model calculations of magnetized QCD in the vacuum.
A natural next step is to extend the spectroscopy to additional channels, in particular vector mesons as well as baryons.
For such studies, Wilson-type fermions (e.g., clover-improved formulations) may be a better choice, as they avoid staggered taste breaking and allow for a more direct treatment of spin-dependent structures.

\section*{Acknowledgments}
This work is supported partly by the National Natural Science Foundation of China under Grants No. 12293064, No. 12293060, and No. 12325508, as well as the National Key Research and Development Program of China under Contract No. 2022YFA1604900 and the Fundamental Research Funds for the Central Universities, Central China Normal University under Grants No. 30101250314 and No. 30106250152. The numerical simulations have been performed on the GPU cluster in the Nuclear Science Computing Center at Central China Normal University ($\mathrm{NSC}^{3}$) and Wuhan Supercomputing Center.

\section*{Data Availability}
The data that support the findings of this article are openly available~\cite{ding_2026_19718601}; embargo periods may apply.

\appendix

\section{Comparison of AICc and Oblique Lanczos Mass Extractions}
\label{app:aicc_lanczos_comparison}
\begin{figure}[!htpb]
    \centering
    \includegraphics[width=0.45\textwidth]{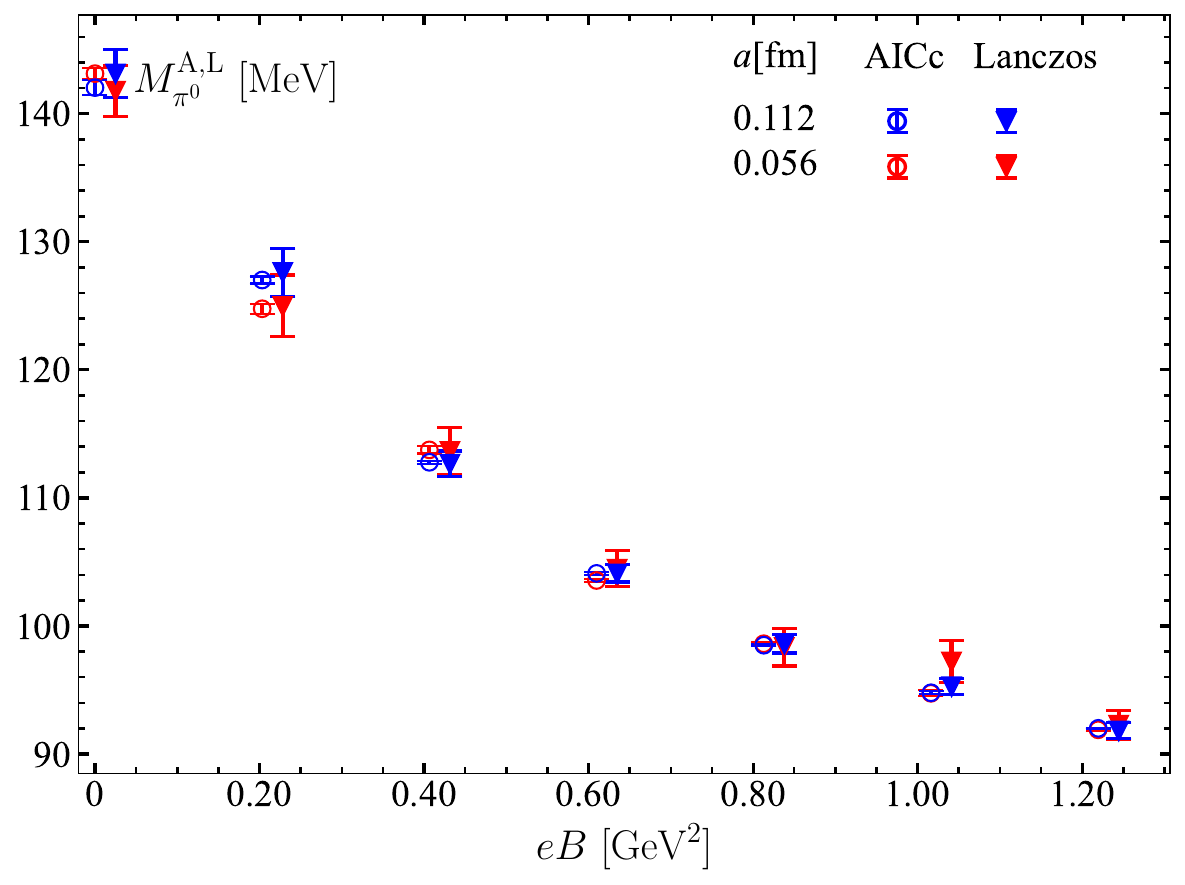}
    \includegraphics[width=0.45\textwidth]{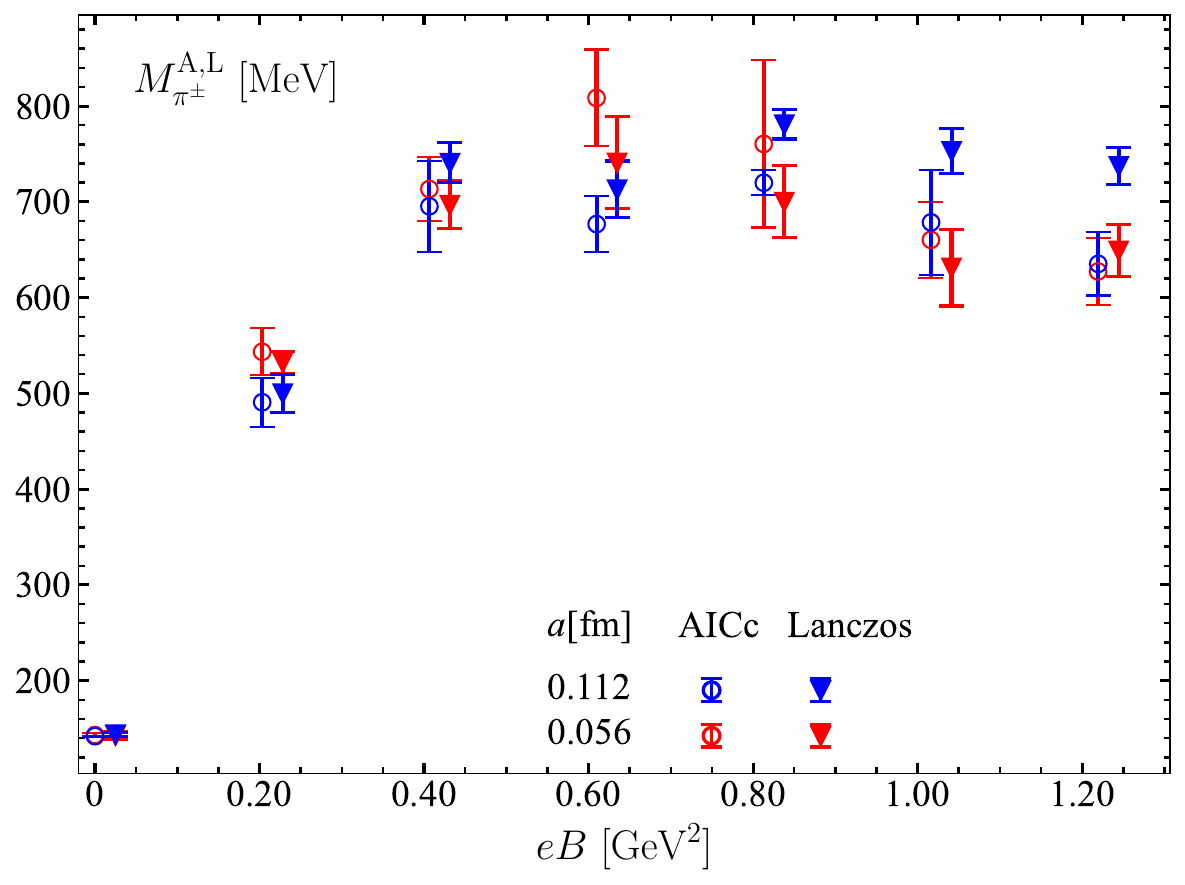}
    \caption{Comparison of pseudoscalar pion masses extracted using the AICc-selected fits (open circles) and the oblique Lanczos method (filled triangles) as a function of the magnetic-field strength $eB$. The top and bottom panels show the neutral ($\pi^0$) and charged ($\pi^\pm$) sectors, respectively. Blue and red symbols denote results from the coarsest ($a=0.112$~fm) and finest ($a=0.056$~fm) lattice spacings. For better visibility, the Lanczos data points are slightly shifted horizontally.}
    \label{fig:lanczos_comparison_pion}
\end{figure}

\begin{figure}[!htpb]
    \centering
    \includegraphics[width=0.45\textwidth]{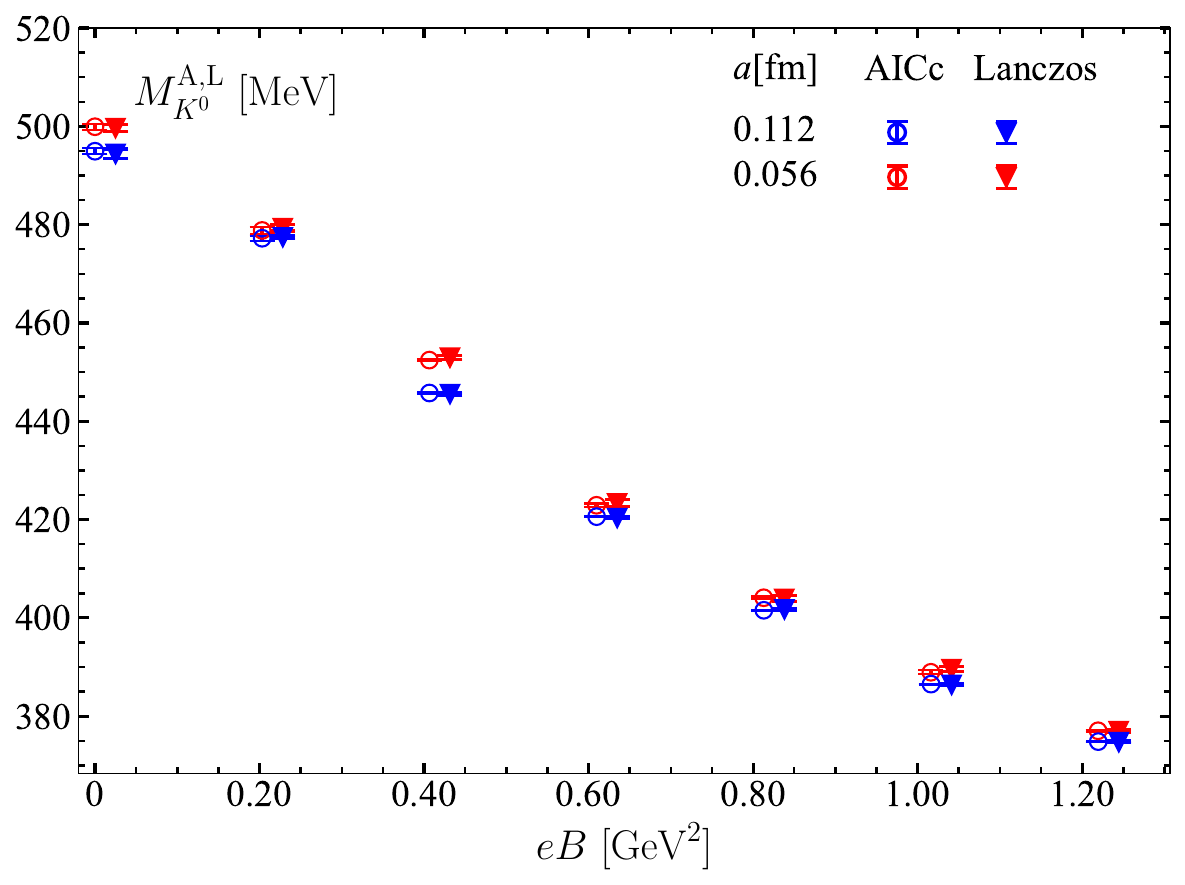}
    \includegraphics[width=0.45\textwidth]{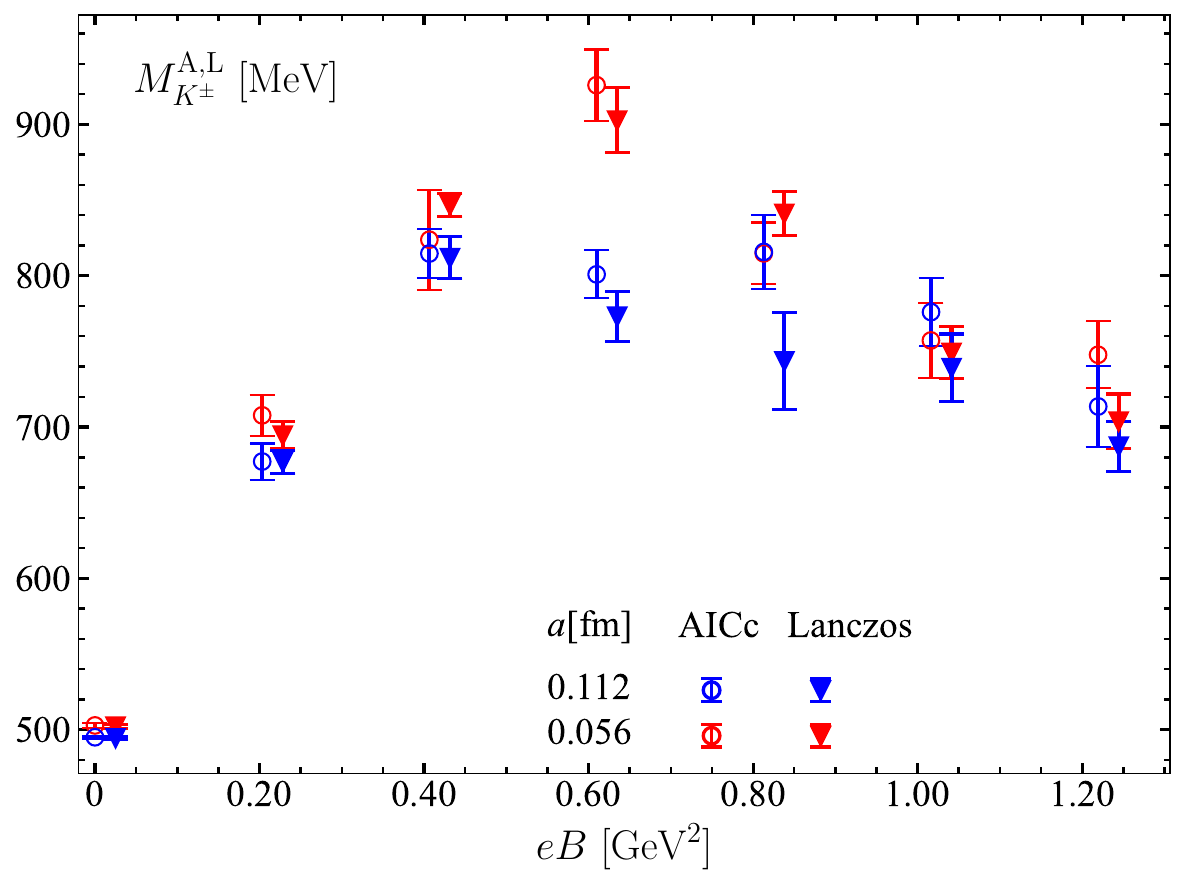}
    \caption{The same as \autoref{fig:lanczos_comparison_pion} but for kaons.}
    \label{fig:lanczos_comparison_kaon}
\end{figure}

This appendix summarizes a systematic comparison of ground-state masses extracted with
(i) AICc-selected multistate fits and (ii) the oblique Lanczos method~\cite{Wagman:2024rid,Ostmeyer:2024qgu}. While a representative comparison of ground-state mass plateaus for $M_{\pi^0_u}$ on a single ensemble at a specific magnetic-field strength is provided in the main text (see~\autoref{fig:plateau_example}), here, we present a comprehensive comparison of the extracted masses for pions and kaons across the full range of magnetic-field strengths. For visual clarity, we show results only for our finest and coarsest lattice spacings, as illustrated in \autoref{fig:lanczos_comparison_pion} for pions and \autoref{fig:lanczos_comparison_kaon} for kaons. Note that the Lanczos data points are slightly shifted in $eB$ for visibility.

Overall, the two methods agree well within uncertainties in both pion and kaon channels.
In the neutral-pion sector, where the correlators have a high signal-to-noise ratio, AICc fits typically yield smaller statistical errors, and the central values from Lanczos are consistent: The relative differences are $\le 0.8\%$ at
$a=0.112~\mathrm{fm}$ and $\le 2.6\%$ at $a=0.056~\mathrm{fm}$ across the shown field range.
In the charged pion sector, correlators become substantially noisier at large $eB$; here the Lanczos method remains effective,
and the two extractions agree in the vast majority of cases. Quantitatively, the mean (median) relative difference is
about $5\%$ ($4.8\%$), with a maximum discrepancy of $16\%$.

The kaon sector shows an analogous pattern. For the neutral kaon, agreement is extremely tight, with an average relative
difference of $0.07\%$ and a maximum of $0.18\%$. For the charged kaon, discrepancies remain modest despite increased noise,
with a mean (median) relative difference of $2.76\%$ ($2.64\%$) and a maximum of $8.84\%$.

The purpose of this comparison is not to require the two methods to have identical statistical precision, but to test the stability of the extracted central values under two qualitatively different analysis strategies. To obtain robust final results, we include the variation between the AICc-selected and oblique Lanczos extractions, together with the continuum-extrapolation variations described in Appendix~\ref{app:cont_extr}, in the combined bootstrap procedure used to define the final quoted result and its uncertainty.

% To obtain robust final results, we treat the variation between the AICc-selected and oblique Lanczos extractions as a
% systematic uncertainty, combined with the continuum-extrapolation systematics described in Appendix～\ref{app:cont_extr}.

\section{Continuum Extrapolation}
\label{app:cont_extr}
\begin{figure}[!htpb]
        \includegraphics[width=0.45\textwidth]{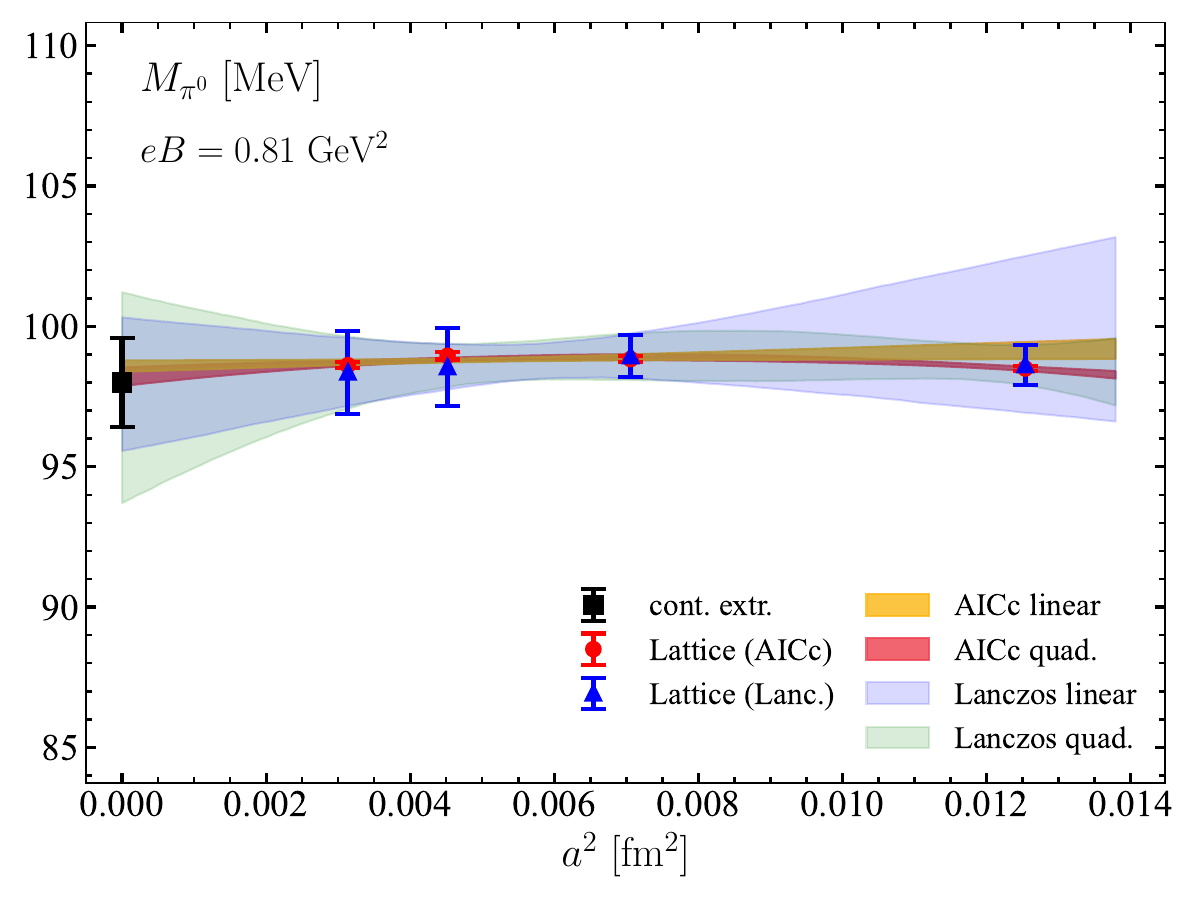} \\
        \caption{Representative continuum extrapolation of the neutral-pion mass $M_{\pi^0}$ at $eB=0.81$~$\mathrm{GeV}^2$. Lattice results are shown for the AICc (red circles) and Lanczos (blue triangles) analysis methods. The shaded bands indicate four individual extrapolation ansätze: AICc quadratic (red), AICc linear (orange), Lanczos quadratic (green), and Lanczos linear (blue). The black marker at $a^2=0$ denotes the final extrapolated result as described in the text.}
        \label{fig:pion_Nb8_cont_extr}
\end{figure}

\begin{figure}[!htpb]
        \includegraphics[width=0.45\textwidth]{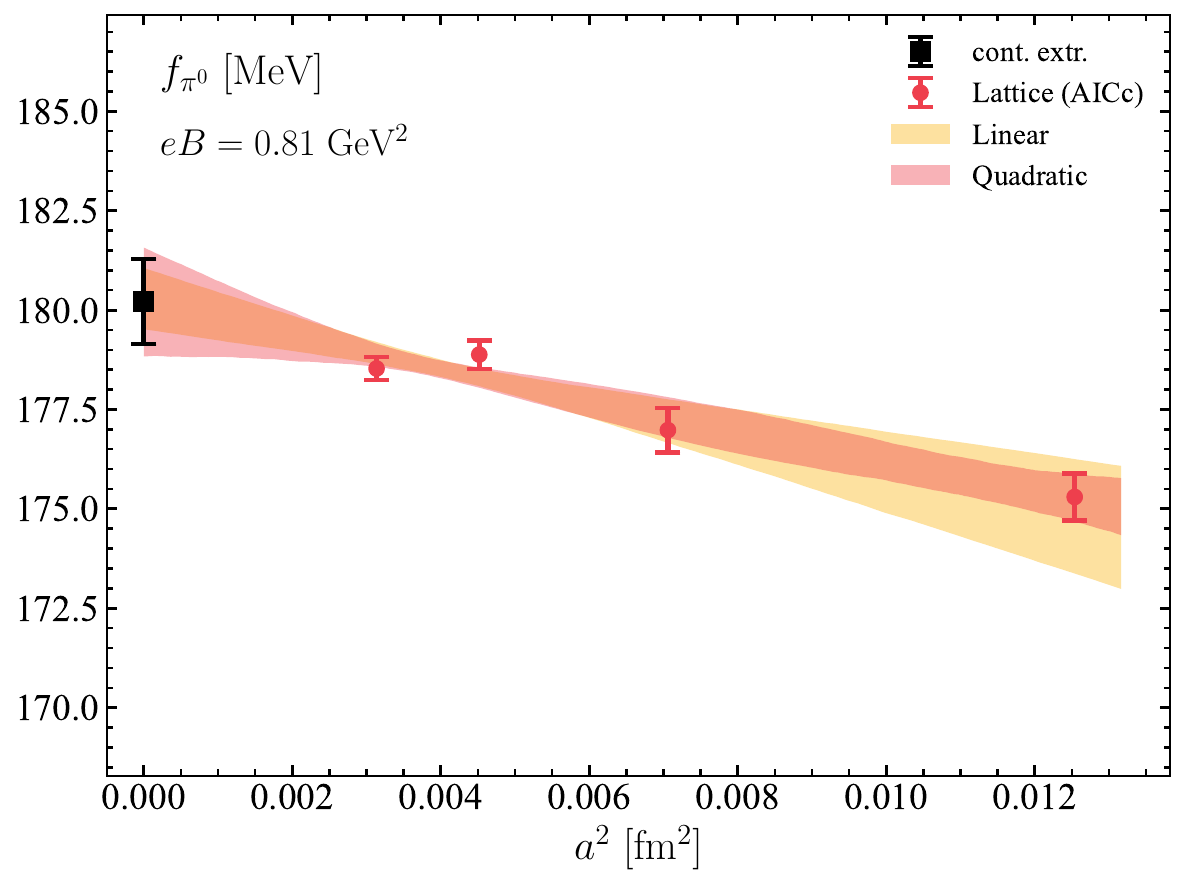} \\
        \caption{Representative continuum extrapolation of the neutral-pion decay constant $f_{\pi^0}$ at $eB=0.81$~$\mathrm{GeV}^2$. Lattice results from the AICc analysis are denoted by red circles. The shaded bands represent the linear (orange) and quadratic (red) extrapolation fits. The final continuum-extrapolated result, shown as a black marker at $a^2=0$, incorporates both ansätze as described in the text.}
        \label{fig:dc_pion_Nb8_cont_extr}
\end{figure}

In this appendix, we detail the procedure used to obtain continuum-limit results for the various observables.

For ground-state meson masses, we use two independent extraction methods: AICc-selected multistate fits and the oblique Lanczos method. 
    In addition, we consider two continuum-extrapolation \textit{Ans\"atze}. Specifically, we use
    \begin{equation}
    O(a^2)=O_{\rm cont}^{\rm lin}+A a^2 ,
    \label{eq:lin_cont}
    \end{equation}
    for a linear fit in $a^2$ using the three finest lattice spacings, and
    \begin{equation}
    O(a^2)=O_{\rm cont}^{\rm quad}+B a^2 + C a^4 ,
    \label{eq:quad_cont}
    \end{equation}
    for a quadratic fit including an $a^4$ term using all four spacings. This leads to four analysis choices in total, which we label by $\alpha=1,\dots,4$.
    
    For each bootstrap sample $b$, we obtain four continuum estimates $O_b^{(\alpha)}$ and define the averaged estimator
    \begin{equation}
    \overline{O}_b=\frac{1}{4}\sum_{\alpha=1}^{4} O_b^{(\alpha)} .
    \label{eq:avg_boot}
    \end{equation}
    Our final central value is taken as the median
    \begin{equation}
    O_{\rm cent} = {\rm median}_b \left(\overline{O}_b\right) .
    \label{eq:cent_boot}
    \end{equation}
    The quoted uncertainty is determined from the central 16th--84th percentile interval of the distribution of $\overline{O}_b$,
    \begin{equation}
    \sigma = \text{max}\{ q_{84}(\overline{O}_b)-O_{\rm cent}, O_{\rm cent}-q_{16}(\overline{O}_b)\}\,,
    \label{eq:tot_err}
    \end{equation}
    where $q_{16}$ and $q_{84}$ denote the 16th and 84th percentiles, respectively.
    
    Since the estimator itself is constructed by combining the accepted analysis choices within each bootstrap sample, this interval is interpreted as the total uncertainty of the final combined estimator. In this way, both the bootstrap fluctuations and the variation among the accepted analysis choices are propagated through the same final distribution. A representative example for $M_{\pi^0}$ at $eB=0.81~\mathrm{GeV}^2$ is shown in~\autoref{fig:pion_Nb8_cont_extr}.
    
    For decay constants, only the AICc-selected analysis is applicable, since the Lanczos method yields eigenvalues but not matrix elements. In this case, only the two continuum-extrapolation \textit{Ans\"atze} contribute to the analysis variation. For each bootstrap sample, we therefore average the linear and quadratic continuum estimates to define the bootstrap estimator. The central value is taken as the median of the resulting bootstrap distribution, and the quoted uncertainty is defined from its 16th--84th percentile interval in the same way as above, again using the larger of the upper and lower deviations from the median when quoting a single symmetric uncertainty. A representative example for $f_{\pi^0}$ at $eB=0.81~\mathrm{GeV}^2$ is shown in \autoref{fig:dc_pion_Nb8_cont_extr}. The renormalized chiral condensates $\Delta\Sigma_{ud}$ and $\Delta\Sigma_{sd}$ are extrapolated using the same procedure as for the decay constants.
    
    For transparency, the continuum central values and bootstrap uncertainties obtained from the individual analysis choices are listed in Appendix~C together with the final quoted results.

\section{Summary of Numerical Results}
\label{app:summary_data_tables}
In this appendix, we summarize the continuum-extrapolated numerical results
for meson masses, decay constants, and the light-quark chiral condensate at
representative values of the magnetic-field strength $eB$. All values reported
in the tables correspond to results in the continuum limit obtained following the
bootstrap-based extrapolation procedure described in
Appendix~\ref{app:cont_extr}.

\begin{table}[!htbp]
	\centering
    \setlength{\tabcolsep}{1pt}
    \footnotesize
    \begin{tabular}{*{6}{c}}
		\toprule  
		\toprule  
        \multirow{2.3}*{$eB[\mathrm{GeV}^2]$} 
	&   \multicolumn{5}{c}{cont.extr. of Mass [MeV].} \\
    \cmidrule(lr){2-6}
    & $\pi^0$ & $K^0$ & $\eta^0_{s\bar{s}}$ & $\pi^\pm$ & $K^\pm$ \\
    \midrule 
    0.0 & 142.4(3.9) & 505.3(1.8) & 697.4(1.1) & 144.7(5.0) & 506.0(2.0) \\
    0.2 & 124.7(2.7) & 479.5(1.0) & 672.2(0.6) & 551.3(28.2) & 703.6(19.7) \\
    0.41 & 115.0(2.1) & 455.0(0.6) & 640.9(0.5) & 747.8(52.4) & 847.1(35.4) \\
    0.61 & 103.5(1.4) & 422.7(0.8) & 604.4(0.5) & 801.9(95.2) & 1054.9(34.0) \\
    0.81 & 98.0(1.6) & 404.0(0.6) & 580.6(0.4) & 830.1(95.6) & 897.6(27.9) \\
    1.02 & 96.1(1.6) & 388.2(0.6) & 560.6(0.3) & 632.8(74.5) & 805.7(32.4) \\
    1.22 & 91.0(1.2) & 375.1(0.5) & 543.9(0.2) & 687.4(65.1) & 706.0(32.0) \\
		\bottomrule  		
	\end{tabular}
    \caption{continuum-extrapolated masses of pseudoscalar mesons at different values of the magnetic-field strength $eB$. Uncertainties include both statistical and systematic effects.}
    \label{tab:meson_masses}
\end{table}

\begin{table}[!htbp]
	\centering
    \footnotesize
    \begin{tabular}{*{6}{c}}
		\toprule  
		\toprule  
        \multirow{2.3}*{$eB[\mathrm{GeV}^2]$} 
	&   \multicolumn{5}{c}{cont.extr. of Decay Constant [MeV].} \\
    \cmidrule(lr){2-6}
    & $\pi^0$ & $K^0$ & $\eta^0_{s\bar{s}}$ & $\pi^0_u$ & $\pi^0_d$ \\
    \midrule 
    0.0 & 85.8(2.9) & 107.0(1.2) & 126.2(0.6) & 85.8(2.6) & 85.8(2.8) \\
    0.2 & 106.7(1.4) & 114.6(0.7) & 130.2(0.3) & 116.0(1.2) & 98.2(1.0) \\
    0.41 & 131.8(1.4) & 128.3(0.4) & 139.3(0.2) & 149.4(1.6) & 114.9(1.2) \\
    0.61 & 157.4(0.7) & 139.8(0.5) & 149.4(0.2) & 179.0(0.9) & 134.4(1.0) \\
    0.81 & 180.2(1.1) & 154.6(0.5) & 162.4(0.2) & 207.3(1.3) & 151.9(1.1) \\
    1.02 & 202.6(1.8) & 168.7(0.6) & 174.5(0.2) & 233.3(1.7) & 169.6(2.6) \\
    1.22 & 220.8(1.4) & 180.9(0.5) & 186.5(0.1) & 256.1(1.4) & 182.1(1.3) \\
		\bottomrule  		
	\end{tabular}
	\caption{continuum-extrapolated decay constants of neutral pseudoscalar mesons as functions of the magnetic-field strength $eB$. Uncertainties include both statistical and continuum-extrapolation systematic effects.}
    \label{tab:decay_constants}
\end{table}

\begin{table}[!htbp]
    \centering
    \footnotesize
    \begin{tabular}{c c c c c}
        \toprule  
        \toprule
        \multirow{2.5}{*}{$eB\ [\mathrm{GeV}^2]$} 
        & \multirow{2.5}{*}{$\Delta \Sigma_{ud}$ }
        & \multirow{2.5}{*}{$\Delta \Sigma_{sd}$ }
        & \multicolumn{2}{c}{Mass [MeV]} \\
        \cmidrule(lr){4-5}
        & & & $M_{\pi^0_u}$ & $M_{\pi^0_d}$ \\
        \midrule 
        0.00 & 0.004(0.055) & -0.001(0.041) & 142.1(4.1) & 142.4(4.0) \\
        0.20 & 0.120(0.045) & 0.108(0.039) & 119.2(3.0) & 131.7(2.4) \\
        0.41 & 0.462(0.065) & 0.373(0.050) & 107.7(2.6) & 125.6(1.8) \\
        0.61 & 0.634(0.052) & 0.507(0.040) & 97.8(1.3) & 110.9(1.7) \\
        0.81 & 0.958(0.049) & 0.796(0.038) & 93.6(1.7) & 106.7(2.0) \\
        1.02 & 1.282(0.055) & 1.091(0.048) & 90.2(1.3) & 104.2(2.5) \\
        1.22 & 1.517(0.047) & 1.297(0.039) & 87.1(1.2) & 97.3(2.0) \\
        \bottomrule  		
    \end{tabular}
    \caption{continuum-extrapolated results of the renormalized chiral condensate $\Delta \Sigma_{ud}$ and $\Delta \Sigma_{sd}$ and the masses of $u$ and $d$ flavored neutral pions at various $eB$.}
    \label{tab:pbp_Mpiud}
\end{table}
\newpage
\begin{table}[!htbp]
    \centering 
    \footnotesize
    \renewcommand{\arraystretch}{1.2} 
    \begin{tabular}{cccccc}
        \toprule
        \multirow{2}{*}{$eB$ [GeV$^2$]} & \multicolumn{2}{c}{AICc (Stat\_err)} & \multicolumn{2}{c}{Lanczos (Stat\_err)} & \multirow{2}[1]{*}{\shortstack{Final quoted\\ results}} \\
        \cmidrule(lr){2-3} \cmidrule(lr){4-5}
         & Quadratic & Linear & Quadratic & Linear & \\
        \midrule
        0.0 & $143.6(   3.7)$ & $141.5(   9.3)$ & $142.6(   2.2)$ & $141.6(   5.5)$ & $142.4(  3.9)$ \\
        0.2 & $124.2(   1.1)$ & $124.7(   6.3)$ & $124.9(   0.7)$ & $125.0(   4.0)$ & $124.7(  2.7)$ \\
        0.41 & $115.7(   1.0)$ & $114.7(   5.4)$ & $114.9(   0.6)$ & $114.5(   3.2)$ & $115.0(  2.1)$ \\
        0.61 & $102.3(   0.3)$ & $104.1(   3.6)$ & $103.1(   0.2)$ & $104.3(   2.5)$ & $103.5(  1.4)$ \\
        0.81 & $98.2(   0.3)$ & $97.4(   3.8)$ & $98.6(   0.2)$ & $97.9(   2.4)$ & $98.0(  1.6)$ \\
        1.02 & $93.7(   0.6)$ & $98.6(   4.0)$ & $94.4(   0.3)$ & $97.6(   2.5)$ & $96.1(  1.6)$ \\
        1.22 & $90.8(   0.3)$ & $90.4(   3.0)$ & $91.5(   0.2)$ & $91.4(   1.9)$ & $91.0(  1.2)$ \\
        \bottomrule
    \end{tabular}
    \caption{Comparison of the continuum-extrapolated masses (in MeV) and their associated statistical errors for the $\pi^0$ state across different extraction methods and extrapolation \textit{Ans\"atze}. The values in the ``final quoted results'' column are the same as those reported for the $\pi^0$ mass in~\autoref{tab:meson_masses}.}
    \label{tab:error_comparison}
\end{table}

\begin{table}[!htbp]
    \centering 
    \footnotesize
    \renewcommand{\arraystretch}{1.2} 
    \begin{tabular}{cccccc}
        \toprule
        \multirow{2}{*}{$eB$ [GeV$^2$]} & \multicolumn{2}{c}{AICc (Stat\_err)} & \multicolumn{2}{c}{Lanczos (Stat\_err)} & \multirow{2}[1]{*}{\shortstack{Final quoted\\ results}} \\
        \cmidrule(lr){2-3} \cmidrule(lr){4-5}
         & Quadratic & Linear & Quadratic & Linear & \\
        \midrule
        0.0 & $507.1(   3.0)$ & $504.4(   3.3)$ & $505.6(   1.9)$ & $504.2(   2.0)$ & $505.3(  1.8)$ \\
        0.2 & $479.0(   1.7)$ & $479.4(   1.7)$ & $479.9(   1.1)$ & $480.1(   1.1)$ & $479.5(  1.0)$ \\
        0.41 & $455.0(   0.9)$ & $455.6(   1.3)$ & $454.5(   0.6)$ & $455.0(   0.8)$ & $455.0(  0.6)$ \\
        0.61 & $421.0(   1.0)$ & $421.7(   1.7)$ & $423.8(   0.6)$ & $424.3(   1.1)$ & $422.7(  0.8)$ \\
        0.81 & $403.0(   0.6)$ & $402.9(   1.3)$ & $404.9(   0.5)$ & $405.2(   0.9)$ & $404.0(  0.6)$ \\
        1.02 & $386.6(   1.0)$ & $387.6(   1.2)$ & $388.9(   0.7)$ & $389.9(   0.8)$ & $388.2(  0.6)$ \\
        1.22 & $374.1(   0.6)$ & $374.3(   1.0)$ & $375.9(   0.4)$ & $376.2(   0.6)$ & $375.1(  0.5)$ \\
        \bottomrule
    \end{tabular}
    \caption{Same as~\autoref{tab:error_comparison} but for $K^0$.}
    \label{tab:error_comparison_sd}
\end{table}

\begin{table}[!htbp]
    \centering 
    \footnotesize
    \renewcommand{\arraystretch}{1.2} 
    \begin{tabular}{cccccc}
        \toprule
        \multirow{2}{*}{$eB$ [GeV$^2$]} & \multicolumn{2}{c}{AICc (Stat\_err)} & \multicolumn{2}{c}{Lanczos (Stat\_err)} & \multirow{2}[1]{*}{\shortstack{Final quoted\\ results}} \\
        \cmidrule(lr){2-3} \cmidrule(lr){4-5}
         & Quadratic & Linear & Quadratic & Linear & \\
        \midrule
        0.0 & $697.4(   1.6)$ & $697.1(   2.2)$ & $697.6(   1.0)$ & $697.5(   1.3)$ & $697.4(  1.1)$ \\
        0.2 & $671.1(   0.7)$ & $671.6(   1.4)$ & $672.9(   0.4)$ & $673.1(   0.8)$ & $672.2(  0.6)$ \\
        0.41 & $640.3(   0.6)$ & $641.0(   1.2)$ & $641.1(   0.3)$ & $641.4(   0.8)$ & $640.9(  0.5)$ \\
        0.61 & $602.9(   0.4)$ & $602.7(   1.1)$ & $606.0(   0.2)$ & $605.9(   0.7)$ & $604.4(  0.5)$ \\
        0.81 & $579.3(   0.5)$ & $579.4(   0.8)$ & $581.8(   0.3)$ & $582.0(   0.5)$ & $580.6(  0.4)$ \\
        1.02 & $558.9(   0.4)$ & $559.6(   0.8)$ & $561.7(   0.2)$ & $562.3(   0.5)$ & $560.6(  0.3)$ \\
        1.22 & $542.7(   0.2)$ & $543.1(   0.6)$ & $544.9(   0.1)$ & $545.0(   0.4)$ & $543.9(  0.2)$ \\
        \bottomrule
    \end{tabular}
    \caption{Same as~\autoref{tab:error_comparison} but for $\eta^0_{s\bar{s}}$.}
    \label{tab:error_comparison_ss}
\end{table}

\begin{table}[!htbp]
    \centering 
    \footnotesize
    \renewcommand{\arraystretch}{1.2} 
    \begin{tabular}{cccccc}
        \toprule
        \multirow{2}{*}{$eB$ [GeV$^2$]} & \multicolumn{2}{c}{AICc (Stat\_err)} & \multicolumn{2}{c}{Lanczos (Stat\_err)} & \multirow{2}[1]{*}{\shortstack{Final quoted\\ results}} \\
        \cmidrule(lr){2-3} \cmidrule(lr){4-5}
         & Quadratic & Linear & Quadratic & Linear & \\
        \midrule
        0.0 & $143.7(   3.4)$ & $141.2(   9.1)$ & $142.6(   2.1)$ & $141.7(   5.5)$ & $142.1(  4.1)$ \\
        0.2 & $115.6(   1.1)$ & $122.3(   7.0)$ & $117.5(   0.6)$ & $120.9(   4.4)$ & $119.2(  3.0)$ \\
        0.41 & $106.8(   0.9)$ & $109.7(   6.2)$ & $106.3(   0.5)$ & $108.2(   3.8)$ & $107.7(  2.6)$ \\
        0.61 & $97.0(   0.4)$ & $98.4(   3.1)$ & $97.2(   0.2)$ & $98.4(   2.0)$ & $97.8(  1.3)$ \\
        0.81 & $92.9(   0.5)$ & $94.2(   4.1)$ & $93.0(   0.3)$ & $94.0(   2.4)$ & $93.6(  1.7)$ \\
        1.02 & $89.5(   0.5)$ & $91.3(   3.1)$ & $89.9(   0.3)$ & $90.3(   2.0)$ & $90.2(  1.3)$ \\
        1.22 & $86.9(   0.4)$ & $86.6(   2.9)$ & $87.5(   0.2)$ & $87.5(   1.8)$ & $87.1(  1.2)$ \\
        \bottomrule
    \end{tabular}
    \caption{Comparison of the continuum-extrapolated masses (in MeV) and their associated statistical errors for the $\pi^0_u$ state across different extraction methods and extrapolation \textit{Ans\"atze}. The values in the ``Final quoted results'' column are the same as those reported for the $\pi^0_u$ mass in~\autoref{tab:pbp_Mpiud}.}
    \label{tab:error_comparison_uu}
\end{table}

\begin{table}[!htbp]
    \centering 
    \footnotesize
    \renewcommand{\arraystretch}{1.2} 
    \begin{tabular}{cccccc}
        \toprule
        \multirow{2}{*}{$eB$ [GeV$^2$]} & \multicolumn{2}{c}{AICc (Stat\_err)} & \multicolumn{2}{c}{Lanczos (Stat\_err)} & \multirow{2}[1]{*}{\shortstack{Final quoted\\ results}} \\
        \cmidrule(lr){2-3} \cmidrule(lr){4-5}
         & Quadratic & Linear & Quadratic & Linear & \\
        \midrule
        0.0 & $143.9(   4.1)$ & $141.4(   9.0)$ & $142.8(   2.5)$ & $141.8(   5.5)$ & $142.4(  4.0)$ \\
        0.2 & $132.1(   1.1)$ & $130.8(   6.1)$ & $132.3(   0.6)$ & $131.8(   3.6)$ & $131.7(  2.4)$ \\
        0.41 & $126.3(   1.0)$ & $126.2(   4.6)$ & $125.0(   0.6)$ & $125.1(   2.8)$ & $125.6(  1.8)$ \\
        0.61 & $109.8(   0.7)$ & $109.8(   4.2)$ & $111.3(   0.5)$ & $112.2(   2.7)$ & $110.9(  1.7)$ \\
        0.81 & $105.3(   0.7)$ & $108.0(   5.0)$ & $106.1(   0.4)$ & $107.5(   3.0)$ & $106.7(  2.0)$ \\
        1.02 & $99.7(   1.2)$ & $109.5(   6.1)$ & $100.9(   0.8)$ & $107.0(   3.8)$ & $104.2(  2.5)$ \\
        1.22 & $96.9(   0.6)$ & $97.4(   4.8)$ & $97.6(   0.4)$ & $97.7(   3.0)$ & $97.3(  2.0)$ \\
        \bottomrule
    \end{tabular}
    \caption{Same as~\autoref{tab:error_comparison_uu} but for $\pi^0_d$.}
    \label{tab:error_comparison_dd}
\end{table}

\begin{table}[!htbp]
    \centering 
    \footnotesize
    \renewcommand{\arraystretch}{1.2} 
    \begin{tabular}{cccccc}
        \toprule
        \multirow{2}{*}{$eB$ [GeV$^2$]} & \multicolumn{2}{c}{AICc (Stat\_err)} & \multicolumn{2}{c}{Lanczos (Stat\_err)} & \multirow{2}[1]{*}{\shortstack{Final quoted\\ results}} \\
        \cmidrule(lr){2-3} \cmidrule(lr){4-5}
         & Quadratic & Linear & Quadratic & Linear & \\
        \midrule
        0.0 & $145.0(   4.2)$ & $145.9(  11.7)$ & $143.4(   2.6)$ & $143.5(   7.0)$ & $144.7(  5.0)$ \\
        0.2 & $560.1(  59.0)$ & $543.5(  35.0)$ & $552.3(  36.5)$ & $544.1(  20.8)$ & $551.3( 28.2)$ \\
        0.41 & $785.4(  90.9)$ & $757.9(  92.3)$ & $744.2(  57.5)$ & $703.8(  51.9)$ & $747.8( 52.4)$ \\
        0.61 & $864.6( 180.7)$ & $731.6( 156.9)$ & $862.3( 109.6)$ & $758.6(  92.9)$ & $801.9( 95.2)$ \\
        0.81 & $915.4( 206.1)$ & $823.1( 117.5)$ & $824.6( 129.3)$ & $740.9(  70.5)$ & $830.1( 95.6)$ \\
        1.02 & $662.2( 148.4)$ & $619.3( 127.7)$ & $644.9(  86.0)$ & $610.0(  75.0)$ & $632.8( 74.5)$ \\
        1.22 & $658.0( 114.7)$ & $766.8( 118.0)$ & $641.1(  68.7)$ & $689.0(  68.4)$ & $687.4( 65.1)$ \\
        \bottomrule
    \end{tabular}
    \caption{Same as~\autoref{tab:error_comparison} but for $\pi^\pm$.}
    \label{tab:error_comparison_du}
\end{table}

\begin{table}[!htbp]
    \centering 
    \footnotesize
    \renewcommand{\arraystretch}{1.2} 
    \begin{tabular}{cccccc}
        \toprule
        \multirow{2}{*}{$eB$ [GeV$^2$]} & \multicolumn{2}{c}{AICc (Stat\_err)} & \multicolumn{2}{c}{Lanczos (Stat\_err)} & \multirow{2}[1]{*}{\shortstack{Final quoted\\ results}} \\
        \cmidrule(lr){2-3} \cmidrule(lr){4-5}
         & Quadratic & Linear & Quadratic & Linear & \\
        \midrule
        0.0 & $507.4(   3.5)$ & $505.8(   3.7)$ & $505.8(   2.2)$ & $505.1(   2.2)$ & $506.0(  2.0)$ \\
        0.2 & $704.4(  44.6)$ & $704.1(  23.1)$ & $705.6(  26.6)$ & $702.0(  14.4)$ & $703.6( 19.7)$ \\
        0.41 & $840.0(  82.4)$ & $880.4(  23.2)$ & $824.2(  51.3)$ & $843.8(  19.6)$ & $847.1( 35.4)$ \\
        0.61 & $1107.6(  67.3)$ & $1058.2(  53.6)$ & $1037.0(  43.3)$ & $1012.7(  32.8)$ & $1054.9( 34.0)$ \\
        0.81 & $914.1(  56.1)$ & $938.9(  38.3)$ & $854.5(  37.4)$ & $885.1(  22.0)$ & $897.6( 27.9)$ \\
        1.02 & $851.8(  67.5)$ & $812.5(  44.2)$ & $782.2(  40.7)$ & $768.6(  28.3)$ & $805.7( 32.4)$ \\
        1.22 & $776.8(  66.6)$ & $627.7(  44.6)$ & $755.4(  40.1)$ & $665.0(  27.2)$ & $706.0( 32.0)$ \\
        \bottomrule
    \end{tabular}
    \caption{Same as~\autoref{tab:error_comparison} but for $K^\pm$.}
    \label{tab:error_comparison_su}
\end{table}

\FloatBarrier
\bibliographystyle{JHEP.bst}  
\bibliography{refs}  

\providecommand{\href}[2]{#2}\begingroup\raggedright\begin{thebibliography}{100}

\bibitem{Kharzeev:2007jp}
D.E.~Kharzeev, L.D.~McLerran and H.J.~Warringa, \emph{{The Effects of
  topological charge change in heavy ion collisions: 'Event by event P and CP
  violation'}},
  \href{https://doi.org/10.1016/j.nuclphysa.2008.02.298}{\emph{Nucl. Phys.}
  {\bfseries A803} (2008) 227}
  [\href{https://arxiv.org/abs/0711.0950}{{\ttfamily 0711.0950}}].

\bibitem{Skokov:2009qp}
V.~Skokov, A.Y.~Illarionov and V.~Toneev, \emph{{Estimate of the magnetic field
  strength in heavy-ion collisions}},
  \href{https://doi.org/10.1142/S0217751X09047570}{\emph{Int. J. Mod. Phys.}
  {\bfseries A24} (2009) 5925}
  [\href{https://arxiv.org/abs/0907.1396}{{\ttfamily 0907.1396}}].

\bibitem{Deng:2012pc}
W.-T.~Deng and X.-G.~Huang, \emph{{Event-by-event generation of electromagnetic
  fields in heavy-ion collisions}},
  \href{https://doi.org/10.1103/PhysRevC.85.044907}{\emph{Phys. Rev.}
  {\bfseries C85} (2012) 044907}
  [\href{https://arxiv.org/abs/1201.5108}{{\ttfamily 1201.5108}}].

\bibitem{Vachaspati:1991nm}
T.~Vachaspati, \emph{{Magnetic fields from cosmological phase transitions}},
  \href{https://doi.org/10.1016/0370-2693(91)90051-Q}{\emph{Phys. Lett.}
  {\bfseries B265} (1991) 258}.

\bibitem{Duncan:1992hi}
R.C.~Duncan and C.~Thompson, \emph{{Formation of very strongly magnetized
  neutron stars - implications for gamma-ray bursts}},
  \href{https://doi.org/10.1086/186413}{\emph{Astrophys. J. Lett.} {\bfseries
  392} (1992) L9}.

\bibitem{Hattori:2023egw}
K.~Hattori, K.~Itakura and S.~Ozaki, \emph{{Strong-field physics in QED and
  QCD: From fundamentals to applications}},
  \href{https://doi.org/10.1016/j.ppnp.2023.104068}{\emph{Prog. Part. Nucl.
  Phys.} {\bfseries 133} (2023) 104068}
  [\href{https://arxiv.org/abs/2305.03865}{{\ttfamily 2305.03865}}].

\bibitem{Endrodi:2024cqn}
G.~Endrodi, \emph{{QCD with background electromagnetic fields on the lattice: A
  review}}, \href{https://doi.org/10.1016/j.ppnp.2024.104153}{\emph{Prog. Part.
  Nucl. Phys.} {\bfseries 141} (2025) 104153}
  [\href{https://arxiv.org/abs/2406.19780}{{\ttfamily 2406.19780}}].

\bibitem{Yamamoto:2021oys}
A.~Yamamoto, \emph{{Overview of external electromagnetism and rotation in
  lattice QCD}},
  \href{https://doi.org/10.1140/epja/s10050-021-00530-8}{\emph{Eur. Phys. J. A}
  {\bfseries 57} (2021) 211}
  [\href{https://arxiv.org/abs/2103.00237}{{\ttfamily 2103.00237}}].

\bibitem{DElia:2010abb}
M.~D'Elia, S.~Mukherjee and F.~Sanfilippo, \emph{{QCD Phase Transition in a
  Strong Magnetic Background}},
  \href{https://doi.org/10.1103/PhysRevD.82.051501}{\emph{Phys. Rev.}
  {\bfseries D82} (2010) 051501}
  [\href{https://arxiv.org/abs/1005.5365}{{\ttfamily 1005.5365}}].

\bibitem{Shovkovy:2012zn}
I.A.~Shovkovy, \emph{{Magnetic Catalysis: A Review}},
  \href{https://doi.org/10.1007/978-3-642-37305-3_2}{\emph{Lect. Notes Phys.}
  {\bfseries 871} (2013) 13} [\href{https://arxiv.org/abs/1207.5081}{{\ttfamily
  1207.5081}}].

\bibitem{Ding:2020inp}
H.-T.~Ding, C.~Schmidt, A.~Tomiya and X.-D.~Wang, \emph{{Chiral phase structure
  of three flavor QCD in a background magnetic field}},
  \href{https://doi.org/10.1103/PhysRevD.102.054505}{\emph{Phys. Rev. D}
  {\bfseries 102} (2020) 054505}
  [\href{https://arxiv.org/abs/2006.13422}{{\ttfamily 2006.13422}}].

\bibitem{Bali:2011qj}
G.S.~Bali, F.~Bruckmann, G.~Endrodi, Z.~Fodor, S.D.~Katz, S.~Krieg et~al.,
  \emph{{The QCD phase diagram for external magnetic fields}},
  \href{https://doi.org/10.1007/JHEP02(2012)044}{\emph{JHEP} {\bfseries 02}
  (2012) 044} [\href{https://arxiv.org/abs/1111.4956}{{\ttfamily 1111.4956}}].

\bibitem{Bali:2012zg}
G.S.~Bali, F.~Bruckmann, G.~Endrodi, Z.~Fodor, S.D.~Katz and A.~Schafer,
  \emph{{QCD quark condensate in external magnetic fields}},
  \href{https://doi.org/10.1103/PhysRevD.86.071502}{\emph{Phys. Rev.}
  {\bfseries D86} (2012) 071502}
  [\href{https://arxiv.org/abs/1206.4205}{{\ttfamily 1206.4205}}].

\bibitem{Ilgenfritz:2013ara}
E.M.~Ilgenfritz, M.~Muller-Preussker, B.~Petersson and A.~Schreiber,
  \emph{{Magnetic catalysis (and inverse catalysis) at finite temperature in
  two-color lattice QCD}},
  \href{https://doi.org/10.1103/PhysRevD.89.054512}{\emph{Phys. Rev.}
  {\bfseries D89} (2014) 054512}
  [\href{https://arxiv.org/abs/1310.7876}{{\ttfamily 1310.7876}}].

\bibitem{Bornyakov:2013eya}
V.G.~Bornyakov, P.V.~Buividovich, N.~Cundy, O.A.~Kochetkov and A.~Sch{\"a}fer,
  \emph{{Deconfinement transition in two-flavor lattice QCD with dynamical
  overlap fermions in an external magnetic field}},
  \href{https://doi.org/10.1103/PhysRevD.90.034501}{\emph{Phys. Rev.}
  {\bfseries D90} (2014) 034501}
  [\href{https://arxiv.org/abs/1312.5628}{{\ttfamily 1312.5628}}].

\bibitem{Bali:2014kia}
G.S.~Bali, F.~Bruckmann, G.~Endr{\"o}di, S.D.~Katz and A.~Sch{\"a}fer,
  \emph{{The QCD equation of state in background magnetic fields}},
  \href{https://doi.org/10.1007/JHEP08(2014)177}{\emph{JHEP} {\bfseries 08}
  (2014) 177} [\href{https://arxiv.org/abs/1406.0269}{{\ttfamily 1406.0269}}].

\bibitem{Tomiya:2019nym}
A.~Tomiya, H.-T.~Ding, X.-D.~Wang, Y.~Zhang, S.~Mukherjee and C.~Schmidt,
  \emph{{Phase structure of three flavor QCD in external magnetic fields using
  HISQ fermions}}, \href{https://doi.org/10.22323/1.334.0163}{\emph{PoS}
  {\bfseries LATTICE2018} (2019) 163}
  [\href{https://arxiv.org/abs/1904.01276}{{\ttfamily 1904.01276}}].

\bibitem{DElia:2011koc}
M.~D'Elia and F.~Negro, \emph{{Chiral Properties of Strong Interactions in a
  Magnetic Background}},
  \href{https://doi.org/10.1103/PhysRevD.83.114028}{\emph{Phys. Rev. D}
  {\bfseries 83} (2011) 114028}
  [\href{https://arxiv.org/abs/1103.2080}{{\ttfamily 1103.2080}}].

\bibitem{Andersen:2014xxa}
J.O.~Andersen, W.R.~Naylor and A.~Tranberg, \emph{{Phase diagram of QCD in a
  magnetic field: A review}},
  \href{https://doi.org/10.1103/RevModPhys.88.025001}{\emph{Rev. Mod. Phys.}
  {\bfseries 88} (2016) 025001}
  [\href{https://arxiv.org/abs/1411.7176}{{\ttfamily 1411.7176}}].

\bibitem{Kojo:2012js}
T.~Kojo and N.~Su, \emph{{The quark mass gap in a magnetic field}},
  \href{https://doi.org/10.1016/j.physletb.2013.02.024}{\emph{Phys. Lett.}
  {\bfseries B720} (2013) 192}
  [\href{https://arxiv.org/abs/1211.7318}{{\ttfamily 1211.7318}}].

\bibitem{Bruckmann:2013oba}
F.~Bruckmann, G.~Endrodi and T.G.~Kovacs, \emph{{Inverse magnetic catalysis and
  the Polyakov loop}},
  \href{https://doi.org/10.1007/JHEP04(2013)112}{\emph{JHEP} {\bfseries 04}
  (2013) 112} [\href{https://arxiv.org/abs/1303.3972}{{\ttfamily 1303.3972}}].

\bibitem{Fukushima:2012kc}
K.~Fukushima and Y.~Hidaka, \emph{{Magnetic Catalysis Versus Magnetic
  Inhibition}},
  \href{https://doi.org/10.1103/PhysRevLett.110.031601}{\emph{Phys. Rev. Lett.}
  {\bfseries 110} (2013) 031601}
  [\href{https://arxiv.org/abs/1209.1319}{{\ttfamily 1209.1319}}].

\bibitem{Ferreira:2014kpa}
M.~Ferreira, P.~Costa, O.~Louren{\c c}o, T.~Frederico and C.~Provid{\^e}ncia,
  \emph{{Inverse magnetic catalysis in the (2+1)-flavor Nambu-Jona-Lasinio and
  Polyakov-Nambu-Jona-Lasinio models}},
  \href{https://doi.org/10.1103/PhysRevD.89.116011}{\emph{Phys. Rev.}
  {\bfseries D89} (2014) 116011}
  [\href{https://arxiv.org/abs/1404.5577}{{\ttfamily 1404.5577}}].

\bibitem{Yu:2014sla}
L.~Yu, H.~Liu and M.~Huang, \emph{{Spontaneous generation of local CP violation
  and inverse magnetic catalysis}},
  \href{https://doi.org/10.1103/PhysRevD.90.074009}{\emph{Phys. Rev. D}
  {\bfseries 90} (2014) 074009}
  [\href{https://arxiv.org/abs/1404.6969}{{\ttfamily 1404.6969}}].

\bibitem{Feng:2015qpi}
B.~Feng, D.~Hou, H.-c.~Ren and P.-p.~Wu, \emph{{Bose-Einstein Condensation of
  Bound Pairs of Relativistic Fermions in a Magnetic Field}},
  \href{https://doi.org/10.1103/PhysRevD.93.085019}{\emph{Phys. Rev. D}
  {\bfseries 93} (2016) 085019}
  [\href{https://arxiv.org/abs/1512.08894}{{\ttfamily 1512.08894}}].

\bibitem{Li:2019nzj}
X.~Li, W.-J.~Fu and Y.-X.~Liu, \emph{{Thermodynamics of 2+1 Flavor
  Polyakov-Loop Quark-Meson Model under External Magnetic Field}},
  \href{https://doi.org/10.1103/PhysRevD.99.074029}{\emph{Phys. Rev.}
  {\bfseries D99} (2019) 074029}
  [\href{https://arxiv.org/abs/1902.03866}{{\ttfamily 1902.03866}}].

\bibitem{Mao:2016lsr}
S.~Mao, \emph{{From inverse to delayed magnetic catalysis in a strong magnetic
  field}}, \href{https://doi.org/10.1103/PhysRevD.94.036007}{\emph{Phys. Rev.}
  {\bfseries D94} (2016) 036007}
  [\href{https://arxiv.org/abs/1605.04526}{{\ttfamily 1605.04526}}].

\bibitem{Gursoy:2016ofp}
U.~G{\"u}rsoy, I.~Iatrakis, M.~J{\"a}rvinen and G.~Nijs, \emph{{Inverse
  Magnetic Catalysis from improved Holographic QCD in the Veneziano limit}},
  \href{https://doi.org/10.1007/JHEP03(2017)053}{\emph{JHEP} {\bfseries 03}
  (2017) 053} [\href{https://arxiv.org/abs/1611.06339}{{\ttfamily
  1611.06339}}].

\bibitem{Xu:2020yag}
K.~Xu, J.~Chao and M.~Huang, \emph{{Effect of the anomalous magnetic moment of
  quarks on magnetized QCD matter and meson spectra}},
  \href{https://doi.org/10.1103/PhysRevD.103.076015}{\emph{Phys. Rev. D}
  {\bfseries 103} (2021) 076015}
  [\href{https://arxiv.org/abs/2007.13122}{{\ttfamily 2007.13122}}].

\bibitem{Abreu:2022cgm}
L.M.~Abreu, E.B.S.~Corr{\^e}a and E.S.~Nery, \emph{{Inverse magnetic catalysis
  and size-dependent effects on the chiral symmetry restoration}},
  \href{https://doi.org/10.1140/epja/s10050-023-01078-5}{\emph{Eur. Phys. J. A}
  {\bfseries 59} (2023) 157}
  [\href{https://arxiv.org/abs/2211.11083}{{\ttfamily 2211.11083}}].

\bibitem{Li:2025wqb}
L.~Li and S.~Mao, \emph{{Mass spectra and Mott transitions of neutral mesons at
  finite temperature and magnetic field in frame of three-flavor
  Polyakov-extended Nambu-Jona-Lasino model}},
  \href{https://arxiv.org/abs/2511.14150}{{\ttfamily 2511.14150}}.

\bibitem{Mao:2024rxh}
S.~Mao, \emph{{Reduction of pseudocritical temperatures of chiral restoration
  and deconfinement phase transitions in a magnetized PNJL model}},
  \href{https://doi.org/10.1103/PhysRevD.110.054002}{\emph{Phys. Rev. D}
  {\bfseries 110} (2024) 054002}
  [\href{https://arxiv.org/abs/2404.05294}{{\ttfamily 2404.05294}}].

\bibitem{Hofmann:2020ism}
C.P.~Hofmann, \emph{{Chiral Perturbation Theory Analysis of the Quark
  Condensate in a Magnetic Field}},
  \href{https://doi.org/10.1103/PhysRevD.102.094010}{\emph{Phys. Rev. D}
  {\bfseries 102} (2020) 094010}
  [\href{https://arxiv.org/abs/2006.07717}{{\ttfamily 2006.07717}}].

\bibitem{DElia:2018xwo}
M.~D'Elia, F.~Manigrasso, F.~Negro and F.~Sanfilippo, \emph{{QCD phase diagram
  in a magnetic background for different values of the pion mass}},
  \href{https://doi.org/10.1103/PhysRevD.98.054509}{\emph{Phys. Rev.}
  {\bfseries D98} (2018) 054509}
  [\href{https://arxiv.org/abs/1808.07008}{{\ttfamily 1808.07008}}].

\bibitem{Endrodi:2019zrl}
G.~Endrodi, M.~Giordano, S.D.~Katz, T.G.~Kov{\'a}cs and F.~Pittler,
  \emph{{Magnetic catalysis and inverse catalysis for heavy pions}},
  \href{https://doi.org/10.1007/JHEP07(2019)007}{\emph{JHEP} {\bfseries 07}
  (2019) 007} [\href{https://arxiv.org/abs/1904.10296}{{\ttfamily
  1904.10296}}].

\bibitem{Bonati:2016kxj}
C.~Bonati, M.~D'Elia, M.~Mariti, M.~Mesiti, F.~Negro, A.~Rucci et~al.,
  \emph{{Magnetic field effects on the static quark potential at zero and
  finite temperature}},
  \href{https://doi.org/10.1103/PhysRevD.94.094007}{\emph{Phys. Rev.}
  {\bfseries D94} (2016) 094007}
  [\href{https://arxiv.org/abs/1607.08160}{{\ttfamily 1607.08160}}].

\bibitem{Gell-Mann:1968hlm}
M.~Gell-Mann, R.J.~Oakes and B.~Renner, \emph{{Behavior of current divergences
  under SU(3) x SU(3)}},
  \href{https://doi.org/10.1103/PhysRev.175.2195}{\emph{Phys. Rev.} {\bfseries
  175} (1968) 2195}.

\bibitem{Jamin:2002ev}
M.~Jamin, \emph{{Flavor symmetry breaking of the quark condensate and chiral
  corrections to the Gell-Mann-Oakes-Renner relation}},
  \href{https://doi.org/10.1016/S0370-2693(02)01951-2}{\emph{Phys. Lett. B}
  {\bfseries 538} (2002) 71}
  [\href{https://arxiv.org/abs/hep-ph/0201174}{{\ttfamily hep-ph/0201174}}].

\bibitem{Bordes:2010wy}
J.~Bordes, C.~Dominguez, P.~Moodley, J.~Penarrocha and K.~Schilcher,
  \emph{{Chiral corrections to the $SU(2)\times SU(2)$ Gell-Mann-Oakes-Renner
  relation}}, \href{https://doi.org/10.1007/JHEP05(2010)064}{\emph{JHEP}
  {\bfseries 05} (2010) 064} [\href{https://arxiv.org/abs/1003.3358}{{\ttfamily
  1003.3358}}].

\bibitem{Bordes:2012ud}
J.~Bordes, C.~Dominguez, P.~Moodley, J.~Penarrocha and K.~Schilcher,
  \emph{{Corrections to the ${\bf SU(3)\times SU(3)}$ Gell-Mann-Oakes-Renner
  relation and chiral couplings $L^r_8$ and $H^r_2$}},
  \href{https://doi.org/10.1007/JHEP10(2012)102}{\emph{JHEP} {\bfseries 10}
  (2012) 102} [\href{https://arxiv.org/abs/1208.1159}{{\ttfamily 1208.1159}}].

\bibitem{Gasser:1984gg}
J.~Gasser and H.~Leutwyler, \emph{{Chiral Perturbation Theory: Expansions in
  the Mass of the Strange Quark}},
  \href{https://doi.org/10.1016/0550-3213(85)90492-4}{\emph{Nucl. Phys. B}
  {\bfseries 250} (1985) 465}.

\bibitem{Boucaud:2007uk}
{\scshape ETM} collaboration, \emph{{Dynamical twisted mass fermions with light
  quarks}}, \href{https://doi.org/10.1016/j.physletb.2007.04.054}{\emph{Phys.
  Lett.} {\bfseries B650} (2007) 304}
  [\href{https://arxiv.org/abs/hep-lat/0701012}{{\ttfamily hep-lat/0701012}}].

\bibitem{Engel:2014cka}
G.P.~Engel, L.~Giusti, S.~Lottini and R.~Sommer, \emph{{Chiral Symmetry
  Breaking in QCD with Two Light Flavors}},
  \href{https://doi.org/10.1103/PhysRevLett.114.112001}{\emph{Phys. Rev. Lett.}
  {\bfseries 114} (2015) 112001}
  [\href{https://arxiv.org/abs/1406.4987}{{\ttfamily 1406.4987}}].

\bibitem{Gasser:1986vb}
J.~Gasser and H.~Leutwyler, \emph{{Light Quarks at Low Temperatures}},
  \href{https://doi.org/10.1016/0370-2693(87)90492-8}{\emph{Phys. Lett.}
  {\bfseries B184} (1987) 83}.

\bibitem{Shushpanov:1997sf}
I.A.~Shushpanov and A.V.~Smilga, \emph{{Quark condensate in a magnetic field}},
  \href{https://doi.org/10.1016/S0370-2693(97)00441-3}{\emph{Phys. Lett.}
  {\bfseries B402} (1997) 351}
  [\href{https://arxiv.org/abs/hep-ph/9703201}{{\ttfamily hep-ph/9703201}}].

\bibitem{Agasian:2001ym}
N.O.~Agasian and I.A.~Shushpanov, \emph{{Gell-Mann-Oakes-Renner relation in a
  magnetic field at finite temperature}},
  \href{https://doi.org/10.1088/1126-6708/2001/10/006}{\emph{JHEP} {\bfseries
  10} (2001) 006} [\href{https://arxiv.org/abs/hep-ph/0107128}{{\ttfamily
  hep-ph/0107128}}].

\bibitem{Adhikari:2024vhs}
P.~Adhikari and B.C.~Tiburzi, \emph{{Chiral Symmetry Breaking and Pion Decay in
  a Magnetic Field}},  \href{https://arxiv.org/abs/2406.00818}{{\ttfamily
  2406.00818}}.

\bibitem{Ding:2019prx}
H.T.~Ding et~al., \emph{{Chiral Phase Transition Temperature in ( 2+1 )-Flavor
  QCD}}, \href{https://doi.org/10.1103/PhysRevLett.123.062002}{\emph{Phys. Rev.
  Lett.} {\bfseries 123} (2019) 062002}
  [\href{https://arxiv.org/abs/1903.04801}{{\ttfamily 1903.04801}}].

\bibitem{Ding:2020rtq}
H.-T.~Ding, \emph{{New developments in lattice QCD on equilibrium physics and
  phase diagram}},
  \href{https://doi.org/10.1016/j.nuclphysa.2020.121940}{\emph{Nucl. Phys. A}
  {\bfseries 1005} (2021) 121940}
  [\href{https://arxiv.org/abs/2002.11957}{{\ttfamily 2002.11957}}].

\bibitem{Bazavov:2017xul}
A.~Bazavov, H.T.~Ding, P.~Hegde, F.~Karsch, E.~Laermann, S.~Mukherjee et~al.,
  \emph{{Chiral phase structure of three flavor QCD at vanishing baryon number
  density}}, \href{https://doi.org/10.1103/PhysRevD.95.074505}{\emph{Phys.
  Rev.} {\bfseries D95} (2017) 074505}
  [\href{https://arxiv.org/abs/1701.03548}{{\ttfamily 1701.03548}}].

\bibitem{Kotov:2021rah}
A.Y.~Kotov, M.P.~Lombardo and A.~Trunin, \emph{{QCD transition at the physical
  point, and its scaling window from twisted mass Wilson fermions}},
  \href{https://doi.org/10.1016/j.physletb.2021.136749}{\emph{Phys. Lett. B}
  {\bfseries 823} (2021) 136749}
  [\href{https://arxiv.org/abs/2105.09842}{{\ttfamily 2105.09842}}].

\bibitem{Aarts:2020vyb}
G.~Aarts et~al., \emph{{Properties of the QCD thermal transition with Nf=2+1
  flavors of Wilson quark}},
  \href{https://doi.org/10.1103/PhysRevD.105.034504}{\emph{Phys. Rev. D}
  {\bfseries 105} (2022) 034504}
  [\href{https://arxiv.org/abs/2007.04188}{{\ttfamily 2007.04188}}].

\bibitem{Bhattacharya:2014ara}
T.~Bhattacharya et~al., \emph{{QCD Phase Transition with Chiral Quarks and
  Physical Quark Masses}},
  \href{https://doi.org/10.1103/PhysRevLett.113.082001}{\emph{Phys. Rev. Lett.}
  {\bfseries 113} (2014) 082001}
  [\href{https://arxiv.org/abs/1402.5175}{{\ttfamily 1402.5175}}].

\bibitem{Umeda:2016qdo}
T.~Umeda, S.~Ejiri, R.~Iwami, K.~Kanaya, H.~Ohno, A.~Uji et~al., \emph{{O(4)
  scaling analysis in two-flavor QCD at finite temperature and density with
  improved Wilson quarks}},
  \href{https://doi.org/10.22323/1.256.0376}{\emph{PoS} {\bfseries LATTICE2016}
  (2017) 376} [\href{https://arxiv.org/abs/1612.09449}{{\ttfamily
  1612.09449}}].

\bibitem{Li:2020wvy}
S.~Li and H.~Ding, \emph{{Chiral Crossover and Chiral Phase Transition
  Temperatures from Lattice QCD}},
  \href{https://doi.org/10.11804/NuclPhysRev.37.2019CNPC65}{\emph{Nucl. Phys.
  Rev.} {\bfseries 37} (2020) 674}.

\bibitem{Hattori:2015aki}
K.~Hattori, T.~Kojo and N.~Su, \emph{{Mesons in strong magnetic fields: (I)
  General analyses}},
  \href{https://doi.org/10.1016/j.nuclphysa.2016.03.016}{\emph{Nucl. Phys.}
  {\bfseries A951} (2016) 1}
  [\href{https://arxiv.org/abs/1512.07361}{{\ttfamily 1512.07361}}].

\bibitem{Avancini:2016fgq}
S.S.~Avancini, R.L.S.~Farias, M.~Benghi~Pinto, W.R.~Tavares and V.S.~Tim\'oteo,
  \emph{{$\pi_0$ pole mass calculation in a strong magnetic field and lattice
  constraints}},
  \href{https://doi.org/10.1016/j.physletb.2017.02.002}{\emph{Phys. Lett. B}
  {\bfseries 767} (2017) 247}
  [\href{https://arxiv.org/abs/1606.05754}{{\ttfamily 1606.05754}}].

\bibitem{Wang:2017vtn}
Z.~Wang and P.~Zhuang, \emph{{Meson properties in magnetized quark matter}},
  \href{https://doi.org/10.1103/PhysRevD.97.034026}{\emph{Phys. Rev.}
  {\bfseries D97} (2018) 034026}
  [\href{https://arxiv.org/abs/1712.00554}{{\ttfamily 1712.00554}}].

\bibitem{Mao:2018dqe}
S.~Mao, \emph{{Pions in magnetic field at finite temperature}},
  \href{https://doi.org/10.1103/PhysRevD.99.056005}{\emph{Phys. Rev.}
  {\bfseries D99} (2019) 056005}
  [\href{https://arxiv.org/abs/1808.10242}{{\ttfamily 1808.10242}}].

\bibitem{Avancini:2018svs}
S.S.~Avancini, R.L.S.~Farias and W.R.~Tavares, \emph{{Neutral meson properties
  in hot and magnetized quark matter: a new magnetic field independent
  regularization scheme applied to NJL-type model}},
  \href{https://doi.org/10.1103/PhysRevD.99.056009}{\emph{Phys. Rev.}
  {\bfseries D99} (2019) 056009}
  [\href{https://arxiv.org/abs/1812.00945}{{\ttfamily 1812.00945}}].

\bibitem{Coppola:2019uyr}
M.~Coppola, D.~Gomez~Dumm, S.~Noguera and N.N.~Scoccola, \emph{{Neutral and
  charged pion properties under strong magnetic fields in the NJL model}},
  \href{https://doi.org/10.1103/PhysRevD.100.054014}{\emph{Phys. Rev.}
  {\bfseries D100} (2019) 054014}
  [\href{https://arxiv.org/abs/1907.05840}{{\ttfamily 1907.05840}}].

\bibitem{Cao:2019res}
G.~Cao, \emph{{Magnetic catalysis effect prevents vacuum superconductivity in
  strong magnetic fields}},
  \href{https://doi.org/10.1103/PhysRevD.100.074024}{\emph{Phys. Rev.}
  {\bfseries D100} (2019) 074024}
  [\href{https://arxiv.org/abs/1906.01398}{{\ttfamily 1906.01398}}].

\bibitem{Xu:2020sui}
K.~Xu, S.~Shi, H.~Zhang, D.~Hou, J.~Liao and M.~Huang, \emph{{Extracting the
  magnitude of magnetic field at freeze-out in heavy-ion collisions}},
  \href{https://doi.org/10.1016/j.physletb.2020.135706}{\emph{Phys. Lett. B}
  {\bfseries 809} (2020) 135706}
  [\href{https://arxiv.org/abs/2004.05362}{{\ttfamily 2004.05362}}].

\bibitem{Kojo:2021gvm}
T.~Kojo, \emph{{Neutral and charged mesons in magnetic fields: A resonance gas
  in a non-relativistic quark model}},
  \href{https://doi.org/10.1140/epja/s10050-021-00629-y}{\emph{Eur. Phys. J. A}
  {\bfseries 57} (2021) 317}
  [\href{https://arxiv.org/abs/2104.00376}{{\ttfamily 2104.00376}}].

\bibitem{Xing:2021kbw}
Z.~Xing, J.~Chao, L.~Chang and Y.-x.~Liu, \emph{{Exposing the effect of the
  p-wave component in the pion triplet under a strong magnetic field}},
  \href{https://doi.org/10.1103/PhysRevD.105.114003}{\emph{Phys. Rev. D}
  {\bfseries 105} (2022) 114003}
  [\href{https://arxiv.org/abs/2110.01245}{{\ttfamily 2110.01245}}].

\bibitem{Mei:2022dkd}
J.~Mei, T.~Xia and S.~Mao, \emph{{Mass spectra of neutral mesons
  K0,{\ensuremath{\pi}}0,{\ensuremath{\eta}},{\ensuremath{\eta}}' at finite
  magnetic field, temperature and quark chemical potential}},
  \href{https://doi.org/10.1103/PhysRevD.107.074018}{\emph{Phys. Rev. D}
  {\bfseries 107} (2023) 074018}
  [\href{https://arxiv.org/abs/2212.04778}{{\ttfamily 2212.04778}}].

\bibitem{Coppola:2023mmq}
M.~Coppola, D.~Gomez~Dumm, S.~Noguera and N.N.~Scoccola, \emph{{Masses of
  magnetized pseudoscalar and vector mesons in an extended NJL model: The role
  of axial vector mesons}},
  \href{https://doi.org/10.1103/PhysRevD.109.054014}{\emph{Phys. Rev. D}
  {\bfseries 109} (2024) 054014}
  [\href{https://arxiv.org/abs/2312.16675}{{\ttfamily 2312.16675}}].

\bibitem{Wen:2023qcz}
R.~Wen, S.~Yin, W.-j.~Fu and M.~Huang, \emph{{Functional renormalization group
  study of neutral and charged pions in magnetic fields in the quark-meson
  model}}, \href{https://doi.org/10.1103/PhysRevD.108.076020}{\emph{Phys. Rev.
  D} {\bfseries 108} (2023) 076020}
  [\href{https://arxiv.org/abs/2306.04045}{{\ttfamily 2306.04045}}].

\bibitem{Dominguez:2018njv}
C.~Dominguez, M.~Loewe and C.~Villavicencio, \emph{{QCD determination of the
  magnetic field dependence of QCD and hadronic parameters}},
  \href{https://doi.org/10.1103/PhysRevD.98.034015}{\emph{Phys. Rev. D}
  {\bfseries 98} (2018) 034015}
  [\href{https://arxiv.org/abs/1806.10088}{{\ttfamily 1806.10088}}].

\bibitem{Ayala:2018zat}
A.~Ayala, R.L.S.~Farias, S.~Hern{\'a}ndez-Ortiz, L.A.~Hern{\'a}ndez, D.M.~Paret
  and R.~Zamora, \emph{{Magnetic field-dependence of the neutral pion mass in
  the linear sigma model coupled to quarks: The weak field case}},
  \href{https://doi.org/10.1103/PhysRevD.98.114008}{\emph{Phys. Rev. D}
  {\bfseries 98} (2018) 114008}
  [\href{https://arxiv.org/abs/1809.08312}{{\ttfamily 1809.08312}}].

\bibitem{Colucci:2013zoa}
G.~Colucci, E.S.~Fraga and A.~Sedrakian, \emph{{Chiral pions in a magnetic
  background}},
  \href{https://doi.org/10.1016/j.physletb.2013.11.028}{\emph{Phys. Lett. B}
  {\bfseries 728} (2014) 19} [\href{https://arxiv.org/abs/1310.3742}{{\ttfamily
  1310.3742}}].

\bibitem{Hidaka:2012mz}
Y.~Hidaka and A.~Yamamoto, \emph{{Charged vector mesons in a strong magnetic
  field}}, \href{https://doi.org/10.1103/PhysRevD.87.094502}{\emph{Phys. Rev.}
  {\bfseries D87} (2013) 094502}
  [\href{https://arxiv.org/abs/1209.0007}{{\ttfamily 1209.0007}}].

\bibitem{Chernodub:2010qx}
M.N.~Chernodub, \emph{{Superconductivity of QCD vacuum in strong magnetic
  field}}, \href{https://doi.org/10.1103/PhysRevD.82.085011}{\emph{Phys. Rev.}
  {\bfseries D82} (2010) 085011}
  [\href{https://arxiv.org/abs/1008.1055}{{\ttfamily 1008.1055}}].

\bibitem{Chernodub:2011mc}
M.N.~Chernodub, \emph{{Spontaneous electromagnetic superconductivity of vacuum
  in strong magnetic field: evidence from the Nambu--Jona-Lasinio model}},
  \href{https://doi.org/10.1103/PhysRevLett.106.142003}{\emph{Phys. Rev. Lett.}
  {\bfseries 106} (2011) 142003}
  [\href{https://arxiv.org/abs/1101.0117}{{\ttfamily 1101.0117}}].

\bibitem{Luschevskaya:2012xd}
E.V.~Luschevskaya and O.V.~Larina, \emph{{The $\rho$ and $A$ mesons in a strong
  abelian magnetic field in $SU(2)$ lattice gauge theory}},
  \href{https://doi.org/10.1016/j.nuclphysb.2014.04.003}{\emph{Nucl. Phys.}
  {\bfseries B884} (2014) 1} [\href{https://arxiv.org/abs/1203.5699}{{\ttfamily
  1203.5699}}].

\bibitem{Luschevskaya:2014lga}
E.V.~Luschevskaya, O.E.~Solovjeva, O.A.~Kochetkov and O.V.~Teryaev,
  \emph{{Magnetic polarizabilities of light mesons in $SU(3)$ lattice gauge
  theory}}, \href{https://doi.org/10.1016/j.nuclphysb.2015.07.023}{\emph{Nucl.
  Phys.} {\bfseries B898} (2015) 627}
  [\href{https://arxiv.org/abs/1411.4284}{{\ttfamily 1411.4284}}].

\bibitem{Luschevskaya:2015cko}
E.~Luschevskaya, O.~Solovjeva and O.~Teryaev, \emph{{Magnetic polarizability of
  pion}}, \href{https://doi.org/10.1016/j.physletb.2016.08.054}{\emph{Phys.
  Lett. B} {\bfseries 761} (2016) 393}
  [\href{https://arxiv.org/abs/1511.09316}{{\ttfamily 1511.09316}}].

\bibitem{Bali:2017ian}
G.S.~Bali, B.B.~Brandt, G.~Endr{\H o}di and B.~Gl{\"a}{\ss}le, \emph{{Meson
  masses in electromagnetic fields with Wilson fermions}},
  \href{https://doi.org/10.1103/PhysRevD.97.034505}{\emph{Phys. Rev.}
  {\bfseries D97} (2018) 034505}
  [\href{https://arxiv.org/abs/1707.05600}{{\ttfamily 1707.05600}}].

\bibitem{Bali:2018sey}
G.S.~Bali, B.B.~Brandt, G.~Endr{\H o}di and B.~Gl{\"a}{\ss}le, \emph{{Weak
  decay of magnetized pions}},
  \href{https://doi.org/10.1103/PhysRevLett.121.072001}{\emph{Phys. Rev. Lett.}
  {\bfseries 121} (2018) 072001}
  [\href{https://arxiv.org/abs/1805.10971}{{\ttfamily 1805.10971}}].

\bibitem{Coppola:2018ygv}
M.~Coppola, D.~Gomez~Dumm, S.~Noguera and N.N.~Scoccola, \emph{{Pion-to-vacuum
  vector and axial vector amplitudes and weak decays of pions in a magnetic
  field}}, \href{https://doi.org/10.1103/PhysRevD.99.054031}{\emph{Phys. Rev.}
  {\bfseries D99} (2019) 054031}
  [\href{https://arxiv.org/abs/1810.08110}{{\ttfamily 1810.08110}}].

\bibitem{Coppola:2019idh}
M.~Coppola, D.~Gomez~Dumm, S.~Noguera and N.N.~Scoccola, \emph{{Magnetic field
  driven enhancement of the weak decay width of charged pions}},
  \href{https://doi.org/10.1007/JHEP09(2020)058}{\emph{JHEP} {\bfseries 09}
  (2020) 058} [\href{https://arxiv.org/abs/1908.10765}{{\ttfamily
  1908.10765}}].

\bibitem{Ding:2020hxw}
H.T.~Ding, S.T.~Li, A.~Tomiya, X.D.~Wang and Y.~Zhang, \emph{{Chiral properties
  of (2+1)-flavor QCD in strong magnetic fields at zero temperature}},
  \href{https://doi.org/10.1103/PhysRevD.104.014505}{\emph{Phys. Rev. D}
  {\bfseries 104} (2021) 014505}
  [\href{https://arxiv.org/abs/2008.00493}{{\ttfamily 2008.00493}}].

\bibitem{cavanaugh1997unifying}
J.E.~Cavanaugh, \emph{Unifying the derivations for the akaike and corrected
  akaike information criteria}, {\emph{Statistics \& Probability Letters}
  {\bfseries 33} (1997) 201}.

\bibitem{Akaike:1974vps}
H.~Akaike, \emph{{A new look at the statistical model identification}},
  \href{https://doi.org/10.1109/TAC.1974.1100705}{\emph{IEEE Trans. Automatic
  Control} {\bfseries 19} (1974) 716}.

\bibitem{Wagman:2024rid}
M.L.~Wagman, \emph{{Lanczos Algorithm, the Transfer Matrix, and the
  Signal-to-Noise Problem}},
  \href{https://doi.org/10.1103/pcvc-734h}{\emph{Phys. Rev. Lett.} {\bfseries
  134} (2025) 241901} [\href{https://arxiv.org/abs/2406.20009}{{\ttfamily
  2406.20009}}].

\bibitem{Ostmeyer:2024qgu}
J.~Ostmeyer, A.~Sen and C.~Urbach, \emph{{On the equivalence of Prony and
  Lanczos methods for Euclidean correlation functions}},
  \href{https://doi.org/10.1140/epja/s10050-025-01495-8}{\emph{Eur. Phys. J. A}
  {\bfseries 61} (2025) 26} [\href{https://arxiv.org/abs/2411.14981}{{\ttfamily
  2411.14981}}].

\bibitem{Kilcup:1986dg}
G.W.~Kilcup and S.R.~Sharpe, \emph{{A Tool Kit for Staggered Fermions}},
  \href{https://doi.org/10.1016/0550-3213(87)90285-9}{\emph{Nucl. Phys.}
  {\bfseries B283} (1987) 493}.

\bibitem{Bazavov:2019www}
A.~Bazavov, S.~Dentinger, H.-T.~Ding et~al., \emph{{Meson screening masses in
  (2+1)-flavor QCD}},
  \href{https://doi.org/10.1103/PhysRevD.100.094510}{\emph{Phys. Rev.}
  {\bfseries D100} (2019) 094510}
  [\href{https://arxiv.org/abs/1908.09552}{{\ttfamily 1908.09552}}].

\bibitem{Fayazbakhsh:2013cha}
S.~Fayazbakhsh and N.~Sadooghi, \emph{{Weak decay constant of neutral pions in
  a hot and magnetized quark matter}},
  \href{https://doi.org/10.1103/PhysRevD.88.065030}{\emph{Phys. Rev.}
  {\bfseries D88} (2013) 065030}
  [\href{https://arxiv.org/abs/1306.2098}{{\ttfamily 1306.2098}}].

\bibitem{Gupta:1997nd}
R.~Gupta, \emph{{Introduction to lattice QCD: Course}},  in \emph{{Les Houches
  Summer School in Theoretical Physics, Session 68: Probing the Standard Model
  of Particle Interactions}}, pp.~83--219, 7, 1997
  [\href{https://arxiv.org/abs/hep-lat/9807028}{{\ttfamily hep-lat/9807028}}].

\bibitem{Johnson:1949wmq}
M.H.~Johnson and B.A.~Lippmann, \emph{{Motion in a Constant Magnetic Field}},
  \href{https://doi.org/10.1103/PhysRev.76.828}{\emph{Phys. Rev.} {\bfseries
  76} (1949) 828}.

\bibitem{Canuto:1968apg}
V.~Canuto and H.Y.~Chiu, \emph{{Quantum theory of an electron gas in intense
  magnetic fields}},
  \href{https://doi.org/10.1103/PhysRev.173.1210}{\emph{Phys. Rev.} {\bfseries
  173} (1968) 1210}.

\bibitem{Follana:2006rc}
{\scshape HPQCD, UKQCD} collaboration, \emph{{Highly improved staggered quarks
  on the lattice, with applications to charm physics}},
  \href{https://doi.org/10.1103/PhysRevD.75.054502}{\emph{Phys. Rev.}
  {\bfseries D75} (2007) 054502}
  [\href{https://arxiv.org/abs/hep-lat/0610092}{{\ttfamily hep-lat/0610092}}].

\bibitem{Luscher:1984xn}
M.~Luscher and P.~Weisz, \emph{{On-Shell Improved Lattice Gauge Theories}},
  \href{https://doi.org/10.1007/BF01206178}{\emph{Commun.Math.Phys.} {\bfseries
  97} (1985) 59}.

\bibitem{Weisz:1982zw}
P.~Weisz, \emph{{Continuum Limit Improved Lattice Action for Pure Yang-Mills
  Theory. 1.}}, \href{https://doi.org/10.1016/0550-3213(83)90595-3}{\emph{Nucl.
  Phys. B} {\bfseries 212} (1983) 1}.

\bibitem{Smit:1986fn}
J.~Smit and J.C.~Vink, \emph{{Remnants of the Index Theorem on the Lattice}},
  \href{https://doi.org/10.1016/0550-3213(87)90451-2}{\emph{Nucl. Phys. B}
  {\bfseries 286} (1987) 485}.

\bibitem{Al-Hashimi:2008quu}
M.H.~Al-Hashimi and U.J.~Wiese, \emph{{Discrete Accidental Symmetry for a
  Particle in a Constant Magnetic Field on a Torus}},
  \href{https://doi.org/10.1016/j.aop.2008.07.006}{\emph{Annals Phys.}
  {\bfseries 324} (2009) 343}
  [\href{https://arxiv.org/abs/0807.0630}{{\ttfamily 0807.0630}}].

\bibitem{Bazavov:2014cta}
A.~Bazavov, F.~Karsch, Y.~Maezawa, S.~Mukherjee and P.~Petreczky,
  \emph{{In-medium modifications of open and hidden strange-charm mesons from
  spatial correlation functions}},
  \href{https://doi.org/10.1103/PhysRevD.91.054503}{\emph{Phys. Rev. D}
  {\bfseries 91} (2015) 054503}
  [\href{https://arxiv.org/abs/1411.3018}{{\ttfamily 1411.3018}}].

\bibitem{Bazavov:2011nk}
A.~Bazavov, T.~Bhattacharya, M.~Cheng, C.~DeTar, H.-T.~Ding et~al., \emph{{The
  chiral and deconfinement aspects of the QCD transition}},
  \href{https://doi.org/10.1103/PhysRevD.85.054503}{\emph{Phys.Rev.} {\bfseries
  D85} (2012) 054503} [\href{https://arxiv.org/abs/1111.1710}{{\ttfamily
  1111.1710}}].

\bibitem{HotQCD:2014kol}
{\scshape HotQCD} collaboration, \emph{{Equation of state in ( 2+1 )-flavor
  QCD}}, \href{https://doi.org/10.1103/PhysRevD.90.094503}{\emph{Phys. Rev. D}
  {\bfseries 90} (2014) 094503}
  [\href{https://arxiv.org/abs/1407.6387}{{\ttfamily 1407.6387}}].

\bibitem{Mazur:2021zgi}
L.~Mazur, \emph{{Topological Aspects in Lattice QCD}}, Ph.D. thesis, Bielefeld
  U., 2021.
\newblock 10.4119/unibi/2956493.

\bibitem{Mazur_2024}
L.~Mazur, D.~Bollweg, D.A.~Clarke, L.~Altenkort, O.~Kaczmarek, R.~Larsen
  et~al., \emph{Simulateqcd: A simple multi-gpu lattice code for qcd
  calculations},
  \href{https://doi.org/10.1016/j.cpc.2024.109164}{\emph{Computer Physics
  Communications} {\bfseries 300} (2024) 109164}.

\bibitem{DElia:2021tfb}
M.~D'Elia, L.~Maio, F.~Sanfilippo and A.~Stanzione, \emph{{Confining and chiral
  properties of QCD in extremely strong magnetic fields}},
  \href{https://doi.org/10.1103/PhysRevD.104.114512}{\emph{Phys. Rev. D}
  {\bfseries 104} (2021) 114512}
  [\href{https://arxiv.org/abs/2109.07456}{{\ttfamily 2109.07456}}].

\bibitem{Ding:2025pbu}
H.-T.~Ding, J.-B.~Gu, S.-T.~Li and R.~Thakkar, \emph{{Chiral condensates and
  screening masses of neutral pseudoscalar mesons from lattice QCD at physical
  quark masses}},
  \href{https://doi.org/10.1103/PhysRevD.111.074513}{\emph{Phys. Rev. D}
  {\bfseries 111} (2025) 074513}
  [\href{https://arxiv.org/abs/2501.11262}{{\ttfamily 2501.11262}}].

\bibitem{Ding:2022tqn}
H.T.~Ding, S.T.~Li, J.H.~Liu and X.D.~Wang, \emph{{Chiral condensates and
  screening masses of neutral pseudoscalar mesons in thermomagnetic QCD
  medium}}, \href{https://doi.org/10.1103/PhysRevD.105.034514}{\emph{Phys. Rev.
  D} {\bfseries 105} (2022) 034514}
  [\href{https://arxiv.org/abs/2201.02349}{{\ttfamily 2201.02349}}].

\bibitem{FlavourLatticeAveragingGroupFLAG:2024oxs}
{\scshape Flavour Lattice Averaging Group (FLAG)} collaboration, \emph{{FLAG
  Review 2024}},  \href{https://arxiv.org/abs/2411.04268}{{\ttfamily
  2411.04268}}.

\bibitem{Miransky:2002rp}
V.A.~Miransky and I.A.~Shovkovy, \emph{{Magnetic catalysis and anisotropic
  confinement in QCD}},
  \href{https://doi.org/10.1103/PhysRevD.66.045006}{\emph{Phys. Rev. D}
  {\bfseries 66} (2002) 045006}
  [\href{https://arxiv.org/abs/hep-ph/0205348}{{\ttfamily hep-ph/0205348}}].

\bibitem{ParticleDataGroup:2024cfk}
{\scshape Particle Data Group} collaboration, \emph{{Review of particle
  physics}}, \href{https://doi.org/10.1103/PhysRevD.110.030001}{\emph{Phys.
  Rev. D} {\bfseries 110} (2024) 030001}.

\bibitem{Bali:2022qja}
{\scshape RQCD} collaboration, \emph{{Leading order mesonic and baryonic SU(3)
  low energy constants from Nf=3 lattice QCD}},
  \href{https://doi.org/10.1103/PhysRevD.105.054516}{\emph{Phys. Rev. D}
  {\bfseries 105} (2022) 054516}
  [\href{https://arxiv.org/abs/2201.05591}{{\ttfamily 2201.05591}}].

\bibitem{Orlovsky:2013gha}
V.D.~Orlovsky and Y.A.~Simonov, \emph{{Nambu-Goldstone mesons in strong
  magnetic field}}, \href{https://doi.org/10.1007/JHEP09(2013)136}{\emph{JHEP}
  {\bfseries 09} (2013) 136} [\href{https://arxiv.org/abs/1306.2232}{{\ttfamily
  1306.2232}}].

\bibitem{Ding:2022uwj}
H.-T.~Ding, S.-T.~Li, J.-H.~Liu and X.-D.~Wang, \emph{{Fluctuations of
  Conserved Charges in Strong Magnetic Fields from Lattice QCD}},
  \href{https://doi.org/10.5506/APhysPolBSupp.16.1-A134}{\emph{Acta Phys.
  Polon. Supp.} {\bfseries 16} (2023) 1}
  [\href{https://arxiv.org/abs/2208.07285}{{\ttfamily 2208.07285}}].

\bibitem{ding_2026_19718601}
H.-T.~Ding and D.~Zhang, \emph{{Dataset for Chiral Properties of (2+1)-Flavor
  QCD in Magnetic Fields at Zero Temperature}},  2026.
\newblock
  \href{https://doi.org/10.5281/zenodo.19718601}{10.5281/zenodo.19718601
  (2026).}

\end{thebibliography}\endgroup

\end{document}